\documentclass{ar-1col}
\usepackage{url}
\usepackage{graphicx}  
\usepackage{subcaption}
\usepackage{xcolor}
\usepackage{hyperref}
\usepackage[numbers,sort&compress]{natbib}
\usepackage{amsmath}
\usepackage{amssymb}
\usepackage{mathrsfs}
\usepackage{simplewick}
\allowdisplaybreaks

\DeclareMathSymbol{\NS}{\mathord}{AMSb}{"4E}

\newcommand{\nord}[1]{\ensuremath{\{\!#1\}}}

\newcommand{\half}{\ensuremath{\tfrac{1}{2}}}

\newcommand{\nuc}[2]{\ensuremath{^{#2}\mathrm{#1}}}

\newcommand{\fm}{\ensuremath{\,\text{fm}}}

\newcommand{\MeV}{\ensuremath{\,\text{MeV}}}

\newcommand{\gen}{\ensuremath{\mathcal{G}}}
\newcommand{\gensum}{\ensuremath{\mathscr{G}}}
\newcommand{\waveop}{\ensuremath{\Omega}}

\begin{document}

\markboth{Stroberg, Hergert, Bogner, and Holt}{Non-Empirical Interactions for the Nuclear Shell Model}

\title{Non-Empirical Interactions for the Nuclear Shell Model: An Update}

\author{ S.~Ragnar~Stroberg$^{1,2,3}$, Heiko~Hergert$^4$, Scott~K.~Bogner$^4$, and Jason~D.~Holt$^1$ 
\affil{$^1$TRIUMF, Vancouver BC, Canada V6T 2A3}
\affil{$^2$Physics Department, Reed College, Portland, OR 97202, USA }
\affil{$^3$Department of Physics, University of Washington, Seattle WA, USA}
\affil{$^4$ Facility for Rare Isotope Beams and Department of Physics \& Astronomy, Michigan State University, East Lansing MI 48824, USA}
}

\begin{abstract}
The nuclear shell model has been perhaps the most important conceptual and computational paradigm for the understanding of the structure of atomic nuclei.
While the shell model has been predominantly used in a phenomenological context, there have been efforts stretching back over a half century to derive shell model parameters based on a realistic interaction between nucleons.
More recently, several ab initio many-body methods---in particular many-body perturbation theory, the no-core shell model, the in-medium similarity renormalization group, and coupled cluster theory---have developed the capability to provide effective shell model Hamiltonians.
We provide an update on the status of these methods and investigate the connections between them and potential strengths and weaknesses, with a particular focus on the in-medium similarity renormalization group approach.
Three-body forces are demonstrated to be an important ingredient in understanding the modifications needed in phenomenological treatments.
We then review some applications of these methods to comparisons with recent experimental measurements, and conclude with some remaining challenges in ab initio shell model theory.

\end{abstract}


\maketitle

\tableofcontents

\section{\label{sec:Introduction}Introduction}

Since its introduction by Goeppert-Mayer and Jensen almost 70 years ago \cite{Mayer1948,Haxel1949,Goeppert-Mayer:1955ns}, the nuclear shell model has provided the primary conceptual framework for the understanding of nuclear structure. Its central idea is that protons and neutrons inside a nucleus move within a self-consistently generated mean field. This mean field can be approximated by Woods-Saxon or harmonic oscillator potentials, provided a strong spin-orbit component is added. With the inclusion of the latter, the spectrum of single-particle energies exhibits a shell structure that explains the experimentally observed magic proton and neutron numbers. In this picture, the low-lying structure of most nuclei results from the interactions between configurations of a few \emph{valence} particles on top of an inert core via a residual nuclear force.

From the beginning, it was hoped that the shell model and its residual force could be derived from basic building blocks, in particular the interaction between free protons and neutrons. Despite early successes, this proved to be an enormously difficult task \cite{Dawson1962,Bertsch1965,Kuo1966,Kuo:1967qf,Kuo:1968ty}. Meanwhile, increasingly elaborate empirical interactions were used with spectacular success to describe a vast array of experimental data~\cite{Cohen:1965it,Wildenthal1984,Caurier1999}. In modern language, this is a clear signal that the shell model contains the relevant degrees of freedom to describe most (if perhaps not all) phenomena observed in low-energy nuclear structure.

In the last two decades, a combination of computational and theoretical advances have provided fresh perspectives and opportunities for the systematic derivation of shell model interactions from realistic nuclear forces\footnote{By ``realistic'', we mean interactions that are rooted to some extent in Quantum Chromodynamics and accurately describe few-body scattering and bound-state data.}, without the need for ad hoc phenomenological adjustments. The primary aim of this review is to describe these developments. The story of the shell model, and of microscopically-derived effective interactions in particular, is long and full of false starts, dead ends, accidental successes, circling back, and rediscovery of old wisdom in new language. We are not equal to the task of writing an authoritative history, and we have not attempted to do so.
Indeed, we barely touch on recent developments in empirical shell model interactions or the computational aspects of configuration interaction calculations. Readers interested in such techniques should consult the existing texts on the subject~\cite{deShalit1963,Ring1980,Heyde1990,Talmi1993}.
For the status of phenomenological approaches, we refer to several thorough reviews~\cite{Brown2001,Talmi:2003kx,Caurier2005,Johnson2013,Brown:2014fk,Sorlin2008,Otsuka:2018bq}. 
For more details on effective interaction theory, we recommend several texts~\cite{Lindgren1986,Kuo1990,Shavitt2009}, and reviews~\cite{Kuo1974,Ellis1977,Hjorth-Jensen1995,Talmi:2003kx,Dean2004a,Coraggio2009}.

\subsection{\label{sec:History}The (Long) Road Towards \emph{Ab Initio} Shell Model Interactions}

Soon after the publication of the seminal works by Goeppert-Mayer and Jensen, first parameterizations of the residual nuclear force based on fits to experimental spectra began to appear (see, e.g., \cite{Talmi:1952lk,Lawson:1957kx,Talmi:1960yt,Talmi:1962wv,Cohen:1965it}). 
Dissatisfied with such approaches because they hide the link between the effective valence-space interaction and the underlying nucleon-nucleon interaction, Brown, Kuo and collaborators pioneered the program to systematically derive and explore this connection~\cite{Bertsch1965,Kuo1966,Kuo:1967qf}. In the 1960s, considerable progress was made in the construction of realistic nucleon-nucleon interactions that described NN scattering phase shifts with high quality \cite{Hamada:1962qt,Reid:1968dq}. Following a similar strategy as theoretical studies of nuclear matter, Kuo and Brown used Brueckner's $G$ matrix formalism \cite{Brueckner:1954qf,Brueckner:1955rw,Day:1967zl} to treat the strong short-range correlations induced by these forces, and included up to second-order diagrams in $G$ to account for important core-polarization effects \cite{Bertsch1965}. Their work culminated in the publication of Hamiltonians for the $sd$ and $pf$ shells \cite{Kuo:1967qf,Kuo:1968ty}. While first applications were successful, Barrett, Kirson, and others soon demonstrated a lack of convergence in powers of the $G$ matrix \cite{Barrett:1970jl,Goode:1974pi}, and more sophisticated treatments with RPA phonons and non-perturbative vertex corrections destroyed the good agreement with experiment \cite{Kirson:1971la,Barrett:1972bs,Kirson:1974oq}. Moreover, Vary, Sauer and Wong found that the use of larger model spaces for intermediate-state summations in diagram evaluations also degraded the agreement with experimental data \cite{Vary:1973dn}. Around the same time, Schucan and Weidenm{\"u}ller~\cite{Schucan1972a,Schucan1973} demonstrated that the presence of low-lying states of predominantly non-valence character (``intruder'' states) render the perturbative expansion for the effective interaction divergent.

Because of these developments, enthusiasm for the perturbative approach to deriving the effective interaction dwindled \cite{Barrett1975}. While Kuo and collaborators pursued the $\hat{Q}$-box (or folded-diagram) resummation of the perturbative series~\cite{Kuo1971,Kuo1974,Krenciglowa1975,Kuo1990,Hjorth-Jensen1995}, the majority of efforts in shell model theory were instead focused on the construction and refinement of empirical interactions~\cite{Whitehead1977,Chung1976,Poves1981,Wildenthal1984,Brown1988,Caurier1994}. Large-scale calculations with such interactions yielded impressive agreement with available nuclear data, and even provided predictive power. An example is the ``gold standard'' universal $sd$ shell interaction (USD)~\cite{Wildenthal1984,Brown1988a,Brown2006a}, which achieves a root-mean-square deviation from experimental levels of merely 130~keV throughout the $sd$ shell.

Meanwhile, serious efforts were undertaken to develop approaches that circumvent the problems plaguing the effective interaction methods by starting from the ``bare'' nuclear interactions and treating all nucleons as active particles. Prominent examples are coordinate-space Quantum Monte Carlo (QMC) techniques \cite{Pieper:2001ve,Carlson:2015lq,Lynn:2019rd}, or the no-core shell model (NCSM) \cite{Navratil:1996fk,Navratil:2000hf,Barrett:2013oq}. The late 1970s also saw a wave of nuclear Coupled Cluster calculations \cite{Kummel:1978jl,Zabolitzky:1979ye,Emrich:1981oj}, which use systematic truncations to solve the Schr\"odinger equation at polynomial cost, as opposed to the exponential scaling of the NCSM. By the mid-1990s, computational advances made quasi-exact calculations for nuclei feasible \cite{Zheng1993,Pudliner1995,Navratil:1996fk,Navratil:2000hf,Pieper:2001ve}. However, such calculations were limited to light nuclei by the sheer numerical cost of coordinate-space QMC, and the slow convergence of configuration-space methods with realistic nuclear interactions. 

\subsection{Fresh Perspectives}
Since the turn of the millennium, nuclear theory has undergone an important philosophical shift with the the adoption of renormalization group (RG) and effective field theory (EFT) concepts. These tools provide a systematic framework for exploring long-existing questions pertaining to the phase-shift equivalency of significantly different nucleon-nucleon interactions or the origin and importance of three-nucleon forces (see, e.g., \cite{Coester:1970tt,Bethe:1971qf}). Most noteworthy for the present work is the clarification of the issues that led to the failure of the aforementioned $G$-matrix based approaches, and the capability to reconcile the shell model, which is based on an (almost) independent-particle picture analogous to that of atomic physics, with the notion that strong correlations are induced by (most) realistic NN interactions.

\subsubsection{\label{sec:EFT}Chiral Effective Field Theory}

The essential idea of chiral effective field theory (EFT) for the nuclear force is that processes relevant for nuclear structure do not resolve the details of short-range interactions between nucleons\footnote{``Short'' in this context refers to distances $r$ where $k_{F}r\lesssim 1$, with $k_F\approx1.4\fm^{-1}$ the Fermi momentum at saturation density.}. There are in fact infinitely-many different potentials, differing at short distances, which all describe low-energy observables equally well. This is good news, because we can take advantage of this arbitrariness and parameterize the short-range physics in a convenient way, e.g., through a series of contact interactions. At long distances, the approximate chiral symmetry that chiral EFT inherits from Quantum Chromodynamics (QCD) dictates that interactions are described by (multi-)pion exchange.

In a pioneering work~\cite{Weinberg1990,Weinberg1991}, Weinberg developed effective Lagrangians to model the interaction between nucleons in terms of pion exchange and contact interactions, with increasingly complicated contributions suppressed by powers of a nucleon's typical momenta $Q\sim k_F$ or  the pion mass $m_\pi$ over the breakdown scale $\Lambda_\chi$ of the EFT, $(Q/\Lambda_{\chi})^n$. This provided a framework to treat two-body forces consistently with three- and higher-body forces, as well as a natural explanation for the relative importance of these terms (see e.g.~\cite{Epelbaum:2009ve,Machleidt:2011bh,Machleidt:2016yo} for recent reviews). Moreover, nuclear transition operators can be derived in a consistent fashion by coupling the chiral Lagrangian to the electroweak fields (see, e.g., \cite{Pastore:2009zr,Pastore:2011dq,Piarulli:2013vn,Kolling:2009yq,Kolling:2011bh}). Despite a number of subtle issues which persist to this day, several families of chiral two- plus three-nucleon interactions have been developed~\cite{Entem2002,Entem:2015qf,Entem:2015hl,Epelbaum:2015gf,Reinert:2018uq,Ekstrom:2015fk,Gezerlis:2014zr,Lynn:2016ec,Piarulli:2015rm,Piarulli:2018xi,Navratil2007,Gazit2009} that reproduce low-energy observables with an accuracy comparable with phenomenological potentials. These interactions have become the standard input for modern nuclear theory.

\subsubsection{\label{sec:srg}The Renormalization Group}

The renormalization group (RG), in particular in the formulation developed by Wilson~\cite{Wilson1974b,Wilson1975,Polchinski1984}, is a natural companion to any effective field theory. As discussed above, an EFT requires a cutoff $\Lambda$ which delineates between ``resolved'' and ``unresolved'' physics. The specific form and location of the cutoff (the ``scheme'' and ``scale'') is arbitrary and observables for momenta $Q\ll\Lambda$ should not depend on this choice. Consequently, there are an infinite number of equivalent theories which differ only in scheme and scale. The RG smoothly connects such equivalent theories.

RG methods debuted in low-energy nuclear physics around the turn of the millenium~\cite{Birse1999,Bogner2001,Barford2002,Bogner2003,Bogner2007,Bogner2007a,Bogner2010,Furnstahl:2013zt}, finally providing a systematic framework which formalized ideas that had been discussed in the nuclear structure community since the 1950s. For instance, both hard- and soft-core NN potentials can be devised which reproduce NN scattering data, but nuclear matter calculations found that soft potentials do not produce empirical saturation properties and so soft potentials were disfavored~\cite{Bethe:1971qf}. The missing piece in the saturation puzzle was the connection between the off-shell NN interaction and the 3N interaction, as demonstrated formally by Polyzou and Gl\"ockle~\cite{Polyzou:1990fk}. Of course, hard-core potentials are much more difficult to handle in many-body calculations, necessitating the use of Brueckner's $G$-matrix~\cite{Brueckner:1955rw,Bethe1956,Goldstone1957,Bethe1963} to deal with correlations due to the short-range repulsion. 

From the RG perspective, the hard- and soft-core potentials are related by an RG transformation which leaves NN scattering observables unchanged, but shifts strength into \emph{induced} 3N (and higher) interactions. Neglecting these induced terms means that observables involving more than two particles will no longer be preserved. This mechanism provides an explanation of the Phillips line~\cite{Phillips1969} and the Tjon line~\cite{Tjon1975}, which describe correlations between few-body observables calculated using different phase-equivalent NN interactions. 

In the context of this work, the RG provides a simple explanation of the observation that --- after being processed by the Brueckner $G$-matrix machinery --- various NN potentials produce very similar spectroscopy~\cite{Rustgi1971,Abzouzi1991}, as long as they reproduce NN scattering data. This can be understood as an indication that the $G$-matrix approximately ``integrates out'' the short-distance physics of the different potentials, leaving the universal long-distance physics. However, Bogner \emph{et al.} have shown that the $G$-matrix can retain significant coupling between off-shell low- and high-momentum modes, rendering it non-perturbative \cite{Bogner2010}. This explains why the historical efforts to construct the effective interaction perturbatively from $G$-matrices were bound to fail. In contrast, methods like the Similarity RG (SRG) \cite{Glazek:1993il,Wegner:1994dk,Bogner2010} achieve a more complete decoupling of the short-distance physics and render the resulting transformed NN+3N interaction suitable for perturbative expansions \cite{Bogner:2006qf,Bogner2010,Hoppe:2017fm,Tichai:2016vl}. The SRG has become the tool of choice in nuclear theory for decoupling low and high momenta because it also provides straightforward means to track induced many-body forces~\cite{Jurgenson2009,Hebeler:2012ly,Wendt:2013ys,Roth:2014fk}, and construct consistently transformed observables \cite{Anderson:2010br,Schuster:2014oq,Parzuchowski2017a}.

\subsection{Current Status of the \emph{ab initio} Shell Model}
Over the past decade, the adaptation of EFT and SRG methods has greatly extended the reach of \emph{ab initio} nuclear many-body theory across the nuclear chart. Simply put, the most convenient scale for formulating a theory of nuclear interactions is often not the most convenient scale for solving that theory. The SRG connects one scale to the other and greatly improves the convergence behavior of nuclear many-body calculations in the process. Large-scale diagonalization methods like the NCSM can be used in the lower $sd-$shell \cite{Barrett:2013oq,Jurgenson:2013fk,Hergert2013a,Roth:2014fk}, and systematically truncated methods like Self-Consistent Green's Functions, Coupled Cluster (CC) and the In-Medium SRG (IMSRG) can even be applied to nuclei as heavy as tin \cite{Binder:2014fk,Hagen:2014ve,Hagen:2016xe,Morris2018}. While SRG-evolved interactions cannot be used easily in QMC due to their nonlocality, new families of local chiral interactions yield encouraging results in such applications \cite{Gezerlis:2014zr,Lynn:2016ec,Piarulli:2018xi,Lynn:2019rd}.

Soon after their introduction to nuclear physics, EFT and RG methods also revitalized efforts to systematically derive shell model interactions \cite{Coraggio:2003gj,Holt:2005mi}. From a practical perspective, this offered a convenient way to confront RG-evolved chiral two- plus three-nucleon interactions with the wealth of available spectroscopic data, using existing shell model codes \cite{Otsuka2010a,Holt:2012fk,Holt:2013cr,Holt:2013fk,Holt:2013hc,Holt:2014vn,Tsunoda2014,Bogner2014,Jansen2014,Simonis:2016uq,Stroberg2016,Stroberg2017,Tsunoda:2017sh}. At the conceptual level, these interactions validate the independent-particle picture underlying the shell model. They provide sufficient binding already at the mean-field level, and allow us to use it as the starting point for the treatment of correlations, either through rapidly converging non-perturbative many-body methods \cite{Hergert2016,Hergert2017,Hagen:2014ve}, or possibly even through finite-order perturbation theory \cite{Tichai:2016vl,Tichai2018}. Furthermore, novel approaches like the valence-space IMSRG (VS-IMSRG) or shell model Coupled Cluster (SMCC), both discussed below, provide both the conceptual framework and practical tools to relate no-core and valence-space methods, as shown by the consistent ground- and excited-state results obtained thus far (see \cite{Stroberg2017,Morris2018} and Sec.~\ref{sec:Applications}). Thus, the end of the long and winding road to \emph{ab initio} shell model interactions appears to be in sight, though challenges remain, see Secs. \ref{sec:Challenges} and \ref{sec:Conclusions}. 

\subsection{Organization of this Work}
This work is organized as follows: In Sec.~\ref{sec:Methods}, we introduce common approaches to the construction of shell model interactions, from the traditional many-body perturbation theory and the Okubo-Lee-Suzuki method to the valence-space IMSRG and shell model CC.
In Sec.~\ref{sec:Comparison}, these approaches are compared within a common formalism to illuminate the relations between them. Section~\ref{sec:3body} discusses the role of three-nucleon forces in the shell model context, and relates modifications of (semi-)empirical interactions that are supposed to capture such effects to the more systematic treatment of these forces in modern approaches.
In Sec.~\ref{sec:Applications}, we highlight selected applications of \emph{ab initio} shell model interactions. Section \ref{sec:Challenges} describes the main challenges we are facing today, and analyzes them primarily from the perspective of the VS-IMSRG. New developments like a direct EFT expansion for shell model interactions and a novel uncertainty quantification effort are touched upon in Sec.~\ref{sec:OtherDev}. Section~\ref{sec:Conclusions} provides concluding remarks alongside a list of take-away messages that summarize the key aspects of modern \emph{ab initio} shell model calculations, and clarify common misconceptions. Certain technical details are collected in the appendices.

\section{\label{sec:Methods}Microscopic effective interactions}
The general problem of effective interaction theory is the following:
Given a Hamiltonian $H$ expressed in a large (typically intractable) Hilbert space $\mathscr{H}$, we wish to obtain an effective Hamiltonian $H_{\mathrm{eff}}$ which acts in a smaller (tractable) Hilbert space $\mathscr{H}_{\mathrm{model}}$, but reproduces a subset of the eigenstates of the large Hilbert space.

\begin{equation}
\underset{\textrm{Full space Schr\"odinger eq.}}{\underbrace{
H|\Psi_n\rangle = E_n |\Psi_n\rangle}}
\hspace{1em}
\Rightarrow
\hspace{1em}
\underset{\textrm{Model space Schr\"odinger eq.}}{\underbrace{
H_{\text{eff}}|\psi_n\rangle = E_n|\psi_n\rangle}}
\end{equation}

In the context of the nuclear shell model, the large Hilbert space will consist of Slater determinants of single-particle states, typically harmonic oscillator eigenstates.
The number of single-particle states included should be sufficient to obtain convergence.
The smaller Hilbert space $\mathscr{H}_{\mathrm{model}}$ is defined by splitting the single-particle states into three categories---core, valence, and excluded\footnote{In chemistry, valence orbits are usually called active states, while excluded orbits are referred to as virtual states.}--- taking the subset of Slater determinants for which all core orbits are occupied, and all excluded orbits are unoccupied.

We note in passing that different partitionings of the Hilbert space can be used for other purposes.
For example, choosing $\mathscr{H}_{\mathrm{model}}$ to be a one-dimensional space corresponds to the single-reference many-body methods for treating the ground state of closed-shell nuclei.
Alternatively, defining $\mathscr{H}_{\mathrm{model}}$ in terms of low-momentum states leads to methods for ``softening'' an interaction, such as $V_{\mathrm{low}k}$~\cite{Bogner2003} or the similarity renormalization group (SRG)~\cite{Bogner2007}.

An important practical requirement on $H_{\mathrm{eff}}$ is that it should obey a rapidly converging cluster expansion, schematically
\begin{equation}\label{eq:ClusterInequalities}
    |V_{2N}|\gg |V_{3N}| \gg |V_{4N}| \ldots
\end{equation}
where $V_{2N}$ is the two-body potential, $V_{3N}$ is the three-body potential, etc. and the vertical bars indicate some measure of size or importance.
This property is essential to the feasibility of large-scale shell model diagonalizations. Nowadays, such calculations can handle basis dimensions upwards of $10^9$ Slater determinants~\cite{Caurier2005,Brown:2014fk,Forssen2018},
which would be impossible if the full matrix needed to be stored.
The limitation to two-body (or possibly three-body) interactions yields a sparse matrix which can be treated efficiently by, e.g., the Lanczos or Davidson methods \cite{Lanczos:1950sp,Arnoldi:1951kk,Davidson:1989pi}.

It is worth taking a moment here to motivate why we would expect the cluster expansion to be valid in nuclei, and to consider where it might run into trouble.
As mentioned in section~\ref{sec:EFT}, chiral effective field theory naturally generates a hierarchy of the type shown in Eq.~ (\ref{eq:ClusterInequalities}), with many-body interactions suppressed by increasing powers of the ratio of low to high scales.
On the other hand, for a system of $A$ nucleons, the importance of an $n$-body term grows combinatorially, accounting for all the different combinations of $n$ particles which can interact, suggesting that for heavy nuclei, many-body forces will dominate.
For large $A$, this grows as $A!/n!(A-n)!\sim A^{n}$.

Fortunately, we are saved by the short range of the nuclear force and the relatively low saturation density of nuclear matter~\cite{Friman2011}.
Roughly, each nucleon does not interact with all other nucleons, but instead only with the other nucleons within some interaction volume ${\mathcal{V}\sim\tfrac{4\pi}{3}r_{\mathrm{int}}^3}$ with $r_\mathrm{int}$ the range of the interaction.
At density $\rho$, the expectation value of an $n$-body force will scale as $\langle V_n\rangle \sim (\rho \mathcal{V})^{n-1}$.
With $\rho\lesssim 0.16\fm^{-3}$, the cluster hierarchy will be maintained so long as $r_{\mathrm{int}}\lesssim1\fm$.
This is satisfied for the short-range terms in a chiral force, which are characterized by a cutoff scale on the order of 0.5$\fm$ or less.
The long pion-exchange tail, with range $r_{\pi}\sim 1.5\fm$, is not obviously suppressed or enhanced by the density, though its contribution to bulk binding is somewhat suppressed due to its spin-isopin structure averaging to zero in symmetric spin-saturated matter. 
However, if we use Wick's theorem to express operators in normal ordered form with respect to a finite-density reference (see sections~\ref{sec:NormalOrdering} and~\ref{sec:ENO}), then the appropriate density is not saturation density $\rho$, but the \emph{quasiparticle} density which will typically be significantly smaller~\cite{Friman2011}.

When we derive an effective interaction, we are eliminating degrees of freedom, namely orbits outside of the valence space.
For high-lying orbits, the relevant interaction matrix element will be dominated by high-momentum (short-distance) components, in which case the above argument holds and the induced terms should still exhibit a cluster hierarchy~\cite{Bogner2008}.
However, for excitations near the fermi surface we have no short-distance argument. Indeed, as we discuss in section~\ref{sec:E2}, low-lying collective excitations can be a source of trouble.

It is important to keep in mind that the above argument only holds if the observable in question can be expressed in terms of connected diagrams (i.e. one can trace a continuous path through the diagram between any two points on it).
If a four-body term consists of two disconnected two-body terms, then there is no reason why all four particles would need to be within some interaction volume.
We return to this point in Sec.~\ref{sec:ValenceCluster}.

In the remainder of this section, we describe the most popular approaches to deriving effective interactions for the nuclear shell model, using the notation that appears in the nuclear physics literature.
In section~\ref{sec:Comparison}, we treat these methods within a more general framework to illuminate the relationships between them.

\subsection{\label{sec:Qbox}Quasidegenerate perturbation theory and the $\hat{Q}$-box resummation}
Let us introduce the projection operator $P$ and its complement $Q$ such that $P\mathscr{H}P=\mathscr{H}_{\mathrm{model}}$ and $P+Q=1$.
If we assume $|\psi_n\rangle = P|\Psi_n\rangle$, that is, that the eigenstates of the effective Hamiltonian are just the projection of the full eigenstates onto the model space, then
the effective Hamiltonian should satisfy
\begin{equation}\label{eq:Decoupling}
    PH_{\mathrm{eff}}P|\Psi_n\rangle = E_n P |\Psi_n\rangle
    \hspace{1em},\hspace{1em}
    QH_{\mathrm{eff}}P = 0.
\end{equation}
Straightforward manipulation then yields the Bloch-Horowitz energy-dependent effective Hamiltonian \cite{Bloch1958,Feshbach1958}
\begin{equation} \label{eq:HBlochHorowitz}
    H_{\mathrm{BH}}(E_n) = PHP + PHQ\frac{1}{E_n-QHQ}QHP.
\end{equation}
One important aspect of Eq.~(\ref{eq:HBlochHorowitz}) is that the effective interaction depends on the eigenvalue $E_n$, and so it must be solved self-consistently.
A second point is that different valence-space eigenstates will in general not be orthogonal, because they are eigenstates of different operators.
The energy dependence can be removed by expanding the denominator about some starting energy $E_0$, yielding ~\cite{Brandow1967,Tsunoda2014}
\begin{equation}\label{eq:HeffEKK}
    H_{\mathrm{eff}} = H_{\mathrm{BH}}(E_0) + \sum_{k=1}^{\infty}\frac{1}{k!} \left[ \frac{d^k}{d E_0^k} H_{\mathrm{BH}}(E_0)\right] (H_{\mathrm{eff}}-E_0)^k.
\end{equation}
The expression (\ref{eq:HeffEKK}) may also be obtained in the context of time-dependent perturbation theory~\cite{Kuo1990}, or by a similarity transformation combined with an iterative solution~\cite{Suzuki1980} for the decoupling condition (\ref{eq:Decoupling}).

A simplification may be obtained by partitioning the Hamiltonian into a zero-order piece and a perturbation $H=H_0+V$, and assuming that the eigenvalues of $H_0$ in the valence space are degenerate, with energy $\epsilon$.
Then one uses Eq.~(\ref{eq:HeffEKK}) with $E_0=\epsilon$, and $(H_{\mathrm{eff}}-\epsilon)=V_{\mathrm{eff}}$.
In this context, a popular approach is to define the ``$\hat{Q}$-box''~\cite{Kuo1971,Krenciglowa1975,Kuo1990}, indicated $\hat{Q}(\epsilon)$.
\begin{equation}\label{eq:Qbox}
    \hat{Q}(\epsilon) = PVP + PVQ\frac{1}{\epsilon-QHQ}QVP.
\end{equation}
The operator $\hat{Q}(\epsilon)$ is not to be confused with the projection operator $Q$.
The effective valence space interaction is, in analogy to Eq.~(\ref{eq:HeffEKK}),
\begin{equation}\label{eq:VeffQbox}
    V_{\mathrm{eff}} = \hat{Q}(\epsilon) + \sum_{k=1}^{\infty} \frac{1}{k!}\frac{d^k\hat{Q}(\epsilon)}{d\epsilon^k} (V_{\mathrm{eff}})^k.
\end{equation}
The commonly adopted strategy for evaluating Eq.~(\ref{eq:VeffQbox}) is to expand the inverse operator in Eq.~(\ref{eq:Qbox}) perturbatively (typically to second or third order in $V$) and to solve Eq.~(\ref{eq:VeffQbox}) self-consistently by iteration, evaluating the derivatives numerically by calculating $\hat{Q}(\epsilon)$ for several starting energies $\epsilon$.

A technical point arises because the $\hat{Q}$-box contains one-body pieces, arising from e.g. core polarization diagrams \cite{Bertsch1965,Kuo1966,Kuo:1967qf}.
For computational convenience, the one-body part is embedded in the two-body part, with an accompanying spectator nucleon.
This leads to disconnected two-body terms in Eq.~(\ref{eq:VeffQbox}) which contain arbitrary numbers of interactions involving one-particle, but no interactions between the two.
These disconnected contributions can be understood as the dressed one-body part of the effective interaction embedded into a two-body interaction.
They can be removed by solving Eq.~(\ref{eq:VeffQbox}) using only the one-body piece of the $\hat{Q}$-box, which is called the $\hat{S}$-box~\cite{Shurpin1977,Shurpin1977a}.
The resulting effective one-body interaction is subtracted off from $V_{\mathrm{eff}}$, leaving only connected terms.
In principle these same diagrams should then be added self-consistently to the degenerate single-particle energies $\varepsilon_i = \langle i | H_0 | i \rangle + \hat{S}_{i}(\varepsilon_i)$.
In practice, it is often discarded and the single particle energies are taken from experiment\footnote{Determining the experimental single-particle energies is ambiguous, as one must choose which experimental state is ``the'' shell model one, or consider an average of multiple states (see, e.g., \cite{Wildenthal1984,Brown1988a})}.
As an exception to this, when three-body forces are included in the normal ordering approximation (see Secs.~\ref{sec:NormalOrdering} and \ref{sec:3body}), and the single-particle energies are computed so that the starting energy $\epsilon$ corresponds to the centroid of the valence-space single-particle energies, no additional adjustments are required~\cite{Holt:2013hc,Holt:2014vn}.

\subsubsection{Extended Kuo-Krenciglowa approach}
In general, one would like to be able to use non-degenerate valence orbits, for example when using a Hartree-Fock basis, or when generating an interaction for two (or more) major harmonic oscillator shells.
Then one should use Eqs.~(\ref{eq:HBlochHorowitz}) and (\ref{eq:HeffEKK}) rather than Eqs.~(\ref{eq:Qbox}) and (\ref{eq:VeffQbox}).
This approach has only recently been explored~\cite{Takayanagi2011,Takayanagi2011a,Tsunoda2014,Tsunoda:2017sh}.
In the case of a valence space spanned by two major oscillator shells, taking harmonic oscillator single-particle energies can lead to zero energy denominators in Eq.~(\ref{eq:VeffQbox}), while the parameter $E$ in Eq.~(\ref{eq:HeffEKK}) may be chosen to avoid zero denominators.
Of course, for a $P$ space with a very large spread in single-particle energies, one would expect that the size of certain contributions to $(H_{\mathrm{eff}}-E_0)^k$ in Eq.~(\ref{eq:HeffEKK}) would be comparable to the corresponding energy denominators, and the Taylor series (\ref{eq:HeffEKK}) might converge slowly, if at all.
However, for a modest spread of energies ($\sim$10 MeV), this does not appear to be a problem~\cite{Tsunoda2014,Tsunoda:2017sh}.

\subsection{\label{sec:OLS}Okubo-Lee-Suzuki transformation}
Another approach to the effective interaction is called the Okubo-Lee-Suzuki\footnote{Perhaps this would more appropriately be called the Okubo-Lee-Suzuki-Okamoto approach.} approach, often employed in conjunction with the NCSM~\cite{Navratil1997,Navratil:2000hf,Barrett:2013oq}, the paradoxical-sounding ``NCSM with a core'' or Double OLS approach \cite{Lisetskiy2008,Lisetskiy2009,Dikmen2015}, or more recently, the coupled cluster method~\cite{Jansen2014,Jansen2016} (``coupled cluster effective interaction'' (CCEI)).
The idea is to obtain a unitary transformation $\mathcal{U}$ which diagonalizes $H$ in the large Hilbert space $\mathscr{H}$, so that
\begin{equation}
    \mathcal{U} H \mathcal{U}^{\dagger} = \mathcal{E}
\end{equation}
where $\mathcal{E}$ is a diagonal matrix.
The effective Hamiltonian $H_{\mathrm{eff}}$ acting in the smaller Hilbert space $\mathscr{H}_{\mathrm{model}}=P\mathscr{H}P$ is given by~\cite{Lisetskiy2008}
\begin{equation}\label{eq:OLS_Heff}
    H_{\mathrm{eff}} = \frac{\mathcal{U}_P^{\dagger}}{\sqrt{\mathcal{U}^{\dagger}_P\mathcal{U}_P}} \mathcal{E} \frac{\mathcal{U}_P}{\sqrt{\mathcal{U}^{\dagger}_P\mathcal{U}_P}}
\end{equation}
where $\mathcal{U}_P\equiv P\mathcal{U}P$ is the projection of the transformation $\mathcal{U}$ onto the model space.
One can easily confirm that the transformation in Eq.~(\ref{eq:OLS_Heff}) is unitary, and so the eigenvalues of $H_{\mathrm{eff}}$ in the model space will be a subset of the eigenvalues of $H$ in the full space.

So far, this does not appear to be a helpful procedure, because the first step was to solve the eigenvalue problem in the large Hilbert space, and the goal of effective interaction theory is to allow applications for which the full solution is not tractable.
The benefit comes when one assumes that the effective interaction $H_{\mathrm{eff}}$ will also provide a good approximation for other systems described in the same model space.
Assuming a cluster expansion, one obtains $H_{\mathrm{eff}}$ for a few active particles in the model space and then applies $H_{\mathrm{eff}}$ to systems with more active particles.

As a concrete example, consider the system of two particles in the no-core shell model.
This can readily be diagonalized in a large space of many harmonic oscillator states, say $N\leq N_{\mathrm{max}}=500$, where $N=2n+\ell$ is the number of oscillator quanta.
On the other hand, even a light nucleus like $^{6}$Li with only six particles cannot possibly be diagonalized in such a space.
Using the OLS transformation (\ref{eq:OLS_Heff}) one can obtain an effective interaction for a manageable model space (say $N_{\mathrm{max}}=10$) which exactly reproduces the low-lying eigenvalues of the two-body calculation in the large space.
The application of this effective interaction to $^{6}$Li then provides a reasonable approximation to the eigenvalues one would obtain in the large space.

\begin{figure}[ht]
    \centering
    \includegraphics[width=\textwidth]{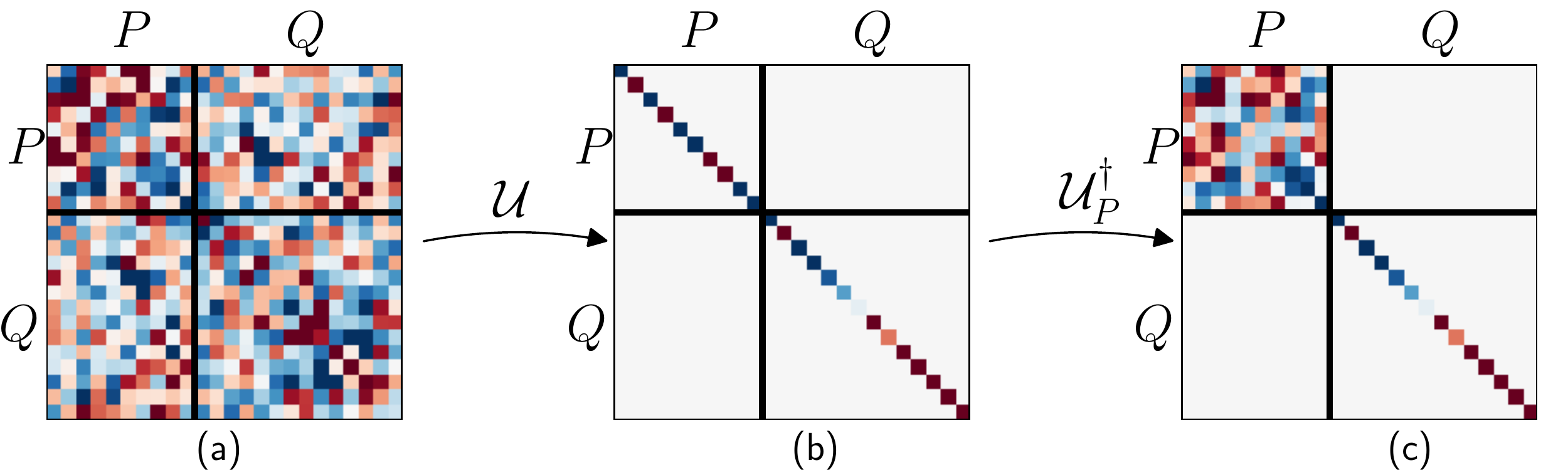}
    \caption{A schematic of how the OLS approach obtains the effective interaction. (a) The original Hamiltonian. (b) The Hamiltonian is diagonalized by transformation $\mathcal{U}$. (c) The transformation $\mathcal{U}_P^{\dagger}$ approximately inverts $\mathcal{U}$ in the $P$ space and yields the effective interaction $H_{\mathrm{eff}}$.}
    \label{fig:OLSschemetic}
\end{figure}

We note that there are essentially two main assumptions here:
 \emph{(i)} the effective interaction one would obtain if one could apply the OLS procedure directly to the six-body system has a rapidly convergent cluster expansion $|V_{2N}|>|V_{3N}|>|V_{4N}|\ldots$, and
\emph{(ii)} the two-body-cluster component $V_{2N}$ of the full effective interaction $H_{\mathrm{eff}}$ for the six-body system is well approximated by the effective interaction obtained for the two-body system.
While both of these assumptions are plausible and encouraging results have been obtained using them, we know of no rigorous proof.
Indeed, as we discuss briefly in Sec.~\ref{sec:ValenceCluster}, there is potential cause for concern related to disconnected diagrams.

\subsection{\label{sec:imsrg}In-medium similarity renormalization group}
In the in-medium similarity renormalization group (IMSRG)\cite{Tsukiyama:2011uq,Tsukiyama:2012fk,Bogner2014,Stroberg2017,Hergert2017}, the effective Hamiltonian is also expressed in terms of a unitary transformation $U$ acting on the initial Hamiltonian
\begin{equation}
    H_{\mathrm{eff}} = U H U^{\dagger}.
\end{equation}
In contrast to the OLS approach, the IMSRG transformation is obtained without solving the eigenvalue problem for a particular many-body system.
Instead, it is parameterized by a continuous flow parameter $s$, and applied to the Hamiltonian through the flow equation
\begin{equation} \label{eq:IMSRGflow}
    \frac{dH(s)}{ds} = [\eta(s),H(s)]\,,
\end{equation}
where the generator $\eta(s)$ is formally defined as
\begin{equation}
    \eta(s) \equiv \frac{dU(s)}{ds}U^{\dagger}(s) = -\eta^{\dagger}(s).
\end{equation}
We split the flowing Hamiltonian $H(s)$ into two pieces, a ``diagonal'' and an ``off-diagonal'' piece
\begin{equation}
    H(s) = H^{d}(s) + H^{od}(s)
\end{equation}
such that
\begin{equation}
  \begin{aligned}
    H^{od}(s) &= PH(s)Q + QH(s)P.
    \end{aligned}
\end{equation}
where the projection operators $P$ and $Q$ have the same meaning as in the previous sections.
Our goal is then to devise a generator $\eta(s)$ such that
\begin{equation}
   \underset{s\rightarrow \infty}{\lim} H^{od}(s) = 0
\end{equation}
and therefore
\begin{equation}
    \underset{s\rightarrow\infty}{\lim} H^{d}(s) = H_{\mathrm{eff}}.
\end{equation}
In the language of the renormalization group, $H_{\mathrm{eff}}$ is a fixed point of the RG flow.

One choice for $\eta(s)$, which is used in the calculations we will describe here is the White generator~\cite{White2002,Hergert2016}
\begin{equation}\label{eq:WhiteGen}
    \eta^{\mathrm{Wh}}(s) \equiv \frac{H^{od}(s)}{\Delta(s)}\,.
\end{equation}
For present and future use, we have introduced a convenient superoperator notation (cf.~\cite{Klein1974}), in which we indicate division of the operator $O$ by a suitably defined energy denominator $\Delta$ is defined as
\begin{equation}
     \langle \phi_i | \frac{O}{\Delta} | \phi_j  \rangle \equiv \frac{\langle \phi_i | O | \phi_j \rangle}{\epsilon_i - \epsilon_j } 
\end{equation}
which can be thought of as element-wise division.
Here $\epsilon_i,\epsilon_j$ are energies associated with the basis states $\phi_i,\phi_j$. The quantity $\frac{O}{\Delta}$ itself is an operator whose Fock-space expression is
\begin{equation}\label{eq:superop}
    \frac{O}{\Delta} = \sum_{ij} \frac{O_{ij}}{\epsilon_i-\epsilon_j} a^{\dagger}_ia_j
    +\frac{1}{4}\sum_{ijkl} \frac{O_{ijkl}}{\epsilon_i+\epsilon_j-\epsilon_k-\epsilon_l} a^{\dagger}_i a^{\dagger}_j a_l a_k + \ldots
\end{equation}

Returning to the flow equation, it is clear that if $H^{od}\rightarrow 0$, then $\eta\rightarrow 0$ and by Eq.~(\ref{eq:IMSRGflow}) we see that $\frac{dH(s)}{ds}\rightarrow 0$, so $H_{\mathrm{eff}}$ is indeed a fixed point of the flow.
One potential issue with the generator (\ref{eq:WhiteGen}) is that a vanishing energy denominator will cause $\eta$ to diverge.
An alternative, also suggested by White~\cite{White2002} (see also~\cite{Suzuki1977}), is
\begin{equation}\label{eq:AtanWhiteGen}
    \eta^{\mathrm{atan}}(s) \equiv \frac{1}{2}\mathrm{atan}\left(\frac{2 H^{od}(s)}{\Delta(s)}\right).
\end{equation}
The arctangent---motivated by the solution of a 2$\times$2 system via Jacobi rotations---regulates the divergent behavior of Eq.~(\ref{eq:WhiteGen}) in the presence of small denominators.
The arctangent and division by the energy denominator in Eq.~ (\ref{eq:AtanWhiteGen}) should be interpreted as operating element-wise, as described above.

\begin{figure}[ht]
    \centering
    \includegraphics[width=\textwidth]{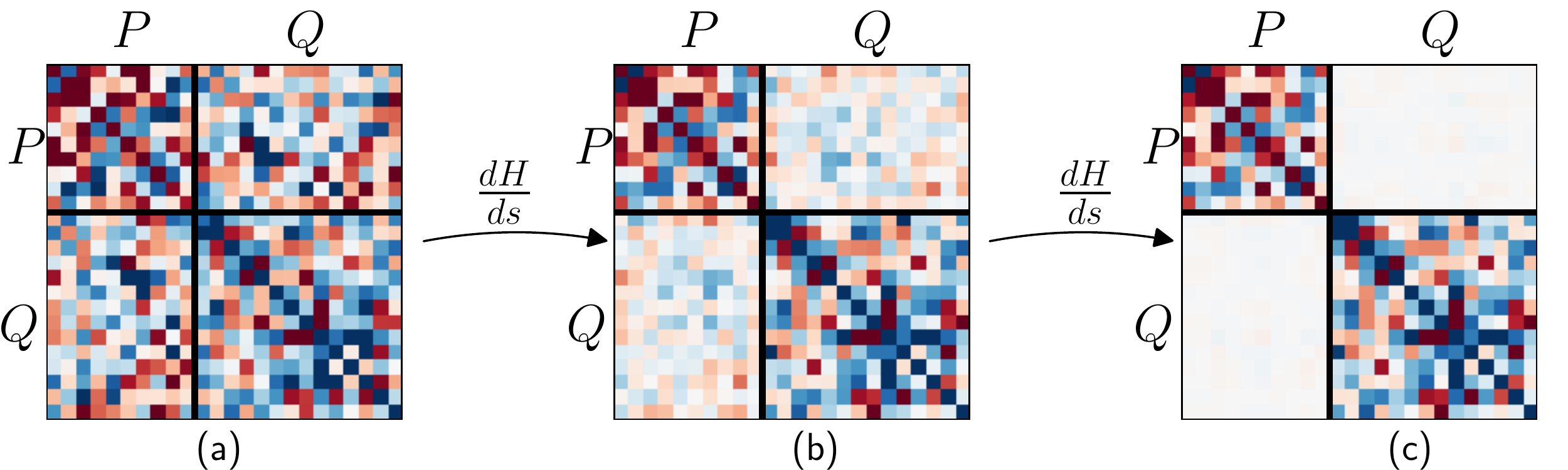}
    \caption{A schematic representing of how the IMSRG approach obtains the effective interaction $H_{\mathrm{eff}}$ by progressively suppressing the off-diagonal terms of $H$. (a)$s=0$, (b)$s=5$, (c)$s=30$}
    \label{fig:IMSRGschemetic}
\end{figure}

The IMSRG is formulated in terms of Fock-space operators, and so its computational cost scales polynomially with the basis size $N$, but not explicitly with the number of particles being treated. In practical applications, we truncate all operators at a consistent particle rank to close the system of flow equations arising from Eq.~(\ref{eq:IMSRGflow}) (see Appendix~\ref{app:IMSRGFlow}). We also set up the decoupling conditions to be minimally invasive to avoid an uncontrolled accumulation of truncation errors, as discussed in detail in Ref.~\cite{Hergert2016}. In VS-IMSRG we perform the decoupling in two stages for this reason, decoupling the reference state from excitations as in a direct ground-state calculation before decoupling the valence space in a second evolution (cf.~Secs.~\ref{sec:Applications} and \ref{sec:Challenges}). 

\subsubsection{\label{sec:NormalOrdering}Normal ordering}
An important feature of the IMSRG method is the use of operators in \emph{normal-ordered form} (see, e.g., \cite{Hergert2016,Hergert2017}).
Starting with the free-space Hamiltonian written as a Fock-space operator with two- and three-body interactions~\footnote{In actual calcualtions, one subtracts off the center-of-mass kinetic energy, and so the kinetic term has a two-body piece~\cite{Hergert2017}. We neglect that here for simplicity.}
\begin{equation}
    H = \sum_{ij}t_{ij}a^{\dagger}_ia_j
    + \tfrac{1}{4}\sum_{ijkl}V_{ijkl}a^{\dagger}_ia^{\dagger}_ja_la_k
    + \tfrac{1}{36}\sum_{\substack{ijk\\lmn}} V_{ijklmn}a^{\dagger}_ia^{\dagger}_ja^{\dagger}_k a_na_ma_l,
\end{equation}
we may use Wick's theorem to express the strings of creation and annihilation operators in normal order with respect to some reference state $|\Phi\rangle$ \cite{Wick:1950fr}.
We denote the normal ordering with braces, and the normal order of a pair of operators is defined so that their expectation in the reference is zero, i.e.
\begin{equation}
    \langle \Phi | \nord{a^{\dagger}_ia_j} | \Phi\rangle = 0.
\end{equation}
Whether the normal order is $a^{\dagger}_ia_j$ or $a_ja^{\dagger}_i$ depends on whether or not the states created and annihilated are present in the reference $|\Phi\rangle$.
If we choose $|\Phi\rangle$ to be a single Slater determinant such as the Hartree-Fock ground state of the system of interest, then application of Wick's theorem allows us to write $H$ as
\begin{equation}
    H = E_0 + \sum_{ij}f_{ij}\nord{a^{\dagger}_ia_j}
    +\tfrac{1}{4}\sum_{ijkl}\Gamma_{ijkl}\nord{a^{\dagger}_ia^{\dagger}_ja_la_k}
    + \tfrac{1}{36}\sum_{\substack{ijk\\lmn}} W_{ijklmn}\nord{a^{\dagger}_ia^{\dagger}_ja^{\dagger}_k a_na_ma_l}
\end{equation}
where the new coefficients can be obtained from the old coefficents by
\begin{equation}\label{eq:NOops}
    \begin{aligned}
    &E_0 = \sum_{a}n_a t_{aa} + \tfrac{1}{2}\sum_{ab}n_an_b V_{abab} + \tfrac{1}{6}\sum_{abc}n_an_bn_cV_{abcabc} \\
    &f_{ij} =t_{ij} + \sum_{a}n_a V_{iaja} + \tfrac{1}{2}\sum_{ab}n_an_bV_{iabjab} \\
    &\Gamma_{ijkl} = V_{ijkl} + \sum_a n_a V_{ijakla}\\
    &W_{ijklmn} = V_{ijklmn}.
    \end{aligned}
\end{equation}
Operators other than the Hamiltonian can be rewritten in the same way.

In Eq.~(\ref{eq:NOops}), $n_a$ is the occupation of orbit $a$ in the reference, i.e. $n_a=\langle \Phi | a^{\dagger}_aa_a | \Phi\rangle$, and for a Slater determinant reference, $n_a$ is either zero or one.
In section~\ref{sec:ENO}, we discuss a different choice of reference for which $a$ can have fractional occupation.
One may also use a correlated reference, constructed out of a linear combination of Slater determinants, in which case one must use the generalized normal ordering of Kutzelnigg and Mukherjee~\cite{Kutzelnigg1997}. This is the basis of the  multireference IMSRG (MR-IMSRG) method that will be used in ground-state energy comparisons in later sections \cite{Hergert2013a,Hergert2017}.

The advantage of expressing operators in normal ordered form is that it puts as much information as possible from the high particle-rank (i.e. many-body) operators into the lower rank operators.
This is evident in Eq.~(\ref{eq:NOops}), where the normal-ordered zero-body term $E_0$ contains contributions from the free one\nobreakdash-, two\nobreakdash- and three-body terms.
If the reference $|\Phi\rangle$ is a good approximation of the exact wave function $|\Psi\rangle$, then the expectation value
${\langle \Psi | \nord{a^{\dagger}a^{\dagger}a^{\dagger}aaa}|\Psi\rangle\approx 0}$, and even formally non-vanishing 3N interactions $W_{ijklmn}$ can be neglected to a good approximation.
Consequently, normal ordering may be thought of as a way to improve the convergence of the cluster expansion described in the beginning of this section. 

\subsubsection{\label{sec:MagnusIMSRG}Magnus formulation}
A particularly convenient formulation of the IMSRG approach relies on the Magnus expansion~\cite{Morris2015,Morris2016}.
The idea is to express the more general unitary IMSRG transformation as the true exponential of the anti-Hermitian Magnus operator $\Omega(s)=-\Omega^{\dagger}(s)$. The evolved Hamiltonian can then be expressed in terms of an infinite series of nested commutators
\begin{equation}\label{eq:MagnusBCH}
\begin{aligned}
    H(s) &= e^{\Omega(s)} H(0) e^{-\Omega(s)} \\
    & = H(0) + [\Omega(s),H(0)] + \frac{1}{2}\left[\Omega(s),[\Omega(s),H(0)]\right] + \ldots
    \end{aligned}
\end{equation}
This formulation of the IMSRG allows for a more transparent comparison with canonical transformation theory \cite{White2002,Yanai2007} and unitary coupled cluster method~\cite{Taube2006} used in quantum chemistry, as well as canonical perturbation theory~\cite{Shavitt1980}, where the expansion in Eq.~(\ref{eq:MagnusBCH}) is evaluated perturbatively.

Considering the flow equation (\ref{eq:IMSRGflow}), we see that under an infinitesimal step $ds$ we may write
\begin{equation}
\begin{aligned}
    H(s+ds) &= H(s) + [\eta(s),H(s)]ds\\
    &= e^{\eta(s)ds} H(s) e^{-\eta(s)ds}\\
    &= e^{\eta(s)ds}e^{\Omega(s)} H(0) e^{-\Omega(s)}e^{-\eta(s)ds}\,.
    \end{aligned}
\end{equation}
Expressing $H(s+ds)$ in Magnus form as well, we obtain an expression for $\Omega(s+ds)$:
\begin{equation}
    e^{\Omega(s+ds)} =  e^{\eta(s)ds}e^{\Omega(s)}.
\end{equation}
Because $\Omega(s)$ and $\eta(s)$ do not in general commute, we use the Baker-Campbell-Hausdorff formula to take the logarithm on both sides and obtain
\begin{equation}\label{eq:MagnusFlow}
    \Omega(s+ds) = \Omega(s) + \eta(s)ds + \tfrac{1}{2}[\eta(s),\Omega(s)]ds+ \tfrac{1}{12}[\Omega(s),[\Omega(s),\eta(s)]]ds
    +\ldots
\end{equation}
This may be expressed in compact form as (suppressing explicit $s$ dependence)
\begin{equation}
    \frac{d\Omega}{ds} = \sum_{k=0}^{\infty} \frac{B_k}{k!}\mathrm{ad}^{(k)}_{\Omega}(\eta)
\end{equation}
where $B_k$ are the Bernoulli numbers, and the adjoint $\mathrm{ad}^{(k)}_{\Omega}(\eta)$ signifies a recursively defined  nested commutator:
\begin{equation}\label{eq:MagnusAdjointFlow}
    \mathrm{ad}^{(k)}_{\Omega}(\eta) = [\Omega,\mathrm{ad}^{(k-1)}_{\Omega}(\eta)]
    \hspace{1em},\hspace{2em}
    \mathrm{ad}^{(0)}_{\Omega}(\eta) = \eta.
\end{equation}
Fortunately, in most practical applications, only the first few terms of the infinite series in Eqs.~(\ref{eq:MagnusBCH}) and (\ref{eq:MagnusFlow}) are important, and so the commutator may be evaluated iteratively until the size of a given term is below some numerical threshold (for an example of an exception to this, see Sec.~\ref{sec:Intruder}).
A major practical advantage of the Magnus method is that by solving for $\Omega(s)$, we can compute arbitrary effective operators besides the Hamiltonian in a consistent and efficient way (see Sec.~\ref{sec:E2} and Refs.~\cite{Morris2015,Parzuchowski2017a}).

\subsection{\label{sec:SMC}Shell model coupled cluster}

\begin{figure}[ht]
    \centering
    \includegraphics[width=\textwidth]{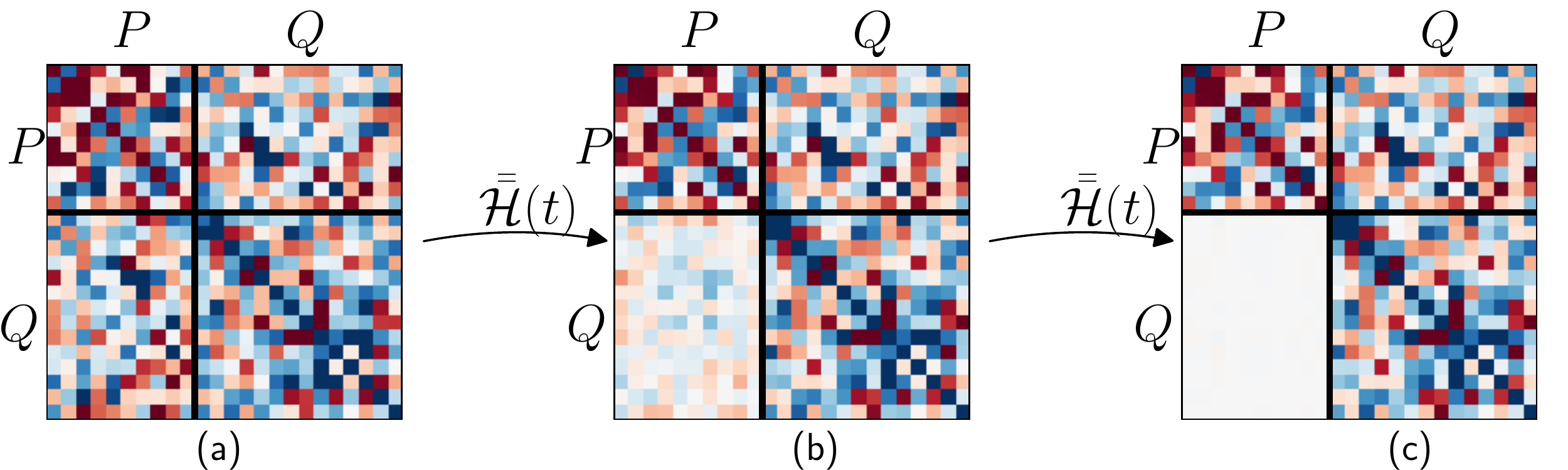}
    \caption{A schematic depicting how the SMCC approach obtains the effective interaction $\bar{\bar{\mathcal{H}}}(t)$. Note that, in contrast to the $H_{\mathrm{eff}}$ of Fig.~\ref{fig:IMSRGschemetic}, $\bar{\bar{\mathcal{H}}}$ is non-Hermitian. (a)$t=0$, (b)$t=5$, (c)$t=30$}
    \label{fig:CCSMschemetic}
\end{figure}

First attempts to derive an effective shell model interaction with coupled cluster methods were similar in spirit to the NCSM-based Double OLS approach (see Sec.~\ref{sec:OLS} and Refs.~\cite{Navratil1997,Lisetskiy2008,Lisetskiy2009,Dikmen2015}). Equation-of-motion CC (EOM-CC) states defined in a space of up to 4p2h excitations are subsequently projected into the shell model space via the OLS method, yielding the so-called coupled cluster effective interaction (CCEI) \cite{Jansen2014,Jansen2016}. The cost of the EOM-CC calculations, however, presented a significant obstacle to widespread applications of this method. A much more efficient alternative is the recently introduced shell model coupled cluster (SMCC) method \cite{Sun2018}, which is formulated in Fock space and can be viewed as a non-unitary cousin of the IMSRG. 

In the standard coupled cluster method~\cite{Shavitt2009,Hagen:2014ve}, a similarity transformation is performed to decouple a single closed-shell reference state $|\Phi\rangle$ from all particle-hole excitations
\begin{equation}
    e^{-T} H e^{T}|\Phi\rangle = E_{\mathrm{corr}} e^{T}|\Phi\rangle 
\end{equation}
where $E_{\mathrm{corr}}$ is the correlation energy and $T$ is the cluster operator which is written as
\begin{equation}
    T = \sum_{ia} t_{i}^{a} \{a^{\dagger}_a a_i\}
    + \tfrac{1}{4}\sum_{abij} t_{ij}^{ab} \{a^{\dagger}_aa^{\dagger}_b a_j a_i\} + \ldots
\end{equation}
Here the indices $a,b,c\ldots$ denote unoccupied orbits and $i,j,k\ldots$ denote occupied orbits.
The similarity-transformed Hamiltonian is written as
\begin{equation}\label{eq:def_Hcc}
    \bar{\mathcal{H}} = e^{-T}He^{T}.
\end{equation}
Shell model coupled cluster extends the idea by performing a similarity transformation which decouples a valence space rather than a single configuration.
Denoting this transformation with an $S$, we have
\begin{equation}
    \bar{\bar{\mathcal{H}}} = e^{-S}\bar{\mathcal{H}}e^{S}
    \hspace{1em},
    \hspace{2em}
    Q \bar{\bar{\mathcal{H}}}P = 0. \label{eq:smcc_heff}
\end{equation}
The operator $S$ is obtained by a flow equation closely mirroring the one used in IMSRG
\begin{equation}
    \frac{dS}{dt} = -\eta(\bar{\bar{\mathcal{H}}}(t))
\end{equation}
where $\eta$ is the generator of the flow. As in the IMSRG, there is considerable freedom for choosing $\eta$ as long as the decoupling condition \eqref{eq:smcc_heff} is realized in the limit $t\to\infty$. In Ref.~\cite{Sun2018}, adapted variants of the White (Eq.~\eqref{eq:WhiteGen}) and arctangent generators (Eq.~\eqref{eq:AtanWhiteGen}) are used.
The essential difference from the IMSRG formulation is that $T$ and $S$ are not anti-Hermitian operators, and so the transformation is not unitary, and the resulting effective Hamiltonian is not Hermitian.
The inconvenience of a non-Hermitian Hamiltonian is compensated by the greater simplicity of the equations which need to be solved.

\section{\label{sec:Comparison}Comparison of various approaches to effective interactions}
To investigate how the methods described in section~\ref{sec:Methods} (and a few others) are related to one another, 
we consider the general structure of effective interactions, and show how the above methods sum the perturbation series.
For more details, we refer readers to Refs.~\cite{Klein1974,Shavitt1980,Kutzelnigg1982,Kutzelnigg1983,Suzuki1983,Suzuki1984}.

\subsection{Formal effective interaction theory}
\subsubsection{\label{sec:GeneralConsiderations}General considerations}
We begin by expressing the effective Hamiltonian in terms of a similarity transformation of the original Hamiltonian, parametrized as the exponential of a generator $\gen$
\begin{equation}\label{eq:BCH}
\begin{aligned}
    H_{\mathrm{eff}} &= e^{\gen}He^{-\gen} \\
   &= H + [\gen,H] + \half\left[\gen,[\gen,H]\right]+\ldots   
    \end{aligned}
\end{equation}
and the decoupling condition
\begin{equation}\label{eq:decoupleHeff}
    QH_{\mathrm{eff}}P =0.
\end{equation}
We partition the original Hamiltonian into an exactly-solvable zero-order part $H_0$ and a perturbation $V$
\begin{equation}
    H = H_0 + V
\end{equation}
and consider an expansion of the generator $\gen$ and the interaction $H_{\mathrm{eff}}$ in powers of $V$
\begin{equation}\label{eq:GperturbativeSeries}
    \gen = \gen^{[1]} + \gen^{[2]} + \gen^{[3]} + \ldots
    \hspace{1em},\hspace{1.5em}
    H_{\mathrm{eff}}=H_{\mathrm{eff}}^{[0]} + H_{\mathrm{eff}}^{[1]}+ H_{\mathrm{eff}}^{[2]}+\ldots
\end{equation}
For convenience, we define the partial sum of the series up to order $n$ as 
\begin{equation}\label{eq:gensum}
    \gensum^{[n]} \equiv \sum_{m=1}^{n}\gen^{[m]}.
\end{equation}
The $n$th order contribution to the effective Hamiltonian is then
\begin{equation}\label{eq:Heffn}
    H_{\mathrm{eff}}^{[n]} =\left( e^{\gensum^{[n]}} H e^{-\gensum^{[n]}} \right)^{[n]}.
\end{equation}
One can easily verify that $\gen^{[n]}$ only contributes to a single term in Eq.~(\ref{eq:Heffn}).
Peeling this term off we have
\begin{equation}
    H_{\mathrm{eff}}^{[n]} = [\gen^{[n]},H_0] + \left( e^{\gensum^{[n-1]}} H e^{-\gensum^{[n-1]}} \right)^{[n]}.
\end{equation}
Enforcing the decoupling condition (\ref{eq:decoupleHeff}) yields an equation for $\gen^{[n]}$ in terms of lower-order contributions
\begin{equation}\label{eq:GnPerturbative}
    Q[H_0,\gen^{[n]}]P = Q\left( e^{\gensum^{[n-1]}} H e^{-\gensum^{[n-1]}}\right)^{[n]}P.
\end{equation}
Equation \eqref{eq:GnPerturbative} is of the general form
\begin{equation}\label{eq:XY_comm}
    [H_0,X] = Y\,.
\end{equation}
If we work in the eigenbasis of $H_0$ such that $H_0|\phi_i\rangle = \epsilon_i|\phi_i\rangle$, the commutator can be easily evaluated in terms of the unperturbed energies:
\begin{equation}
    \langle \phi_i | [H_0,X] | \phi_j \rangle = (\epsilon_i-\epsilon_j) \langle \phi_j | X | \phi_j\rangle \equiv \Delta_{ij} \langle \phi_j | X | \phi_j\rangle.
\end{equation}
This suggests that a solution to Eq.~\eqref{eq:XY_comm} can be written as
\begin{equation}\label{eq:XYZ}
    X = \frac{Y}{\Delta} + Z\,,
\end{equation}
where $Z$ is some arbitrary function which commutes with $H_0$, and the superoperator notation introduced in Eq.~\eqref{eq:superop} is used for brevity.

Following this line of reasoning we solve Eq.~(\ref{eq:GnPerturbative}) as\footnote{The additional term like $Z$ in Eq.~(\ref{eq:XYZ}), which commutes with $H_0$, will vanish when sandwiched between $Q$ and $P$, so we need not include that term here.}
\begin{equation}\label{eq:QGPperturbative}
    Q\gen^{[n]}P = Q\frac{\left( e^{\gensum^{[n-1]}} H e^{-\gensum^{[n-1]}}\right)^{[n]}}{\Delta}P.
\end{equation}
As we can see, the decoupling condition applies to the $Q\gen P$ block of the generator $\gen$, and we have some freedom to choose the rest of $\gen$, namely $P\gen P$, $P\gen Q$, $Q\gen Q$.
The various choices, which we will outline below, result in different effective Hamiltonians.
There are a few important consequences of these choices.

First, in order for the transformation to be unitary, we need the generator to be anti-Hermitian: $\gen=-\gen^{\dagger}$. Consequently, the popular choice $P\gen Q=0$ will result in a non-unitary transformation, and a non-Hermitian effective Hamiltonian.
All else being equal, a Hermitian effective Hamiltonian is preferable, but the significant simplifications that come with taking $P\gen Q=0$ can make this choice attractive.

\begin{figure}[ht]
    \centering
    \begin{subfigure}{0.18\textwidth}
    \centering
    \includegraphics[width=1.0\textwidth]{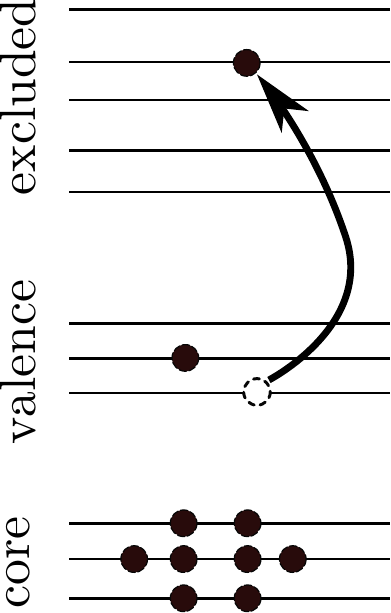}
    \end{subfigure}
    \hspace{0.15\textwidth}
    \begin{subfigure}{0.18\textwidth}
    \centering
    \includegraphics[width=1.0\textwidth]{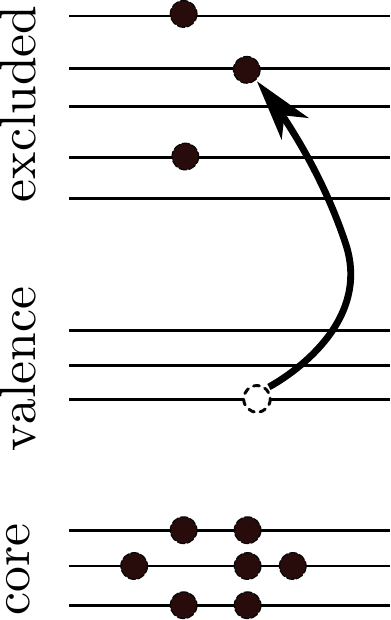}
    \end{subfigure}
    \caption{A schematic illustration showing that the same operator (indicated by the arrow) can connect \emph{Left:} a $P$ space configuration to a $Q$ space configuration as well as \emph{Right:} a $Q$ space configuration to a $Q$ space configuration.}
    \label{fig:PtoQshells}
\end{figure}

Second, the choice $Q\gen Q=0$ cannot be enforced in a Fock-space formulation, and so this choice can only be made when working directly in the $A$-body Hilbert space.
To understand this point~\footnote{To our knowledge, this point was first made in the chemistry literature by Kutzelnigg~\cite{Kutzelnigg1982,Kutzelnigg1983}, but it has not been explicitly stated in the nuclear physics literature.} consider a one-body Fock-space operator which excites a particle from the valence space to the excluded space.
As illustrated in Fig.~\ref{fig:PtoQshells}, if this operator acts on a configuration which belongs to the $P$ space, it will generate a configuration which belongs to the $Q$ space.
However, if that same operator acts on a $Q$-space configuration which already has some other particle-hole excitation, it will generate a distinct $Q$-space configuration.
So the operator also connects $Q$-space configurations to $Q$-space configurations.
In order to enforce $Q\gen Q=0$ while allowing $Q\gen P\neq 0$, the operator $\gen$ needs to ensure that it acts only on $P$-space states, which means $\gen$ must be an $A$-body operator.

Third, if we choose both $P\gen P=0$ and $P\gen Q=0$, then we have
\begin{equation}\label{eq:GenBlochP}
    P e^{\gen} = P(1+\gen + \tfrac{1}{2}\gen^2+\ldots) = P.
\end{equation}
If we denote the eigenstate of the full Hamiltonian by $|\Psi_i\rangle$ and the corresponding eigenstate of the effective Hamiltonian $|\psi_i\rangle$, these are related by the similarity transformation
\begin{equation}
    |\Psi_i\rangle = e^{-\gen} |\psi_i\rangle,
\end{equation}
and since $|\psi_i\rangle$ exists entirely in the $P$ space, Eq.~ (\ref{eq:GenBlochP}) implies that
\begin{equation}
    |\psi_i\rangle = P|\Psi_i\rangle.
\end{equation}
Physically, the eigenstate of the effective Hamiltonian is just given by the projection of the full-space wave function to the $P$ space.
This means that $|\Psi_i\rangle$ and $|\psi_i\rangle$ cannot be simultaneously normalized to one, and we must employ an ``intermediate normalization'' $\langle \Psi_i|\psi_i\rangle =1.$
On the other hand, if we do not require $P\gen Q=0$, then the above argument no longer holds and the eigenstates of the effective Hamiltoninan will in general not be simply projections of the full-space eigenstates.

Finally, before investigating various choices of $\gen$, we consider an iterative method for summing the perturbative series to all orders.
To do this, we notice that selecting out the $n$th order contribution on the right-hand side of Eq.~(\ref{eq:QGPperturbative}) quickly leads to complicated formulas (see, e.g., section III of Ref.~\cite{Klein1974}).
Overall, the right-hand side of Eq.~(\ref{eq:QGPperturbative}) is of order $n$ and higher, since the lower order terms in $\gen$ have been selected to eliminate the undesired components of $H_{\mathrm{eff}}$ to their respective orders.
This means that if we do not specifically select the $n$th order terms on the right hand side, but instead take everything, we obtain a contribution $\gen^{[n]}$ which suppresses the $n$th order term in $H_{\mathrm{eff}}$, as well as some contribution from higher order terms.
These higher order terms will be suppressed during later iterations.
Because this is no longer a strict order-by-order perturbative expansion, we use a subscript to denote the $n$th iteration, to distinguish it from the superscript indicating the $n$th-order contribution:
\begin{equation}
    Q\gen_n P = Q\frac{e^{\gensum_{n-1}} H e^{-\gensum_{n-1}}}{\Delta}P.
\end{equation}
Here, we have defined $\gensum_{n} \equiv \sum_{m=1}^{n}\gen_m$ in analogy with Eq.~(\ref{eq:gensum}).
Defining the transformed Hamiltonian after $n$ iterations as $H_n\equiv e^{\gensum_n}He^{-\gensum_n}$, this may be written as
\begin{equation}
    Q\gen_n P = Q \frac{H_{n-1}}{\Delta} P.
\end{equation}
or
\begin{equation}\label{eq:GnIter}
    Q \gensum_{n}P = Q\gensum_{n-1}P + Q\frac{H_{n-1}}{\Delta} P.
\end{equation}
Beginning with $\gensum_0=0$ and iterating Eq.~(\ref{eq:GnIter}) successively eliminates contributions to $QH_{\mathrm{eff}}P$ of increasing powers of the perturbation $V$, and for $n\rightarrow\infty$ yields the exact generator: $\gensum_{\infty}=\gen$.

In the following subsections, we consider choices found in the literature for fully specifying $\gen$, and the consequences of these choices.

\subsubsection{\label{sec:LeeSuzuki}Lee-Suzuki}
We begin with the most restrictive combination $Q\gen Q=P\gen P=P\gen Q =0$, which allows for the greatest simplification.
Following the notation of Suzuki and Lee~\cite{Suzuki1980}, for this choice we write the generator as $\gen=-\omega$.
The effective Hamiltonian is
\begin{equation} \label{eq:omegadecouple}
    H_{\mathrm{eff}} = e^{-\omega}He^{\omega}
\end{equation}
and we have the great simplification that $\omega^2=0$ so that $e^{\omega}=1+\omega$.
As we remarked above, this choice will yield a non-Hermitian effective Hamiltonian which is necessarily formulated in the $A$-body Hilbert space.
The decoupling condition is
\begin{equation}\label{eq:LeeSuzuki1}
\begin{aligned}
    QH_{\mathrm{eff}}P &= (1-\omega)H(1+\omega) \\
    &= QVP +QH\omega P - Q\omega H_0P -Q\omega VP - Q\omega V \omega P = 0,
    \end{aligned}
\end{equation}
where in the second line we have used $H=H_0+V$.
If we take a degenerate $P$ space with energy $\epsilon$ so that $Q\omega PH_0P=\epsilon Q\omega P$,
we may rearrage to solve for $\omega$
\begin{equation}\label{eq:LeeSuzukiOmega}
    \omega = \frac{1}{\epsilon-QHQ}\left(QVP - \omega (PVP + PVQ\omega) \right).
\end{equation}
As shown by Suzuki and Lee~\cite{Suzuki1980},
defining
\begin{equation}
    H_{\mathrm{eff}} = PH_{0}P + PVP + PVQ\omega
\end{equation}
and iteratively inserting Eq.~(\ref{eq:LeeSuzukiOmega}) into itself yields the $Q$-box folded diagram expansion (\ref{eq:VeffQbox}) or, with a different iteration scheme,
\begin{equation}
    R_{n} = \frac{1}{1-\hat{Q}_1-\sum\limits_{m=2}^{n-1}\hat{Q}_{m} \prod\limits_{k=n-m+1}^{n-1}R_k}\hat{Q}
\end{equation}
where $V_{\mathrm{eff}}=R_{\infty}$, $\hat{Q}$ is the same $\hat{Q}$-box defined in Eq.~(\ref{eq:Qbox}), and $\hat{Q}_m \equiv \frac{d^m}{d\epsilon^m}(\hat{Q}(\epsilon)$.

As we have pointed out in section~\ref{sec:GeneralConsiderations}, the requirement $Q\omega Q=0$ implies that this approach must be formulated in the $A$-body space.
However, as we see next, a Fock-space formulation yields the same effective interaction.

\subsubsection{\label{sec:GenBloch}Generalized Bloch equation}
If we drop the requirement $Q\gen Q=0$, we have
$P\gen Q = P \gen P = 0$.
Following convention, we express the transformation in terms of the M{\o}ller wave operator and its inverse\footnote{We use $\Omega$ for the M{\o}ller wave operator for consistency with existing literature in the present subsection, but caution that it should not be confused with the Magnus operator used in the IMSRG.}
\begin{equation}
    \waveop = e^{-\gen}\hspace{1em},\hspace{1em} \waveop^{-1}=e^{\gen}.
\end{equation}

Noting that $Q\waveop^{-1}(1-\waveop P)=Q\waveop^{-1}$,
we can satisfy the decoupling condition (\ref{eq:decoupleHeff})
if
\begin{equation}
    (1-\waveop P)H\waveop P=0.
\end{equation}
Again, splitting up $H=H_0+V$ and rearranging, this may be written as
\begin{equation}\label{eq:GenBloch}
    Q[\waveop,H_0]P = QV \waveop P -Q\waveop PV\waveop P
\end{equation}
which is the generalized Bloch equation~\cite{Lindgren1974a}.

The effective interaction in the $P$ space is given by
\begin{equation}\label{eq:GenBlochHeff}
    PH_{\mathrm{eff}}P = PH\waveop P = PH_0P + PV\waveop P.
\end{equation}
Expanding Eqs.~(\ref{eq:GenBloch}) and (\ref{eq:GenBlochHeff}) in powers of $V$ yields a linked expansion for the Rayleigh-Schr\"odinger perturbation series~\cite{Brandow1967,Lindgren1986}.
In the nuclear case, the order-by-order convergence of this series is questionable (cf. Sec.~\ref{sec:Introduction}).

The second term on the right hand side of Eq.~(\ref{eq:GenBloch}) can be represented by ``folded'' diagrams~\cite{Brandow1967,Lindgren1986}.
The $\hat{Q}$-box method described in section~\ref{sec:Qbox} amounts to a perturbative expansion of the first term in Eq.~(\ref{eq:GenBloch}), followed by a summation of certain higher-order terms in the folded diagram series.
We note that if we express the wave operator as $\waveop=P+\chi$, where $\chi$ is called the ``correlation operator'', then Eq.~(\ref{eq:GenBloch}) is equivalent to Eq.~(\ref{eq:LeeSuzuki1}) with $\chi=\omega$.
Interestingly, while the Lee-Suzuki approach required $Q\omega Q=0$, and therefore could not be expressed in a Fock-space formulation, the Bloch equation approach does not use that constraint and so may be formulated in Fock-space.
Evidently, the Bloch equation approach does not make any reference to $Q\gen Q$, and so setting it to zero does not alter the resulting effective Hamiltonian.
For a discussion of the differences between the Lee-Suzuki scheme and the $\hat{Q}$-box approach, see e.g. Ref.~\cite{Hjorth-Jensen1995}.

\subsubsection{\label{sec:SMCC}Shell model coupled cluster}
Instead of working with the wave operator, we may directly work with the generator $\gen$.
Following the notation of Ref.~\cite{Sun2018}, we write $\gen$ as $-S$.
We employ the iterative procedure laid out in section~\ref{sec:GeneralConsiderations} to obtain an iterative expression for $S$:
\begin{equation}\label{eq:SMCCiter}
    S_n = S_{n-1} - Q \frac{H_{n-1}}{\Delta}P
\end{equation}
where we have denoted the similarity-transformed Hamiltonian\footnote{In practical applications, the procedure is applied to the CC Hamiltonian \eqref{eq:def_Hcc}, (c.f. Sec.~\ref{sec:SMC})} at the $n$th step $H_n = e^{-S_n}He^{S_n}$.
The effective Hamiltonian is then $H_{\mathrm{eff}} = H_{\infty}$.
In arriving at Eq.~(\ref{eq:SMCCiter}), we have implicitly assumed that the terms in $QSQ$ are only those which also contribute to $Q S P$.
Consequently, if a term in $S$ only connects $Q$ configurations, then it is taken to be zero.

Often, the iterations are better behaved with the help of a convergence factor, which we denote $dt$ to connect with the formulation of Ref.~\cite{Sun2018}. Multiplying the second term of Eq.~(\ref{eq:SMCCiter}) and taking the limit $dt\rightarrow 0$, we can reinterpret it as a flow equation:
\begin{equation}
    \frac{dS}{dt} = -\eta(t) \equiv -Q\frac{H(t) }{\Delta}P.
\end{equation}
Taking the $P$ space to be a single Slater determinant, Eq.~(\ref{eq:SMCCiter}) yields an iteration scheme to solve the coupled cluster equations, while defining the $P$ space in terms of a valence space yields the shell model coupled cluster approach~\cite{Sun2018} described in section~\ref{sec:SMCC}.
As discussed above, the requirement $P S Q=0$ means that $H_{\mathrm{eff}}$ is not Hermitian in this approach.
Additionally, we see that the SMCC effective interaction is equivalent to the other two non-Hermitian effective interactions discussed in sections~\ref{sec:LeeSuzuki} and~\ref{sec:GenBloch}, as long as no approximations are made.

\subsubsection{\label{sec:nspaceCanonical}Canonical perturbation theory}
We next drop the condition $Q \gen P = 0$, enabling us to enforce $\gen^{\dagger}=-\gen$ so that the transformation is unitary and $H_{\mathrm{eff}}$ is Hermitian.
We retain the condition $Q\gen Q=0$, with the consequence that we cannot express the theory in terms of Fock-space operators.
However, this restriction greatly simplifies the analysis.
For consistency with the literature, we write the generator as $\gen = -G$.
It can be shown~\cite{Shavitt1980,Suzuki1982a} that the operator $G$ is related to the operator $\omega$ from section~\ref{sec:LeeSuzuki} by\footnote{Here, the hyperbolic arctangent of an operator is defined in terms of its Taylor series expansion~\cite{Shavitt1980}.}
\begin{equation}
    G = \mathrm{arctanh}\left(\omega-\omega^{\dagger}\right)
\end{equation}
and the transformation is~\cite{Suzuki1982a}
\begin{equation}
    e^{G}=(1+\omega-\omega^{\dagger})(1+\omega \omega^{\dagger} +\omega^{\dagger}\omega)^{-1/2}.
\end{equation}
The effective Hamiltonian obtained in this approach is the Hermitized version of the effective Hamiltonian resulting from Eq.~(\ref{eq:LeeSuzuki1}).
To connect this result with the OLS approach presented in section~\ref{sec:OLS}, we write that transformation out explicitly:
\begin{equation}
    H_{\mathrm{eff}} = \frac{\mathcal{U}_P^{\dagger}}{\sqrt{\mathcal{U}_P^{\dagger}\mathcal{U}_P}}\mathcal{U}H\mathcal{U}^{\dagger}\frac{\mathcal{U}_P}{\sqrt{\mathcal{U}_P^{\dagger}\mathcal{U}_P}}.
\end{equation}
By inserting a sum over the eigenstates of $H$, and using $|\Psi_i\rangle=(1+\omega)P|\Psi_i\rangle$, one can show~\cite{Navratil1997,Lisetskiy2008} that $\mathcal{U}^{\dagger}\mathcal{U}_P=(1+\omega)P$, and the OLS effective interaction is equivalent to $Pe^{-G}He^{G}P$.

An iterative scheme very similar to the one described in section \ref{sec:GeneralConsiderations} was proposed by Suzuki~\cite{Suzuki1977}, but not directly pursued further.
The unitary model operator approach (UMOA)~\cite{Suzuki1982a} follows this formalism, with the valence cluster expansion (see section~\ref{sec:ValenceCluster}) carried out on the generator, rather than on the effective Hamiltonian.
So far, studies with the UMOA have focused on ground state energies of closed-shell nuclei~\cite{Miyagi2017}, so we will not discuss it further here.

\subsubsection{Fock-space canonical perturbation theory\label{sec:FockCanonicalPT}}
If we desire a Hermitian effective operator with a Fock-space decomposition, then we should drop the restriction $Q\gen Q=0$, leaving only $P\gen P=0$.
Writing out Eq.~(\ref{eq:QGPperturbative}) order-by-order with the requirement $\gen^{\dagger}=-\gen$ yields the canonical perturbation theory of Primas~\cite{Primas1963} and Klein~\cite{Klein1974}.
Interestingly~\cite{Kutzelnigg1983}, the resulting expansion is different from the expansion obtained in section \ref{sec:nspaceCanonical}, i.e. the Fock-space and $A$-body space formulations are not equivalent, in contrast to what was found for the non-Hermitian formulation.
This approach has not been pursued in the nuclear physics literature.

\subsubsection{Unitary coupled cluster}

Alternatively, we may take $P\gen P=0$ and $\gen^{\dagger}=-\gen$ and follow the iterative procedure of section~\ref{sec:GeneralConsiderations} to obtain
\begin{equation}
    \gen_n = \gen_{n-1} + Q\frac{H_{n-1}}{\Delta}P + P\frac{H_{n-1}}{\Delta}Q.
\end{equation}
This yields a unitary coupled cluster expansion for the effective interaction.
As with the SMCC solution, this may be recast as a differential equation (here we use $s$ instead of $t$)
\begin{equation}\label{eq:UCCflow}
    \frac{d \gen(s)}{ds} = Q\frac{H(s)}{\Delta}P + P\frac{H(s)}{\Delta}Q.
\end{equation}
This approach has also not been persued in nuclear physics, although it is very closely related to the Magnus formulation of the IMSRG, as we will show below.

\subsubsection{Magnus IMSRG}
Finally, we may drop the constraint $P\gen P=0$ and instead specify $\gen$ by the requirement that it should reproduce the flowing Hamiltonian $H(s)$ along its entire trajectory.
Following Ref.~\cite{Morris2015}, we write $\gen=\Omega(s)$, which we call the Magnus operator, and require
\begin{equation}
    e^{\Omega(s)} H e^{-\Omega(s)} = H(s)
\end{equation}
where $H(s)$ is the solution of the flow equation (\ref{eq:IMSRGflow}).
This may be reorganized as a flow equation for the Magnus operator $\Omega(s)$, as described in section~\ref{sec:MagnusIMSRG}.
Considering the first few terms in the series, we have
\begin{equation}\label{eq:OmegaFirstTwo}
    \frac{d\Omega(s)}{ds} = \eta(s) - \frac{1}{2}[\Omega(s),\eta(s)] + \ldots
\end{equation}
If we choose the White generator (\ref{eq:WhiteGen}), which may be written as
\begin{equation}
    \eta^{\mathrm{Wh}}(s) = Q\frac{H(s)}{\Delta}P + P\frac{H(s)}{\Delta}Q,
\end{equation}
and neglect all terms on the right hand side of Eq.~(\ref{eq:OmegaFirstTwo}) aside from the first one, we recover the unitary coupled cluster equation (\ref{eq:UCCflow}).
The difference between Magnus IMSRG and unitary coupled cluster then lies in the commutator terms of Eq.~(\ref{eq:OmegaFirstTwo}).
A perturbative analysis reveals that the leading-order effect of the first commutator term is to induce contributions to $P\Omega P$ and $Q\Omega Q$ at third order.
If the transformation is evaluated exactly, these terms of course have no effect on the resulting observables.
However, if approximations are made---as they inevitably must be---then these terms may produce a different result.
This has not yet been investigated in detail.

Integrating Eq.~(\ref{eq:OmegaFirstTwo}) numerically with a step size $ds=1$ ---again neglecting all but the first term--- we find, following the discussion leading to Eq.~(\ref{eq:GnIter}), that the first integration step yields a generator which satisfies the decoupling condition to first order in perturbation theory (cf.~Appendix~\ref{app:PerturbativeMagnus}).
Likewise, the second step in $ds$ satisfies the decoupling condition to second order, and the $n$th step satisfies decoupling to $n$th order.
Thus, numerical integration of the flow equation with step size $ds=1$ corresponds
to an order-by-order summation of the perturbation series.
If instead we take a smaller step size, $ds=0.5$, then after the first integration step we will have only suppressed half of the first-order term in the decoupling condition.
After the second integration step, taking us to $s=1$, we have suppressed half of the \emph{remaining} first order term, as well as half of the second order term.
Taking the continuous limit $ds\rightarrow 0$, we find the first-order off-diagonal piece suppressed as $e^{-s}$, with the higher-order terms also suppressed at the same rate.

In light of this discussion, we can view the numerical integration of the flow equation (\ref{eq:OmegaFirstTwo}) with some finite step-size $ds$ as a summation of the perturbative expansion \cite{Hergert2016,Hergert2017}, with the step size specifying anything from an order-by-order summation ($ds=1)$ to all orders at once ($ds\rightarrow 0$).
This connection may have important consequences in cases where the perturbative expansion does not converge (see section~\ref{sec:Intruder}).

\subsection{Approximation schemes}
An exact evaluation of the formulas for $H_{\mathrm{eff}}$ presented in the preceding subsections will inevitably be at least as expensive as a direct diagonalization of the Hamiltonian in the full Hilbert space---precisely the task we set out to avoid.
The utility of the effective interaction framework is that it facilitates approximations which greatly reduce the required effort while minimally impacting the accuracy of the computed quantities of interest, namely observables related to low-lying eigenstates. Within the shell-model context, this implies the need for some sort of cluster truncation.

Perhaps the most straightforward approximation scheme is a truncation in in powers of the residual interaction $V$, i.e. perturbation theory.
Unfortunately, in nuclear physics the effective interaction often converges slowly in powers of $V$, and as discussed in section \ref{sec:Intruder}, the intruder-state problem suggests that in most cases the perturbation series is divergent.

Nonperturbative truncation schemes have been made essentially along two lines: either a cluster truncation is imposed within a Fock-space formulation, as for IMSRG and CC (cf.~Secs.~\ref{sec:MagnusIMSRG} and \ref{sec:SMCC}), or the problem is solved directly in the $A$-body system for a few valence particles, followed by a cluster expansion.

\subsubsection{\label{sec:FockSpaceCluster}Fock-space cluster truncation}
When working in a Fock-space formulation, it is natural to perform a cluster truncation on the generator $\gen$ or on all operators, typically limiting them to consist of zero-, one- and two-body pieces.
From a practical point of view, such a truncation is a necessity; keeping three-body terms is unpleasant but feasible, while the need for e.g. six-body terms would be sufficiently onerous to render the method useless.

For the special case of a one-dimensional $P$ space (i.e. a single-reference calculation), using the non-hermitian formulation of section \ref{sec:SMCC}, truncating $\gen$ to one- and two-body operators is equivalent to coupled cluster with singles and doubles (CCSD)~\cite{Shavitt2009}.
Here, we find the desirable feature that the Baker-Campbell-Hausdorff expansion (\ref{eq:BCH}) formally truncates after a finite number of nested commutators (four in the CCSD approximation, if $H$ has at most two-body terms).

For the case of interest in the context of the shell model, with the dimension of the $P$ space greater than one, the Baker-Campbell-Hausdorff expansion does not formally truncate~\cite{Sun2018}.
One approach to this issue is to truncate the series at a finite order of perturbation theory, or else at a finite power of $\gen$ (see, e.g.~\cite{Taube2006}).
Alternatively, one can specify a form for the Fock-space operators, e.g. retaining one- and two-body terms while discarding the rest, allowing the series to be evaluated iteratively~\cite{White2002,Morris2016}.
While the series remains infinite with this truncation, in most cases of interest it is found that the series converges so that for a given precision only a finite number of nested commutators must be evaluated.
Importantly, this truncation scheme retains only connected diagrams, and so maintains size extensivity.
This approach is used in the VS-IMSRG and SMCC described above, and in the canonical transformation theory of Yanai et al~\cite{Yanai2007,Watson:2016eu}.

When operators are normal ordered with respect to a finite-density reference, many-body operators can feed back into fewer-body operators through the commutators in the Baker-Campbell-Hausdorff expansion.
However, the reduction in particle-rank of an operator always comes with an occupation number (see the flow equation in Appendix~\ref{app:IMSRGFlow}), corresponding to a factor of the density, and so the discussion in Sec.~\ref{sec:Methods} about the cluster hierarchy justifies this truncation.

\subsubsection{\label{sec:ValenceCluster}Valence cluster expansion}
The other approximation scheme is to work within the $A$-body Hilbert space formulation and build up the effective interaction in order of increasing cluster rank~\cite{Lisetskiy2008}. One diagonalizes the $A_{\mathrm{core}}, A_{\mathrm{core}}+1, A_{\mathrm{core}}+2$ systems successively, and extracts the consistent core energy, single-particle energies, and two-body matrix elements by subtracting the contributions of lower particle rank. One could continue this procedure to obtain higher-body effective interaction, with rapidly increasing effort.
Instead, assuming that the effective interaction has a sufficiently convergent cluster expansion, the effective interaction obtained in the zero-, one-, and two-valence-particle systems can then be applied to systems with more valence particles.
For self-bound systems like nuclei, one must take care to properly treat the mass-dependence of the intrinsic kinetic energy in the construction procedure for the effective interaction ~\cite{Lisetskiy2008,Dikmen2015,Jansen2016}, although this effect becomes less important for heavier systems.

This scheme is used in the Okubo-Lee-Suzuki approaches based on coupled cluster (CCEI)~\cite{Jansen2014,Jansen2016} and the no-core shell model~\cite{Navratil1997,Lisetskiy2008,Dikmen2015}, as well as the $\hat{Q}$-box approach of Kuo and collaborators~\cite{Kuo1974,Shurpin1977,Coraggio2009}.
One potential drawback of the valence cluster expansion is that the optimal cluster decomposition for two valence particles might be significantly different from the optimal decomposition for many valence particles.
Considering the $sd$ shell as a specific example, the appropriate mean field for an empty valence space, corresponding to $^{16}$O, will be quite different from the appropriate mean field for a filled valence space, corresponding to $^{40}$Ca, and so one would expect that different single-particle energies would be optimal.

Another potentially more serious issue arises from a perturbative analysis.
As was discussed in Sec.~\ref{sec:GeneralConsiderations}, the requirement $Q\gen Q=0$ cannot be enforced in a Fock-space formulation.
If it is enforced in the $A$-body formulation, an analysis reveals that disconnected diagrams arise~\cite{Brandow1975,Kutzelnigg1983}.
For example, in a system with four valence particles and a purely two-body interaction, a disconnected four-body contribution to the effective interaction arises at fourth order.
Such a contribution --- essentially two-body interactions between two independent pairs --- is not subject to the arguments at the beginning of Sec.~\ref{sec:Methods} about short-range interactions at low density, because it does not depend on the separation between the clusters.
We may expect such terms to be combinatorially enhanced, which would be a serious problem --- this calls for further investigation.

The above point may provide some explanation for the surprising finding in Ref.~\cite{Jansen2016}, where a single interaction for the $sd$ shell obtained with the CCEI method produced impressive agreement with experimental binding energies throughout the shell.
While at first glance such a result is cause for celebration, a closer look suggests trouble.
As a specific example, other \emph{ab initio} calculations~\cite{Hergert2013,Binder:2014fk,Soma:2014eu,Stroberg2017} (including coupled cluster) using the same input interaction find $^{40}$Ca to be over-bound by nearly 40~MeV, with relatively small variation among the calculations, while the CCEI result is under-bound by just 2~MeV. The supplemental material of Ref.~\cite{Jansen2016} makes note of this, since it
is unreasonable to expect that the CCEI method should be more accurate than the coupled cluster method upon which it is based. In the present context, we may speculate that the combination of \emph{(i)} the over-binding inherent in the input force, \emph{(ii)} missing valence many-body effects (see Sec.~\ref{sec:MassDependence}) enhanced by the effect of disconnected diagrams, and \emph{(iii)} truncation errors in high-lying eigenvalues from the equation-of-motion coupled cluster method~\cite{Jansen2013} incidentally conspire to cancel out in the $sd$ shell. However, one can and should not rely on such a cancellation in general.

\section{\label{sec:3body}Three-body forces and the connection with phenomenological adjustments}

Shortly after Yukawa's formulation of the nuclear interaction in terms of pion exchange, it was pointed out~\cite{Primakoff1939} that a description of a quantum field theory in terms of an instantaneous (or, equivalently, energy-independent) potential inevitably leads to three-body and higher-body forces.
The connection between these many-body forces and nuclear saturation was also suspected early on~\cite{Wentzel1942,Drell1953,Bethe:1971qf}, although the calculations were necessarily schematic.
Likewise, mean-field calculations using a Skyrme~\cite{Skyrme1959,Vautherin1972} or Gogny~\cite{Decharge1980} parameterization of the force require a three-body, or density-dependent two-body, term. 

\begin{figure}[th]
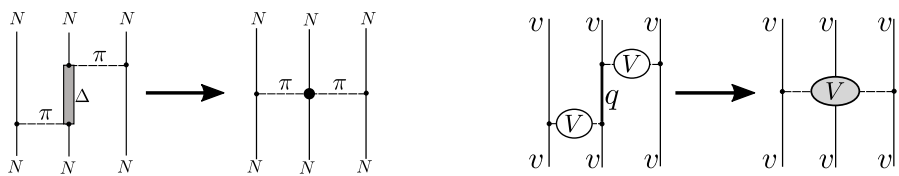

    \centering
    \begin{subfigure}{0.38\textwidth}
    \centering
    \includegraphics[width=1.0\textwidth]{DeltaTo3NDiagram.pdf}
    \end{subfigure}\hspace{0.15\textwidth}
    \begin{subfigure}{0.38\textwidth}
    \centering
    \includegraphics[width=1.0\textwidth]{QTo3NDiagram.pdf}
    \end{subfigure}
    \caption{Three-body forces generated from \emph{Left:} elimination of the $\Delta$ isobar degree of freedom, and \emph{Right:} elmination of an excitation to a $Q$-space configuration.}
    \label{fig:3Ndiagrams}
\end{figure}

Of course, even if the initial interaction were solely of a two-body nature, the effective interaction in the valence space will still in general contain three-body and higher-body forces.
In fact, these ``effective'' or ``induced'' 3N forces and the ``genuine'' 3N forces are essentially of the same origin: the elimination of degrees of freedom.
This is illustrated in Fig.~\ref{fig:3Ndiagrams}, where the elmination of the $\Delta$ isobar degree of freedom and the elimination of an excitation toa $Q$-space configuration both lead to effective 3N interactions.
Some previous studies have found the induced and ``genuine'' 3N forces are of comparable magnitude~\cite{Polls1983,Rath1984}, although this is depends on the renormalization scheme and scale of the interaction.

As a practical matter---due to the ambiguity in producing a three-body term consistent with the two-body interaction, as well as the difficulty in handling a three-body term in a many-body calculation---explicit three-body forces have historically been neglected in shell model treatments, although there were some exploratory calculations (see e.g.~\cite{Osnes1968,Goldhammer1968,Singh1973,Polls1983,Rath1984}).
There have also been more recent calculations evaluating 3N forces in the valence space either perturbatively~\cite{Caesar2013} or explicitly~\cite{Sun2018}.
Zuker and collaborators~\cite{Zuker2003,Caurier2005} argued that the main effect of the three-body forces should be to modify the monopole (i.e. diagonal, $J$-averaged) component of the effective interaction.
This argument simultaneously justified the omission of explicit three-body terms as well as the phenomenological adjustment of monopole terms in the effective interaction, which resulted in excellent reproduction of the experimental data.
Further supporting this point of view were \emph{(i)} the fact that the various realistic NN interactions produced similar shell model matrix elements\footnote{This observation is easily understood from the RG/EFT point of view---the various potentials differ in their high-momentum content but reproduce the same low-momentum physics~\cite{Bogner2003a}.}, allowing little room for improvement, and \emph{(ii)} the observed improvement in spectroscopy of light nuclei obained of quantum Monte Carlo and no-core shell model calculations when explicit three-body forces were included~\cite{Pieper2002,Navratil2007a}.

An important demonstration of the effect of three-body forces in the shell model was a calculation showing that three-body forces could help explain the location of the neutron dripline in oxygen~\cite{Otsuka2010a}, followed by an explanation of the $N=28$ magic number in the calcium isotopes~\cite{Holt:2013cr}.
These calculations used a normal-ordering approximation (see Sec.~\ref{sec:NormalOrdering}) for the three-body force and obtained essentially the monopole effect described by Zuker, although they used empirical single-particle energies and scaled the two-body matrix elements by $A^{1/3}$ as in phenomonological calculations. The same effect was soon confirmed in \emph{ab initio} calculations without phenomenological adjustments~\cite{Cipollone2013,Hergert2013a,Hagen2012a}.

The first VS-IMSRG calculations of the oxygen isotopes did not obtain the correct dripline~\cite{Bogner2014}, even though three-body forces were included in the normal-ordered approximation.
More troubling, the heavier oxygen isotopes were systematically overbound by approximately 10 MeV.
The issue was that the normal ordering of the Hamiltonian used the core wave function as a reference state in these initial VS-IMSRG calculations,
hence the effects of three-body interactions between valence nucleons were not properly captured.
This deficiency was remedied by the use of ensemble normal ordering (ENO)~\cite{Stroberg2017}, which enables an approximate treatment of the effect of three-body forces that does not degrade as valence particles are added.
This echoes the results of previous investigations of the effects of three-body forces in the shell model~\cite{Nomura1979,Rath1984,Polls1983,Dean2004a}.
Since only a brief account of ENO has been given in the literature~\cite{Stroberg2017}, we provide a more detailed description in the following section.

\subsection{\label{sec:ENO}Ensemble normal ordering}
When performing a VS-IMSRG calculation, a natural choice for the normal ordering reference $|\Phi\rangle$ is the core of the valence space.
This allows an approximate treatment of 3N forces in which a sum over particles in the core yields effective one-body and two-body forces in the valence space.
However, VS-IMSRG calculations of the oxygen isotopic chain using chiral NN+3N forces overpredicted the binding energy of neutron-rich oxygen nuclei compared to an earlier MR-IMSRG study with the same interactions~\cite{Bogner2014,Hergert2013a}. Calculations involving both protons and neutrons in the valence space yielded even more significant overbinding~\cite{Stroberg2016}.

This discrepancy was essentially due to the fact that the normal ordering in the MR-IMSRG calculation is performed directly with respect to the system of interest, not with respect to the core of the valence space. This meant that, in $^{24}$O for example, the MR-IMSRG was better capturing the 3N interactions between the 8 valence neutrons. Indeed, taking the normal ordering reference to be the nearest closed-shell nucleus brought the VS-IMSRG binding energies back in line with the MR-IMSRG results~\cite{Stroberg2016}.

\begin{figure}[ht]
    \centering
    \includegraphics[width=\textwidth]{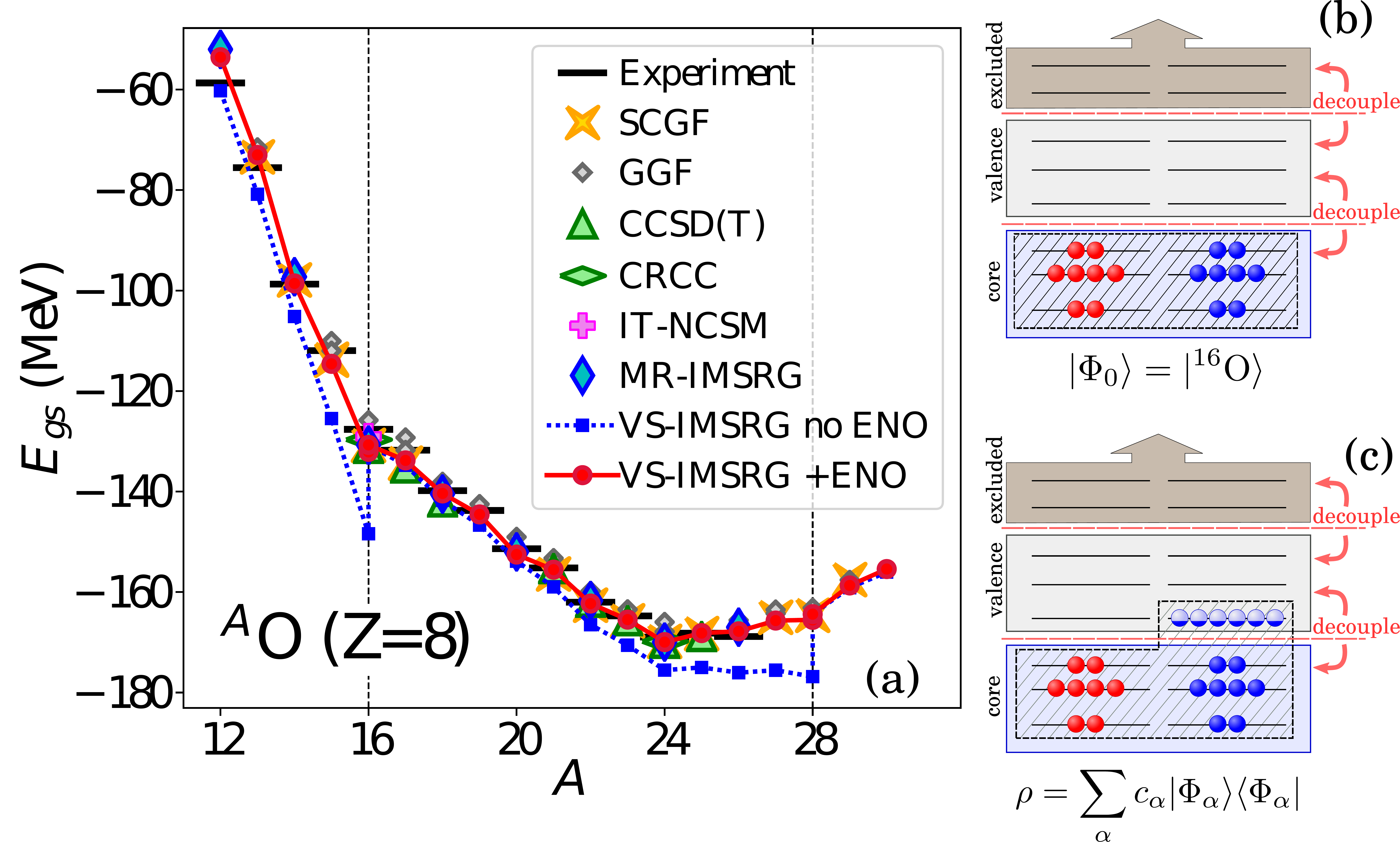}
    \caption{(a) The binding energy as a function of mass number $A$ for the oxygen isotopes, calculated with a variety of many-body methods\cite{Hergert2013a,Hergert2014,Jansen2014,Binder:2014fk,Cipollone2015}. The blue curve labeled IMSRG(SM) corresponds to the scheme in (b) with the core taken as the normal-ordering reference, while the red curve labeled IMSRG(ENO) corresponds to (c) with an ensemble reference. Adapted from~\cite{Stroberg2017}.}
    \label{fig:ENOoxygen}
\end{figure}

This approach was then generalized to treat systems that are not close to any closed sub-shell, by allowing fractional occupation numbers.
As an example, consider $^{19}$O, which in a naive shell model picture has three neutrons in the $0d_{5/2}$ orbit on top of a closed $^{16}$O core.
Equivalently, it could be considered as three neutron holes in $^{22}$O.
When using an $^{16}$O reference, the occupation number for the neutron $0d_{5/2}$ orbit would be 0, while it would be 1 for an $^{22}$O reference.
The $^{16}$O reference will underestimate the missing three-body effects, while the $^{22}$O reference will overestimate them.
The compromise is then to take the occupation to be 0.5, i.e. filling the orbit half way.
This strategy is frequently used in mean-field theory, and it is known as the ``equal-filling approximation'' or simply the ``filling approximation''.
(For an application in chemistry, see~\cite{Watson:2016eu}).

The question then arises: what reference state (if any) is actually being used when we select fractional occupation numbers?
As explained in Ref.~\cite{Perez-Martin2007,Perez-Martin2008} the equal filling approximation can be framed in terms of a mixed state or ensemble, in the sense of quantum statistical mechanics, specified by a density matrix\footnote{Not to be confused with the one-body density matrix $\rho_{pq}=\langle\Psi| a^{\dagger}_pa_q |\Psi\rangle$.}
\begin{equation}
    \rho = \sum_\alpha c_\alpha |\Phi_{\alpha}\rangle\langle\Phi_{\alpha}|
\end{equation}
for some set of coefficients $c_{\alpha}$.
(Here $\alpha$ labels different Slater determinants).
The expectation value of an operator $\mathcal{O}$ in the ensemble is obtained by a trace over the density matrix: $\langle \mathcal{O}\rangle = \mathrm{Tr}\left[\mathcal{O}\rho\right]=\sum_{\alpha}c_\alpha \langle \Phi_{\alpha}|\mathcal{O}|\Phi_\alpha \rangle$.

As discussed in Sec.~\ref{sec:imsrg}, for a single reference $|\Phi\rangle$, the normal order of a pair of creation and annihilation operators is the one which gives zero expectation value in the reference. 
The normal order of a string of more than two creation/annihilation operators can be chosen so that every pair of operators in the string is in normal order.

Wick's theorem \cite{Wick:1950fr}, which expresses a string of creation/annihilation operators in terms of normal-ordered strings and contractions, was extended to more general reference states by Kutzelnigg and Mukherjee~\cite{Kutzelnigg1997}.
In this case, the normal ordering is still defined so that the reference expectation value of a normal ordered string of creation/annihilation operators vanishes.
However, the concept of a contraction now involves one-body, two-body, and higher-body density matrices, which encode the correlations contained in the reference.
This generalized Wick's theorem is used, for instance, to formulate the MR-IMSRG~\cite{Hergert2013a,Hergert2017}.

As shown by Kutzelnigg and Mukherjee, Wick's theorem also applies to a mixed-state, or ensemble, reference:
\begin{equation}
    \langle \nord{a^{\dagger}_pa_q} \rangle
    = \mathrm{Tr}\left[ \nord{a^{\dagger}_pa_q}\rho \right]
    = \sum_\alpha c_{\alpha}\langle \Phi_{\alpha} | \nord{a^{\dagger}_pa_q} |\Phi_{\alpha}\rangle =0.
\end{equation}
This is an extension of the original formulation of the finite-temperature Wick's theorem~\cite{Matsubara1955,Gaudin1960,Fetter2003}, which only applied to the \emph{expectation value} of an operator in the ensemble.
The formulation of Kutzelnigg and Mukherjee, on the other hand, is an operator identity, just like the original zero-temperature Wick's theorem.

Our goal is then to find an ensemble such that contractions have the same form as in the single-reference case, except that the occupation of an orbit may have some non-integer value between 0 and 1.
That is, we want all two-body and higher-body irreducible density matrices, as well as the off-diagonal one-body density matrix, to vanish.
Such an ensemble must necessarily contain a variable number of particles.
This may be easily understood by considering a single particle placed in two levels.
If we require that there always be a fixed total number of particles (like in canonical or micro-canonical ensembles, for instance), then the occupation of one level implies that the other level must be empty and so the occupations are correlated, leading to an irreducible two-body density matrix.
A grand canonical ensemble, on the other hand will meet our needs. 

The ensemble that has been used in VS-IMSRG calculations published thus far corresponds to the zero-temperature limit of a finite-temperature Hartree-Fock calculation~\cite{Fetter2003}, with the chemical potential chosen to fix the average particle number.
To illustrate the application of Wick's theorem, we consider a single level with degeneracy $d=2j+1$, so a configuration has $\mathcal{N}$ particles with $0\leq \mathcal{N}\leq 2j+1$.
The contraction of two operators is given by
\begin{equation}
\begin{aligned}
    \contraction{}{a}{^{\dagger}_p}{a} a^{\dagger}_pa_q = \langle a^{\dagger}_pa_q\rangle &= \delta_{pq} \frac{1}{\mathcal{Z}}\sum_{\mathcal{N}=0}^{2j+1}
    \begin{pmatrix}2j \\ \mathcal{N}-1 \end{pmatrix} e^{\beta(\epsilon-\mu)\mathcal{N}} \\
    &= \delta_{pq} \frac{1}{\mathcal{Z}}\sum_{\mathcal{N}=0}^{2j+1}
    \begin{pmatrix}2j+1 \\ \mathcal{N} \end{pmatrix}\frac{\mathcal{N}}{2j+1} e^{\beta(\epsilon-\mu)\mathcal{N}} \\
    &= \delta_{pq} \frac{1}{2j+1}\langle \mathcal{N} \rangle\\
    &= \delta_{pq} n_p.
\end{aligned}
\end{equation}
Here $\beta$ is the inverse temperature, $\mu$ is the chemical potential, $\epsilon$ is the energy of the level, $\mathcal{Z}$ is the partition function, and the binomial coefficients count how many of the configurations with $\mathcal{N}$ particles will have orbit $p$ occupied.
Of course, other ensembles may be selected, and they need not be thermal ensembles, i.e. multiple levels could be fractionally occupied and there need not be a connection between the energy of a level and its occupation~\cite{Perez-Martin2008}.
Another very reasonable choice of occupations is to use natural orbitals, or a perturbative approximation of them~\cite{Tichai2018}.

In actual calculations, this ensemble need not be explicitly constructed; we only use the corresponding definition of the contraction when we use Wick's theorem.
In fact, there is another reference which can be constructed to produce the same fractional occupations.
Instead of employing an ensemble state, we may use a single-determinant reference built from a single-particle basis which is slightly different from the one used in the calculation.
To fractionally fill an orbit $p$, we admix in some other ``inert'' orbit $\mathcal{Q}$ which is orthogonal to all of the single-particle states used in our calculation
\begin{equation}
   \begin{pmatrix} |p\rangle \\ |\mathcal{Q}\rangle \end{pmatrix}
   \rightarrow
   \begin{pmatrix} |\bar{p}\rangle \\ |\bar{\mathcal{Q}}\rangle \end{pmatrix}
   =
   \begin{pmatrix}\sqrt{n} &\sqrt{1-n} \\ -\sqrt{1-n} & \sqrt{n} \end{pmatrix}
   \begin{pmatrix} |p\rangle \\ |\mathcal{Q}\rangle \end{pmatrix}
\end{equation}
where $0\leq n \leq 1$.
If we choose a reference in which the orbit $\bar{p}$ is filled, i.e. $\langle \Phi| a^{\dagger}_{\bar{p}}a_{\bar{p}}|\Phi\rangle=1$, then the occupation in terms of the original orbit $p$ is $\langle \Phi | a^{\dagger}_p a_p | \Phi\rangle = n$.
Because the reference $\Phi$ is a single Slater determinant, all higher-body density matrices vanish automatically.
In addition, there will be a non-zero occupation of the inert orbit $\mathcal{Q}$: $\langle \Phi | a^{\dagger}_{\mathcal{Q}} a_{\mathcal{Q}}|\Phi\rangle=(1-n)$, as well as off-diagonal one-body densities $\langle \Phi | a^{\dagger}_{p}a_{\mathcal{Q}}| \Phi\rangle = \sqrt{n(1-n)}$, which are not desirable.
However, we have asserted that the orbit $\mathcal{Q}$ is inert.
By this we mean that $a^{\dagger}_{\mathcal{Q}}$ and $a_{\mathcal{Q}}$ do not appear in any operator we consider, and we may neglect terms involving orbit $\mathcal{Q}$ without changing the physics.
While the introduction of inert orbits might seem contrived, it is no more contrived than the ensemble with variable particle number.
Indeed, we could say that the inert orbits live somewhere in the reservoir that supplies the additional particles.

Again, for practical purposes, it is irrelevant whether the reference is an ensemble or is constructed with an inert orbit mixed in.
What matters is that we may use Wick's theorem with fractional occupation numbers, and that this procedure constitutes an exact rewriting of our operators---given an operator which is normal ordered with fractional occupations, we can reconstruct the operator normal-ordered with respect to the true vacuum.
The importance of this point is that by employing fractionally-filled orbitals we have not introduced an additional approximation.
If we retain all the induced operators up to $A$-body operators, then the IMSRG calculation is exact.
What the fractional filling does is reduce the impact discarding the residual three-body terms has on the low-lying states.
Because standard shell model codes typically work with valence particles (not valence holes), after the IMSRG decoupling, we again use Wick's theorem to rewrite all operators in normal order with respect to the core (which is a single Slater determinant)\footnote{This ``re-normal-ordering'' is easily achieved by using (\ref{eq:NOops}), replacing $n_a \rightarrow (n_a^{\mathrm{new}}-n_a^{\mathrm{old}})$.}.

Certainly, an uncorrelated ensemble reference is a crude approximation of the exact wave function, and one might envision that a correlated reference state as used in the MR-IMSRG could do better~\cite{Hergert2013a,Hergert2017}.
However, as argued in Ref.~\cite{Watson:2016eu}, it is not clear that this is the best way to proceed in a valence space context.
If correlation effects are included in the reference in order to better approximate a particular state, then this might well deteriorate the description of other low-lying states (which contain different correlations), leading to a worse overall description of the spectroscopy.

\subsection{\label{sec:MassDependence}Mass dependence of the effective interaction}
A significant consequence of the ensemble normal ordering (ENO) procedure is that a different valence space interaction is obtained for each nucleus.
It is important to emphasize here that because the procedure does not involve any fitting to data, there is \emph{no loss of predictive power}.
The ENO should be considered as a technique for reducing the impact of the truncation to two-body operators.
In terms of computational effort, the need to generate a new interaction for each nucleus makes a study of the full $sd$ shell more laborious, but still manageable.
For nuclei in middle of the $pf$ shell, the exponential scaling of the valence space diagonalization catches up with the polynomial scaling of the VS-IMSRG and so generating the effective interaction takes about as long as the shell model calculation that uses it.

The need for some mass-dependence of the effective interaction has been known for a long time.
The $sd$ shell interactions of Kuo and collaborators~\cite{Kuo1966,Kuo:1967qf} yielded a good description of spectroscopy for a few valence particles or valence holes, but agreement deteriorated for mid-shell systems~\cite{Cole1975}.
Investigations by Chung and Wildenthal~\cite{Chung1976,Wildenthal1984} suggested that a single phenomenological adjustment could not remedy the situation, and a scaling of the two-body matrix elements according to $A^{0.3}$ was introduced.
This prescription has been adopted in many later treatments~\cite{Caurier1999,Holt:2013cr,Coraggio2009}.
The scaling is typically justified in terms of the increasing nuclear radius changing the optimal harmonic oscillator frequency~\cite{Wildenthal1984,Brown1988a,Caurier1999,Tsunoda:2017sh}.
While such an argument would suggest that the core and single-particle energies should also change with mass, these effects could in principle be absorbed into the scaling of the two-body matrix elements~\cite{Wildenthal1984}.
On the other hand, the need for mass-dependence of two-body matrix elements could be interpreted as a signal of non-negligible three-body terms in the effective interaction, and indeed this has been suggested a number of times~\cite{Nomura1979,Polls1983,Rath1984}.

\begin{figure}[ht]
    \centering
     \includegraphics[width=1.0\textwidth]{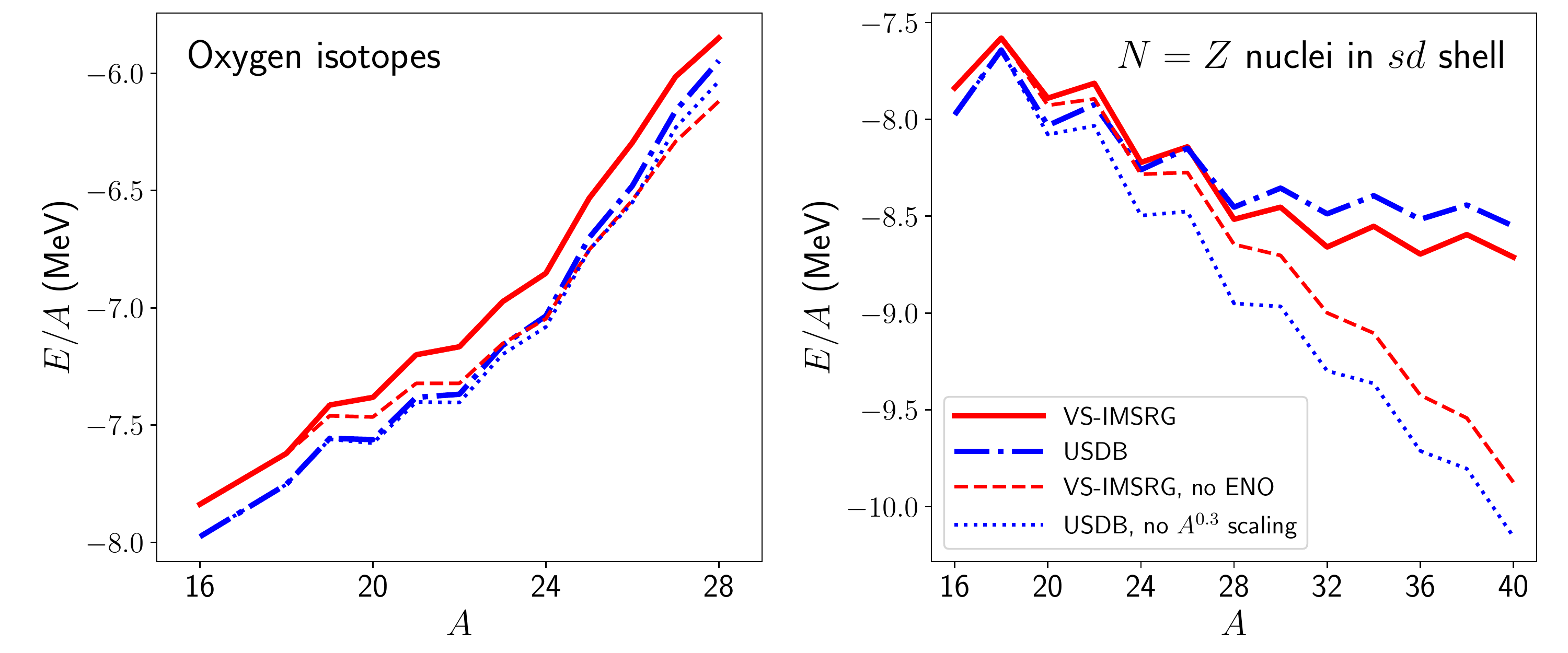}
    \caption{Energy per nucleon for \emph{Left} the oxygen isotopes $16\leq A \leq 28$, and \emph{Right} the $N=Z$ nuclei in the $sd$ shell, obtained with VS-IMSRG using the EM1.8/2.0 interaction compared with the results obtained with the USDB interaction. The thinner lines indicate the effect of turning off the ensemble normal ordering (ENO) in the VS-IMSRG calculation, or turning off the $A^{0.3}$ scaling of two-body matrix elements in the USDB interaction.}
    \label{fig:AdepvsUSD}
\end{figure}

We may expect that ensemble normal ordering should capture both the effects of a changing mean field and of the residual three-body effective interaction\footnote{Indeed, these effects are not entirely distinct; the induced three-body interaction depends on the choice of reference. }. 
Fig.~\ref{fig:AdepvsUSD} displays binding energies per nucleon obtained for oxygen isotopes and $N=Z$ nuclei in the $sd$-shell nuclei with the USDB interaction, both with and without the mass scaling of the TBMEs.
These are compared to the binding energies of the same nuclei calculated using the VS-IMSRG with and without ensemble normal ordering.
It is evident that the ensemble normal ordering has qualitatively the same effect as the scaling of the TBMEs, athough an investigation of the VS-IMSRG TBMEs reveals no such smooth scaling (the effect is largely captured in the core and single-particle energies).

\begin{figure}[ht]
    \centering
    \includegraphics[width=1.0\textwidth]{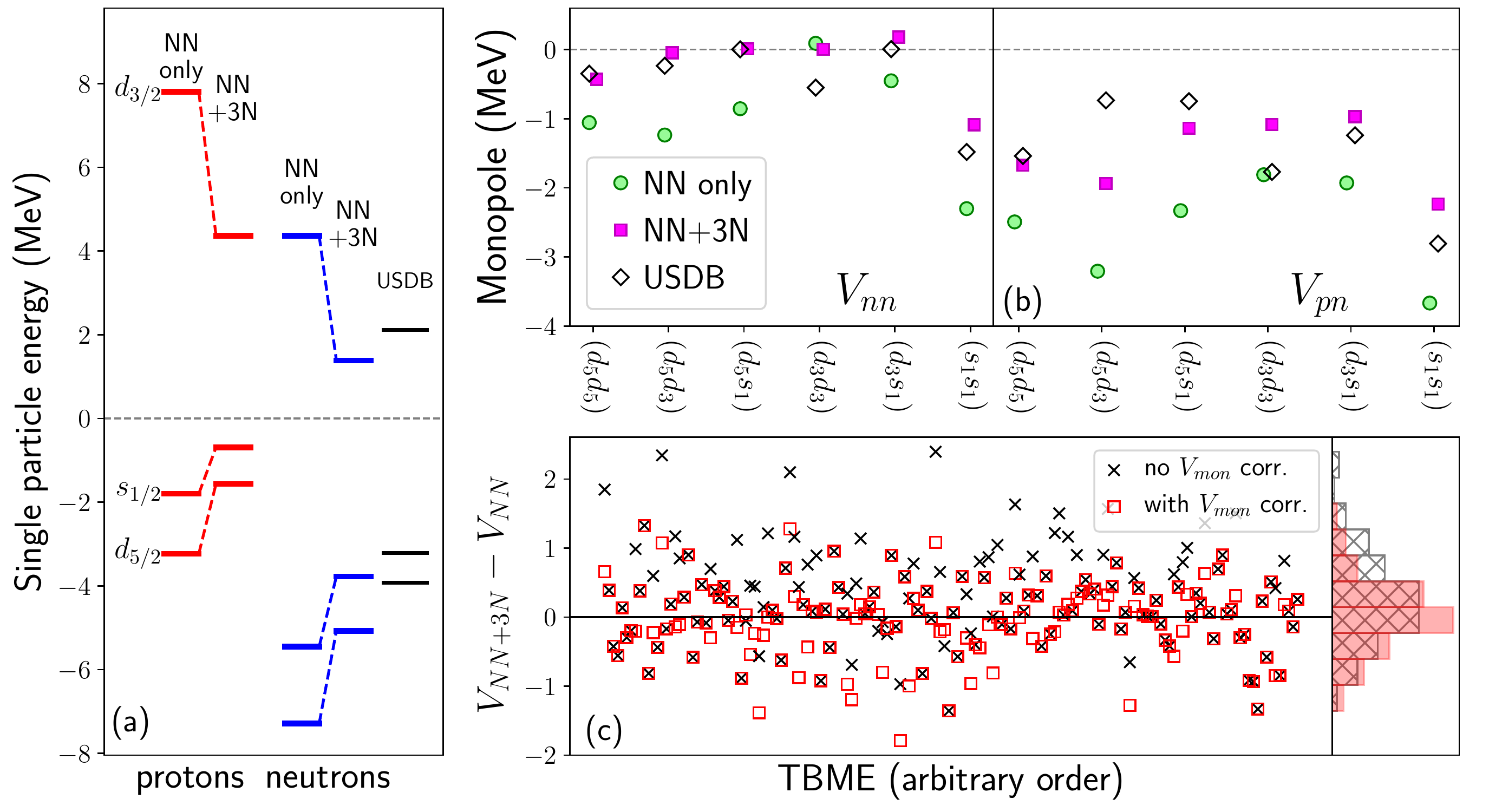}
    \caption{(a) Single particle energies in the $sd$ shell obtained with the VS-IMSRG normal ordered with respect to a $^{28}$Si reference, using an NN interaction, with or without the 3N piece. (b)Neutron-neutron and proton-neutron monopoles of the two-body matrix elements, with and without the 3N force. (c) The difference between matrix elements obtained with NN only and NN+3N, with and without a monopole correction. All calculations use the EM 1.8/2.0 interaction of Ref.~\cite{Hebeler2011}.}
    \label{fig:Si28monopoles}
\end{figure}

Figure \ref{fig:Si28monopoles} shows the single-particle energies and monopoles of two-body matrix elements (TBMEs) obtained for a $^{28}$Si reference with and without explicit 3N forces. Including the 3N force has a significant impact on the single-particle energies; indeed, the neutron SPEs are shifted closer to the USD~\cite{Brown2006a} values (USD does not include the Coulomb interaction).
In panel (b) of Fig.~\ref{fig:Si28monopoles}, we see that the effect of the 3N interaction on the TBME monopoles is repulsive, as expected from binding energy calculations, and that they shift the monopoles towards the USDB values.
In panel (c) of Fig.~\ref{fig:Si28monopoles} we show the difference between each of the two-body matrix elements obtained with the NN only and NN+3N interactions (black crosses) and the difference when the NN only monopoles have been shifted to the NN+3N values.
The monopole shift does not yield perfect agreement---there is still some scatter in the red markers---but the remaining discrepancy is approximately Gaussian and centered on zero.
It is not unreasonable that there would be moderate cancellation between the remaining terms, and that the monopole correction would approximately account for the missing 3N forces, as claimed by Zuker et al.~\cite{Zuker2003}.
For a related approach using density functionals to inform the monopole correction, see Ref.~\cite{Brown2011}.

Based on our discussion, we can conclude that the ``standard'' phenomenological adjustments made to shell model interactions can be understood essentially in terms of the effect of missing (normal-ordered) three-body forces (see e.g. the discussion in section 8 of Ref.~\cite{Brandow1967}, as well as Refs.~\cite{Dirim1975,Nomura1979,Polls1983,Rath1984}).
\begin{itemize}
\item Historically, the core energy was taken from experiment (typically no effort was made to calculate it consistently from the input force), and modern \emph{ab initio} calculations have confirmed the importance of three-body forces to binding energies~\cite{Roth2011,Cipollone2013,Hergert2013}.
\item Likewise, single-particle energies were typically taken from experiment, as the ones obtained from the NN interaction did not reproduce the spectra of one-valence-particle systems.
The normal-ordered contribution of three-body forces to the single-particle energies essentially accounts for this discrepancy.
\item Even with the core and single-particle energies taken from experiment, realistic NN forces typically did not give good spectroscopy, and needed phenomenological adjustment.
Zuker~\cite{Zuker2003} argued that the most important adjustment was of the two-body monopoles, and that this shift should be understood in terms of missing three-body forces.
Indeed, when the three-body contribution to the normal-ordered two-body interaction is taken into account, no phenomenological shifts are needed.
Moreover, as shown in Figure~\ref{fig:Si28monopoles}, the bulk of the discrepancy between an interaction derived from only NN forces and one including 3N effects can be corrected by a shift of the monopoles.
The remaining discrepancy (the ``multipole'' terms) is approximately Gaussian and centered on zero, so that the net effect will be in general small.
\item Finally, the $\sim A^{1/3}$ scaling of two-body matrix elements can be understood as a way to capture the bulk effects of three-body forces among valence particles.
This same physics is captured by employing ensemble normal ordering (see also~\cite{Caesar2013,Fukui2018}).
\end{itemize}

\section{\label{sec:Applications}Applications}

As discussed previously, shell model calculations with \emph{ab initio} interactions allow us to confront our starting point, the underlying chiral two- plus three-nucleon force of our choice, with a wealth of available experimental data. Until recently, applications have been primarily focused on ground- and excited state energies, with very encouraging results. In the following, we will highlight selected examples.

\subsection{Ground and Excited States of $sd-$Shell Nuclei}
In Sec.~\ref{sec:3body}, we discussed the importance of chiral three-nucleon forces for the correct description of nuclear shell structure. Their effect on the location of the oxygen dripline was one of the first high-profile applications of \emph{ab initio} interactions in the nuclear shell model \cite{Otsuka2010a}, which has subsequently been confirmed in more consistent calculations with both valence-space and no-core methods (cf.~Fig.~\ref{fig:ENOoxygen}). 

Multiple studies for $sd$-shell nuclei with a progressively more consistent perturbative construction of the valence-space interaction followed \cite{Holt:2013cr,Holt:2013fk,Holt:2013hc,Caesar2013,Simonis:2016uq}, including first works for the derivation of multi-shell interactions \cite{Tsunoda2014,Tsunoda:2017sh}, until the non-perturbative VS-IMSRG and CCEI/SMCC emerged \cite{Bogner2014,Jansen2014,Stroberg2016,Stroberg2017,Sun2018}. The VS-IMSRG, in particular, has been widely used to compute ground and excited-state energies \cite{Stroberg2016,Stroberg2017,Morris2018,Leach:2016fb,Brodeur:2017mi,Crawford:2017jo,Garnsworthy:2017tx,Reiter:2017ud,Steppenbeck:2017db,Henderson2018,Izzo:2018cs,Leistenschneider:2018mh,Liu:2018mt,Michimasa:2018qe,Mougeot2018,Reiter:2018ay,Randhawa:2019qd}, although theoretical uncertainties stemming from the method still prove challenging (see Sec.~\ref{sec:Challenges}).

\begin{figure}[t]
    \centering
    \includegraphics[width=1.0\textwidth]{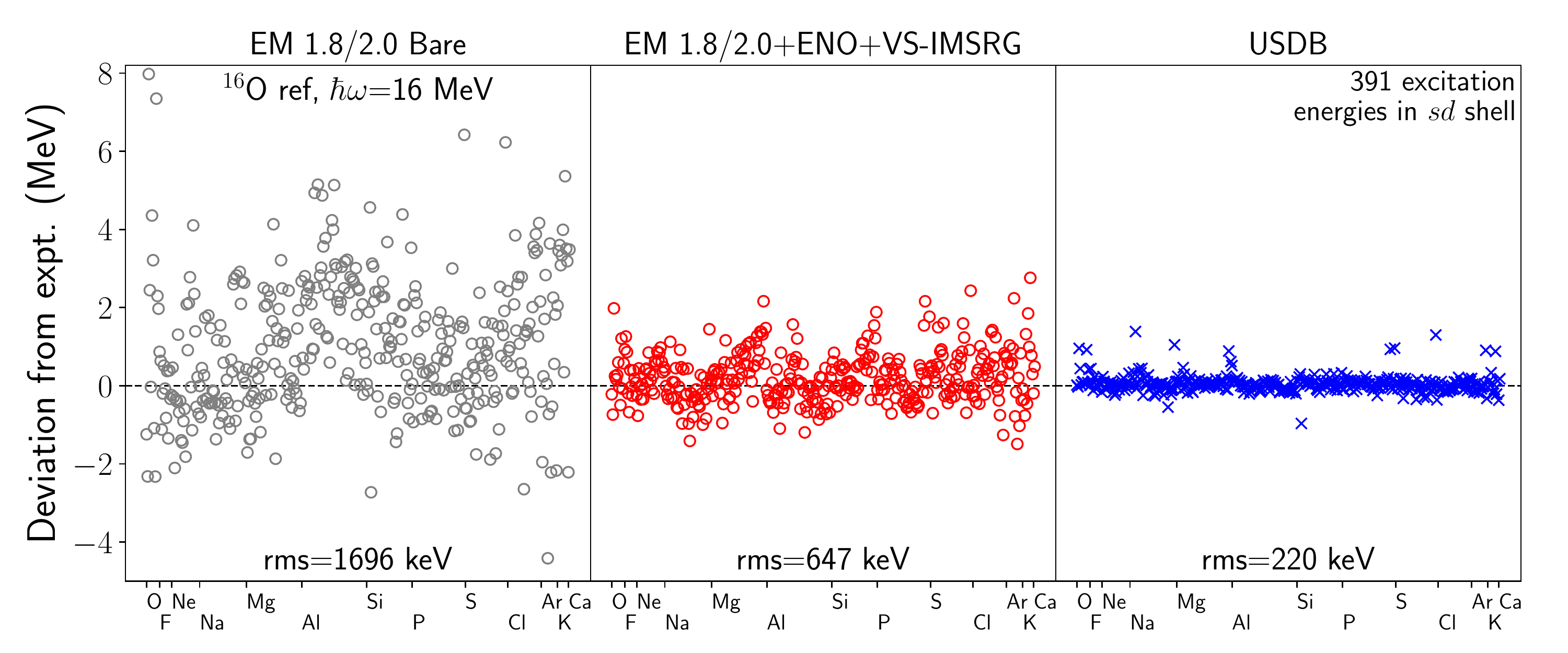}
    \caption{Deviation from experiment for excited states throughout the $sd$ shell, obtained with \emph{Left:} the EM 1.8/2.0 interaction without transformation, \emph{Middle:} EM 1.8/2.0 transformed with the VS-IMSRG using ensemble normal ordering, \emph{Right:} the USDB interaction.}
    \label{fig:rmsBare}
\end{figure}

In Fig.~\ref{fig:rmsBare}, we show results from a VS-IMSRG survey of 391 levels in the $sd$-shell, starting from the EM1.8/2.0 chiral two- plus three-nucleon interaction \cite{Hebeler2011}. The points indicate the deviation between the computed and experimental energies for all of these levels, which contribute to the specified cumulative root-mean-squared deviation between theory and experiment. In the left panel of Fig.~\ref{fig:rmsBare}, we have simply used the ``bare'' matrix elements of the EM1.8/2.0 interaction in the $sd$-shell valence space, while the center panel shows the results obtained by applying the VS-IMSRG with ensemble normal ordering, as described in Secs.~\ref{sec:Methods} and \ref{sec:ENO}. Since our starting interaction has been evolved to a low resolution scale, correlations due to the strong short-range repulsion and the tensor force have largely been accounted for. Thus, the shell model picture is reasonable: Low-lying nuclear states are bound and excitation energies are at least of the correct order of magnitude, with a sizable rms deviation of 1696~keV. 

The deviations from experiment are reduced significantly when we use the VS-IMSRG to decouple the $sd$-shell valence space from other excitations, accounting for core polarization and other types of long-range, many-body correlations (see Sec.~\ref{sec:Methods}). With an rms deviation of 647 keV, we are not doing as well as the gold-standard USDB interaction for which the deviation is merely 220~keV for the selected levels (and only $\sim130$~keV for all 600+ measured $sd$-shell levels). This is not unexpected: USDB is essentially the best possible fit to experimental data under the chosen model assumptions, including the choice of a valence space containing only the $1s_{1/2}, 0d_{3/2}$ and $0d_{5/2}$ orbitals, the mass-dependence of the two-body matrix elements, and the omission of residual three- and higher-body effective interactions. The accuracy of the VS-IMSRG results is subject to the uncertainties of the input interaction and the truncation used in the method. Both can and will be systematically improved in future applications. 

\subsection{The Calcium Region}

\begin{figure}[t]
    \centering
    \includegraphics[width=1.0\textwidth]{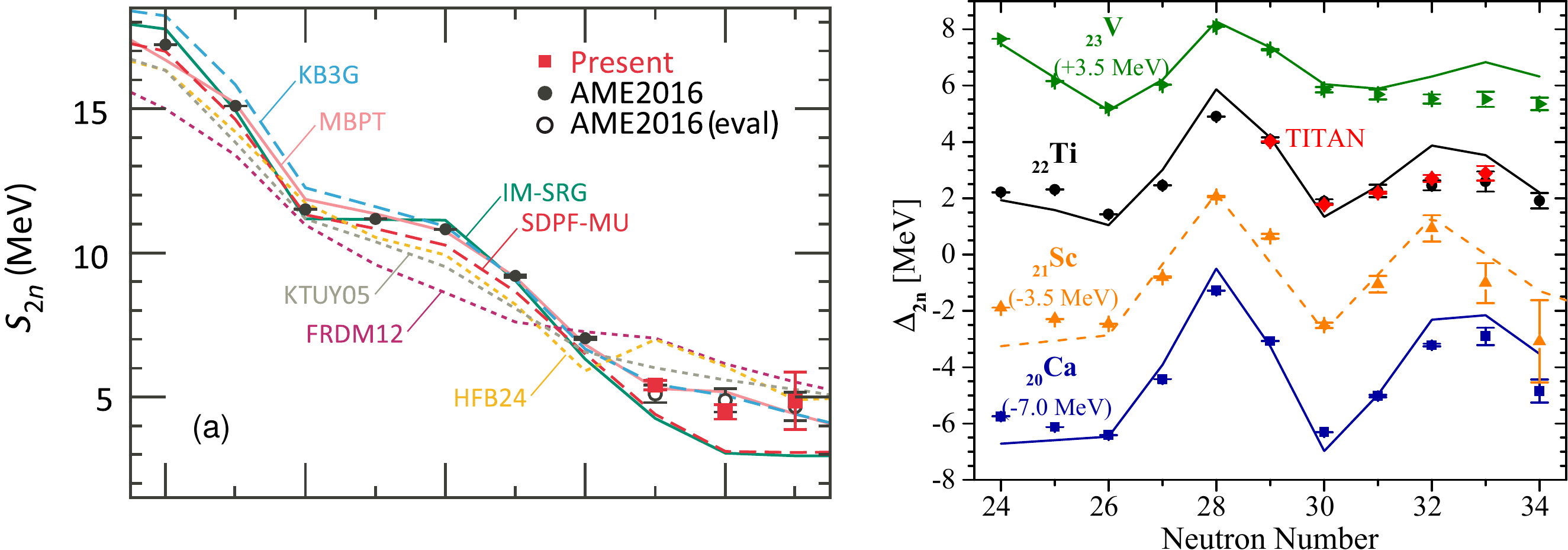}
    \caption{\emph{Left:} Two-neutron separation energies of neutron-rich calcium isotopes from recent measurements at RIKEN, compared to VS-IMSRG and MBPT results obtained with the EM1.8/2.0 interaction, as well as results for phenomenological interactions. Adapted from \cite{Michimasa:2018qe}. \emph{Right:} VS-IMSRG (solid lines) results for three-point energy differences in the calcium isotopes and neighboring chains, compared to both AME data and new titanium measurements at TITAN. VS-IMSRG results used the EM1.8/2.0 interaction, while Gor'kov Green's Function results for the scandium chain (dashed line) use a different chiral interaction. Reprinted from \cite{Leistenschneider:2018mh}. See original references for additional details.}
    \label{fig:Ca_d2N}
\end{figure}

Soon after the successful application of perturbatively constructed shell model interactions in the $sd$-shell, first results for the calcium isotopes followed, including a successful prediction of the two-neutron separation energies at the sub-shell closure in $\nuc{Ca}{52}$ \cite{Holt:2012fk,Wienholtz:2013bh,Holt:2014vn}. More recently, the masses of $\nuc{Ca}{55-57}$ were measured at RIKEN, showing the onset of a flat trend in the separation energies beyond $\nuc{Ca}{54}$ that would be consistent with the filling of the neutron $0f_{5/2}$ shell (see Fig.~\ref{fig:Ca_d2N}). Such a trend had also been found in MR-IMSRG and Gor'kov Greens Function (GGF) ground-state calculations using chiral interactions \cite{Hergert2014}, although absolute  two-neutron separation values ($S_{2n}$) could not be determined precisely because of theoretical uncertainties in the interactions and the many-body methods.

Recent high-precision mass measurements of the titanium isotopes at TITAN aimed to shed new light on the evolution of the $N=32$ shell closure  \cite{Leistenschneider:2018mh}. The right panel of Fig.~\ref{fig:Ca_d2N} shows three-point energy differences $\Delta_{2n}\equiv S_{2n}(N,Z)-S_{2n}(N+2,Z)$ extracted from the new data alongside AME data, in comparison with results from the VS\nobreakdash-IMSRG for isotopic chains in the lower $pf$-shell and GGF calculations for the scandium chain. While the theoretical $\Delta_{2n}$ compare favorably with experimental data overall, the strength of the $N=32$ closure is overestimated with increasing $Z$. This artificial enhancement of shell closures is frequently observed in calculations with current chiral interactions (see, e.g., \cite{Soma:2014eu,Hergert2014,Lapoux2016a}), and might provide important clues toward the refinement of next-generation forces. 

\subsection{Heavy Nickel and Light Tin}
For sufficiently soft interactions, IMSRG and CC calculations for nuclei in the upper $pf$ and lower $sdg$ shells can be converged \cite{Binder:2014fk,Hagen:2016xe,Morris2018}.
The limiting factor is a truncation in the three-body matrix elements $e_1+e_2+e_3\leq E_{3\mathrm{max}}$ where $e=2n+\ell$. Memory constraints have restricted calculations to $E_{3\mathrm{max}}\leq 18$. The dimension of the valence space also becomes an issue during the diagonalization of the effective interaction, but approaches like the Monte Carlo shell model \cite{Otsuka2001a} or importance-truncated configuration interaction (IT-CI) \cite{Stumpf:2016fk} can be used to tackle this problem.

\begin{figure}[t]
    \centering
    \includegraphics[width=0.7\textwidth]{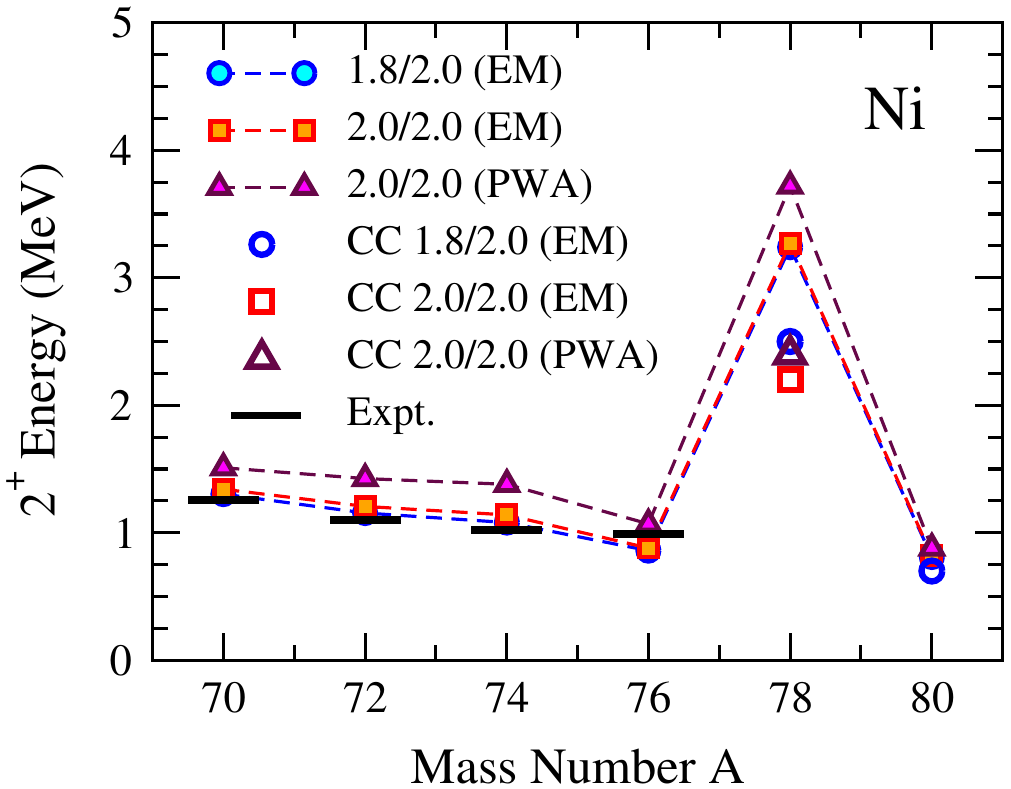}
    \caption{Energies of the first excited $2^+$ states in Ni isotopes from VS-IMSRG and Coupled Cluster calculations including triples corrections \cite{Hagen:2016xe}, using EM1.8/2.0 and other chiral two- plus three-nucleon interactions as input (see \cite{Hebeler2011} for details). }
    \label{fig:Ni78}
\end{figure}

In Figure \ref{fig:Ni78}, we show the evolution of the first excited $2^+$ state in neutron-rich nickel isotopes, which serves as a strong indicator for (sub-)shell closures. The jump in the $2^+$ energy at $\nuc{Ni}{78}$ suggests that this nucleus is indeed doubly magic. The VS-IMSRG reproduces the available experimental data \cite{NNDC} well, and the energies are insensitive under (admittedly small) variations of the interaction's resolution scales or low-energy constants --- see \cite{Hebeler2011} for more details on these Hamiltonians. 

Recently, Hagen \emph{et al.} also computed the $2^+$ \cite{Hagen:2016xe} energies of $\nuc{Ni}{78,80}$ using the Equation-of-Motion Coupled Cluster approach. In Fig.~\ref{fig:Ni78}, we include their excitation energies from the so-called EOM-CCSD(T) method, which are about 1 MeV lower than the VS-IMSRG results with the corresponding interactions. This difference can be traced back to the effects of triples (i.e., 3p3h) correlations and continuum effects that are currently not included in the VS-IMSRG.

\begin{figure}[t]
\centering
\includegraphics[width=1.0\textwidth]{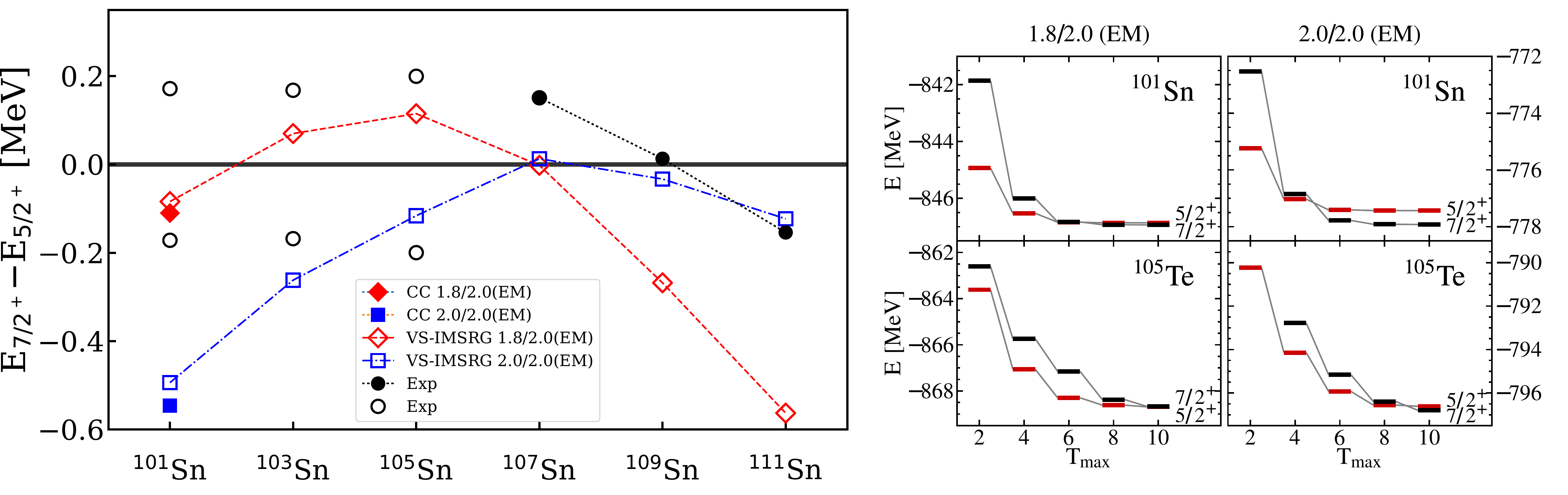}
    \caption{VS-IMSRG and EOM-CC results for the ground and first excited states of odd-mass tin isotopes and $\nuc{Te}{105}$, using chiral NN+3N interactions \cite{Hebeler2011}. The right panel shows the convergence of the states as a function of the model space truncation in the IT-CI diagonalization. Figure adapted from Ref.~\cite{Morris2018}.}
    \label{fig:Sn}
\end{figure}

Moving to even heavier nuclei, the structure of the lightest tin isotopes was the subject of a recent joint EOM-CC and VS-IMSRG study \cite{Morris2018}. Figure \ref{fig:Sn} shows results for the energy gap between the two lowest-lying states in light odd-mass tin isotopes and $\nuc{Te}{105}$. The no-core EOM-CC and the VS-IMSRG results for $\nuc{Sn}{101}$ are consistent, and the VS-IMSRG produces a near-degeneracy of the $J^\pi=5/2^+$ and $J^\pi=7/2^+$ states that is compatible with experiment, the systematic uncertainties of the method must be properly quantified --- and, most likely, reduced --- before one can make spin assignments with confidence. 

\section{\label{sec:Challenges}Current challenges}
While great strides have been made in deriving effective interactions for the shell model, challenges remain. Here we focus on two in particular, and analyze them from the perspective of the VS-IMSRG.

First, electric quadrupole (E2) observables which are sensitive to low-lying collective excitations and which historically have been treated phenomenologically by introducing effective charges, are not captured well with present techniques.
This can qualitatively be understood in the context of the cluster expansion discussed at the beginning of Sec.~\ref{sec:Methods}. It is precisely the low-lying \emph{collective} modes that are expected to violate the cluster hierarchy upon which the IMSRG relies.

Second, several regions of the nuclear chart---e.g. the ``islands of inversion''~\cite{Warburton1990a,Caurier2014}, or the charge radii of the calcium isotopes~\cite{Caurier2001,GarciaRuiz2016}---display features which suggest that a naive valence space of a single major harmonic oscillator shell is not an appropriate first-order description. However, the derivation of effective interactions for non-standard valence spaces leads to difficulties related to the well-known intruder-state problem.

\subsection{\label{sec:E2}Electromagnetic transitions}

The first attempt at a microscopic treatment of electric quadrupole ($E2$) observables was the work of Horie and Arima~\cite{Horie1955}, investigating the role of configuration mixing on quadrupole moments.
A series of investigations by Seigel and Zamick~\cite{Siegel1969,Siegel1969a,Siegel1970} demonstrated the importance of terms beyond first order in perturbation theory.
Specifically, they investigated the impact of Tamm-Dancoff (TDA) and random-phase approximation (RPA) graphs to the effective charge, with the physical interpretation that the effective charge comes largely from a coupling to the giant quadrupole resonance.
A subsequent calculation by Kirson~\cite{Kirson:1974oq} indicated that a self-consistent treatment including screening effects essentially canceled the effect obtained with RPA.
For a discussion, see Ref.~\cite{Ellis1977}.

An important development came with the application of the OLS approach to an effective interaction for $^{6}$Li in the $p$-shell~\cite{Navratil1997}, where the resulting effective $E2$ operator could be reasonably approximated by the use of effective charges.
This approach was investigated in more detail over a decade later~\cite{Lisetskiy2009}, showing that a nonperturbative treatment could produce the collective effects of $E2$ observables.
Of course, it is not always clear what lessons learned in the $p$ shell carry over to heavier masses.

As discussed in Sec.~\ref{sec:MagnusIMSRG}, the Magnus formulation of the IMSRG provides a straightforward way to construct effective valence space operators for general observables. All operators are consistently transformed according to
\begin{equation}\label{eq:MagnusOeff}
    \mathcal{O}_{\mathrm{eff}} = e^{\Omega}\mathcal{O}e^{-\Omega}
    = \mathcal{O} + [\Omega,\mathcal{O}]+\tfrac{1}{2}\left[\Omega,[\Omega,\mathcal{O}]\right]+\ldots
\end{equation}
A first application of this approach was to electromagnetic transitions in light and medium mass nuclei~\cite{Parzuchowski2017a}, where it was found that the observables were well-converged with respect to the model space truncation (i.e. frequency and number of major shells included in the initial harmonic oscillator basis).
However, the computed values for collective observables like magnetic moments or electric quadrupole and octupole transitions were substantially smaller than experimental data.

\begin{figure}[ht]
    \centering
    \includegraphics[width=1.0\textwidth]{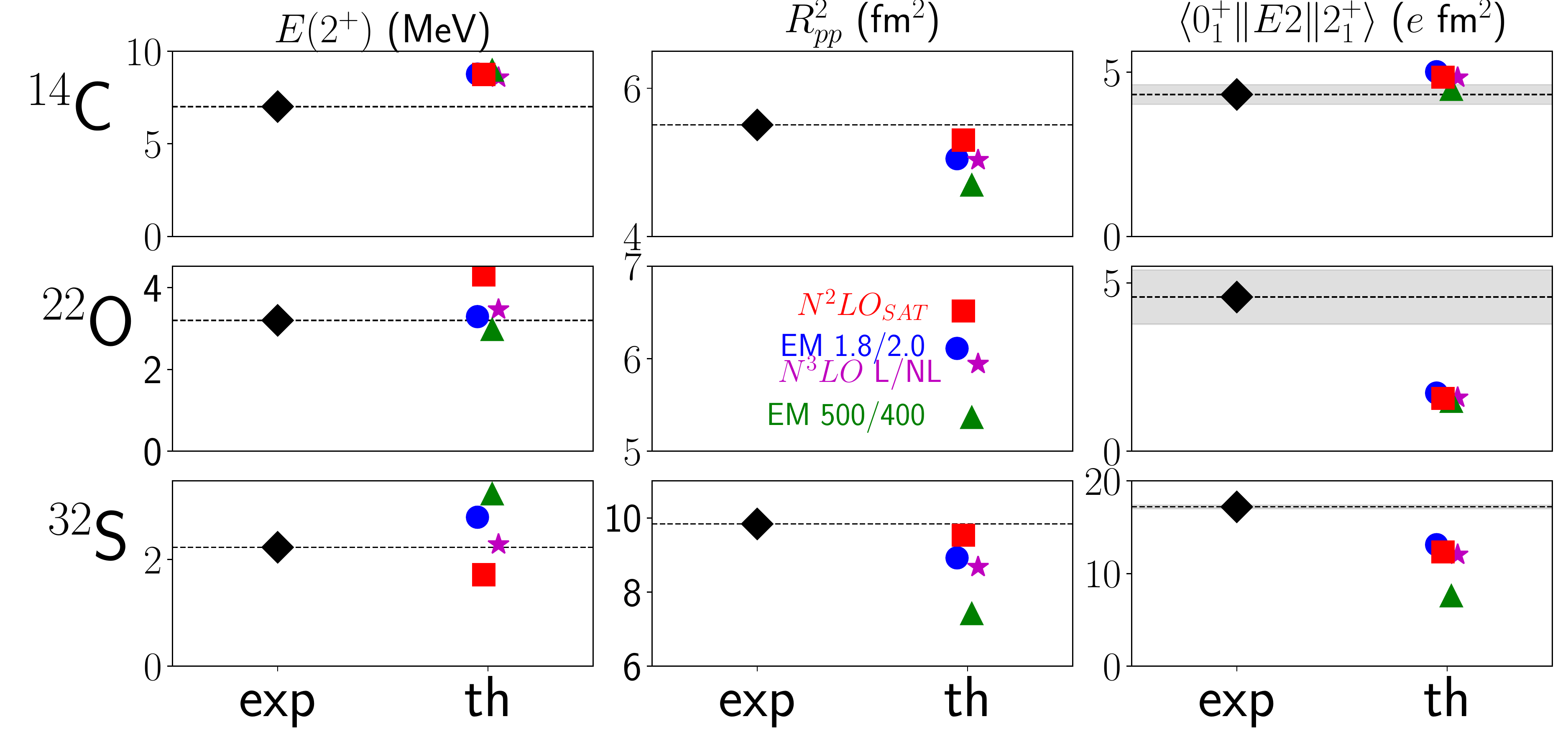}
    \caption{Electric quadrupole transition matrix element $\langle 0^{+}\| E2 \| 2^{+}\rangle$ in $^{14}$C, $^{22}$O, and $^{32}$S computed using the VS-IMSRG with several choices of input chiral interaction. Also shown are the energy of the $2^+_1$ state and the point proton radius squared. Experimental radii are from~\cite{Angeli2013}, energies and transition matrix elements are from~\cite{NNDC}.}
    \label{fig:E2compare}
\end{figure}

The possible explanations for this discrepancy are that either that the truncation of Eq.~(\ref{eq:MagnusOeff}) to two-body operators is insufficient to capture this type of collectivity, or else the input chiral interactions are deficient in some way.
Most likely, both are in effect to some degree.
The interaction used in Ref.~\cite{Parzuchowski2017a} is known to underpredict charge radii in these same nuclei~\cite{Lapoux2016a}. Given that the electric quadrupole operator is proportional to $r^2$, where $r$ is the point proton radius, and that the transition strengths $B(E2)$ go as $r^4$, one would naturally expect some underestimation of the quadrupole strength.
However, as demonstrated in Fig.~\ref{fig:E2compare}, this cannot be the whole story.

The point proton radius squared, indicated $R^2_{pp}$ in Fig.~\ref{fig:E2compare}, is underpredicted at approximately the same level in $^{14}$C and $^{32}$S.
On the other hand, while the $E2$ strength for $^{14}$C is reasonably reproduced, in $^{32}$S it is underpredicted by $\sim$25-50\%, and the strength in $^{22}$O is underpredicted by $\sim$65\%.
Clearly, the underprediction of $E2$ strength in $^{32}$S cannot be explained solely by the radius deficiencies.

Supporing this interpretation, unpublished calculations in a small space where exact diagonalizations are possible show unambiguously that the IMSRG(2) truncation misses a significant fraction of the $E2$ strength, and that capturing the full strength requires inclusion of correlated many-particle many-hole excitations.
In addition, symmetry-adapted no-core shell model calculations of $sd$ shell nuclei yielded significantly larger $E2$ strength using the same starting interaction~\cite{Henderson2018}.

\begin{figure}
    \centering
    \begin{subfigure}{0.15\textwidth}
    \includegraphics[width=\textwidth]{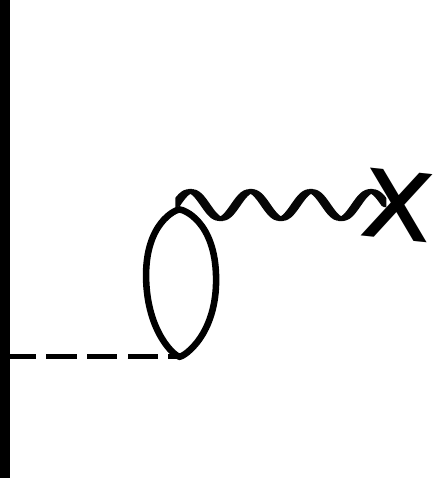}
    \caption{Core pol.}
    \end{subfigure}
    \hspace{5mm}
    \begin{subfigure}{0.18\textwidth}
    \includegraphics[width=\textwidth,]{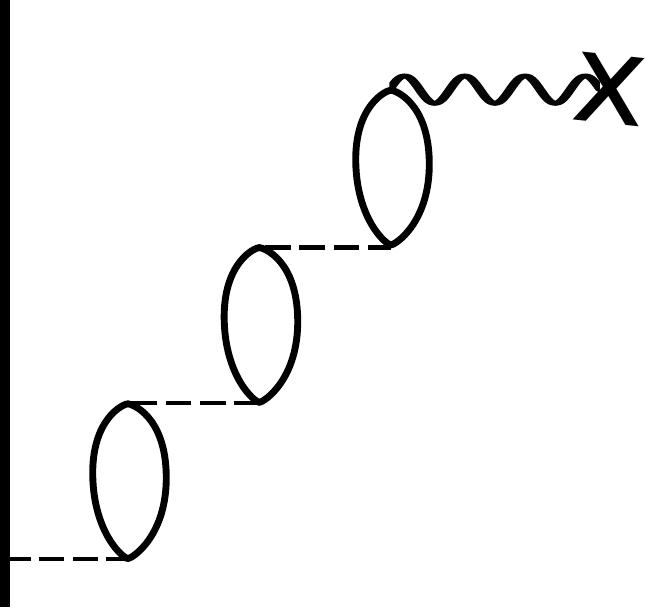}
    \caption{TDA}
    \end{subfigure}
    \hspace{5mm}
    \begin{subfigure}{0.18\textwidth}
    \includegraphics[width=\textwidth]{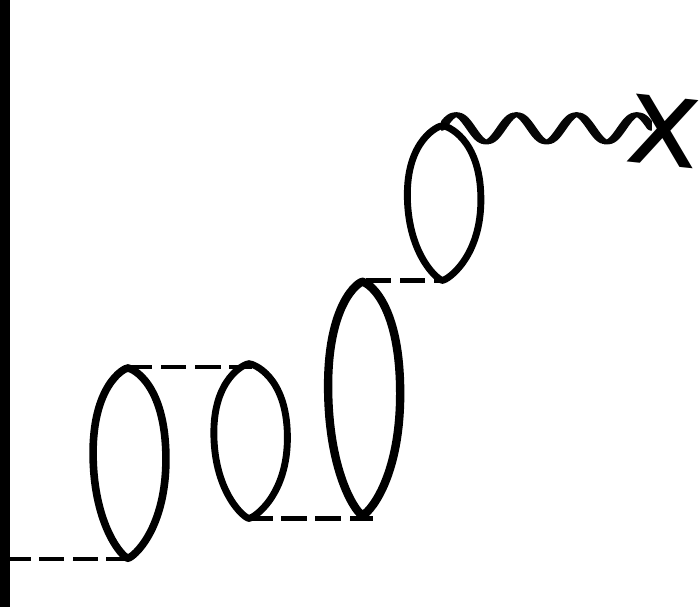}
    \caption{RPA}
    \end{subfigure}
    \hspace{5mm}
    \begin{subfigure}{0.20\textwidth}
    \includegraphics[width=\textwidth,]{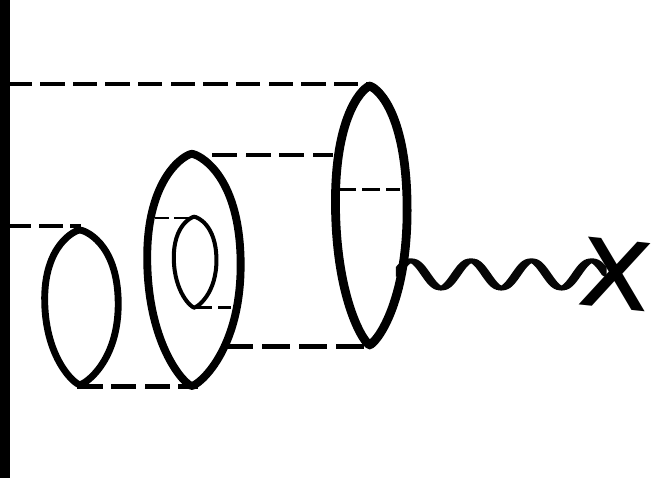}
    \caption{\label{fig:E2DiagramIMSRG}IMSRG}
    \end{subfigure}
    \caption{Exaples of diagrams contributing to the one-body part of the effective $E2$ operator. The X represents the bare operator.}
    \label{fig:E2diagrams}
\end{figure}

It is interesting to compare the IMSRG results with the previously-mentioned approach of Siegel and Zamick~\cite{Siegel1970}.
They considered three levels of approximation for the $E2$ operator, first-order core-polarization, TDA, and RPA.
Typical diagrams contributing in these approximations are shown in Fig.~\ref{fig:E2diagrams}.
The effective operator generated by the VS-IMSRG via Eq.~(\ref{eq:MagnusOeff}) contains TDA and RPA graphs to all orders. The VS-IMSRG also sums higher-order diagrams like Fig.~\ref{fig:E2DiagramIMSRG}, but because of the truncation to two-body operators, certain types of  diagrams are undercounted or missing altogether (see Refs.~\cite{Hergert2016,Morris2015,Morris2016} for more details).

\begin{table}[th]
    \caption{Neutron effective charges for the $sd$ shell obtained with first-order core polarization, including TDA and RPA graphs to all orders, and from IMSRG. These results are obtained with the EM1.8/2.0 interaction~\cite{Hebeler2011}, in a Hartree-Fock basis constructed from an oscillator basis with $e_{max}=10$, $\hbar\omega=16\MeV$.}
    \label{tab:neutron_effcharge_emax10}
    \centering
    \begin{tabular}{r|ccccc}
    \hline\hline
    & $d_{5/2}d_{5/2}$ & $d_{3/2}d_{3/2}$ & $d_{5/2}d_{3/2}$ & $d_{5/2}s_{1/2}$ & $d_{3/2}s_{1/2}$ \\
    \hline 
    Core. pol. & 0.110 & 0.035 & 0.064 & 0.034 & 0.026  \\
    TDA            & 0.121 & 0.037 & 0.062 & 0.040 & 0.031  \\
    RPA            & 0.119 & 0.037 & 0.061 & 0.038 & 0.030  \\
    IMSRG          & 0.202 & 0.098 & 0.222 & 0.163 & 0.093 \\
    \hline\hline
    \end{tabular}
\end{table}

Table~\ref{tab:neutron_effcharge_emax10} presents the effective charge for a neutron in the $sd$ shell in these various levels of approximation.
The orbit-dependent effective charge is obtained as~\cite{Siegel1970}
\begin{equation}
    e_{ab} = \frac{\langle a \| \mathcal{O}^{E2}_{\mathrm{eff}}\| b\rangle}{\langle \pi a \| \mathcal{O}^{E2}_{\mathrm{bare}} \| \pi b\rangle}
\end{equation}
where in the denominator, we take the matrix element for the corresponding proton orbit.
We work in a Hartree-Fock basis constructed from an oscillator basis with frequency $\hbar\omega=16\MeV$ and $e_{max}=10$.
In this basis, we obtain a bare proton matrix element $\langle \pi d_5 \| \mathcal{O}^{E2}_{\mathrm{bare}}\| \pi d_5\rangle=-9.03e\fm^2$, and we see that in order to reproduce the experimental quadrupole moment of $^{17}$O ($Q=-2.56(2) e\fm$~\cite{Stone2016}), 
we require an effective neutron charge of\footnote{With our definition of the $E2$ operator, $Q=\sqrt{16\pi/5}\langle J,M=J | \mathcal{O}^{E2}|J,M=J\rangle $} $e_n\approx0.37$.
Likewise, the bare proton matrix element $\langle \pi d_5 \| \mathcal{O}^{E2}_{\mathrm{eff}}\|\pi s_1\rangle=-9.29e\fm^2$, and so to reproduce the experimental transition strength $B(E2;\tfrac{1}{2}^{+}\rightarrow\tfrac{5}{2}^{+})=6.21(8)e^2\fm^4$ we require an effective neutron charge $e_n\approx0.38$.
We find that while the IMSRG generates a larger neutron effective charge than the other methods, the result is still well below the experimental value.

These IMSRG effective charges are essentially the same as those found in a previous study~\cite{Parzuchowski2017a} with a different chiral interaction.
That study also found proton effective charges close to 1, i.e. with almost no renormalization.
As discussed in Ref.~\cite{Ellis1977}, this can potentially be understood by considering that in order to ``dress'' a valence nucleon, that nucleon must excite a proton out of the core.
A valence neutron can do this through the $T=0$ channel, while a valence proton must act in the weaker $T=1$ channel.

\subsection{\label{sec:Intruder}The Intruder-state Problem}

Over the last few decades, experimental investigations of nuclei far from stability have revealed the existence of several ``islands of inversion'', where nuclei near traditional shell closures have ground states that indicate significant deformation or correlated particle-hole excitations out of the closed shell~\cite{Wildenthal1980,Warburton1990a,Caurier2014}.
The classic examples are $^{31}$Na and $^{32}$Mg, both with $N=20$.
$^{31}$Na has a ground state spin-parity of $\tfrac{3}{2}^+$ , while shell model calculations predicted $\tfrac{5}{2}^+$, and $^{32}$Mg has a $2^+$ excitation energy of 885 keV, far lower than expected for a closed neutron shell.
Both have greater binding energies than predicted in the shell model.
If these correlations are sufficiently important, then it is possible that the ground state will not be among the subset of eigenstates reproduced in the valence space diagonalization.
Indeed, Watt et al.~\cite{Watt1981} found that by explicitly allowing neutron excitations out of the standard $sd$ shell and into the $f_{7/2}$ shell, the discrepancies for $^{31}$Na and $^{32}$Mg could be understood.
Of course, even if the correlated ground state is formally among the valence states, it is likely that incorporating the correlated excitations into an effective Hamiltonian would require large many-body forces.

It is therefore desirable to be able to produce an effective interaction for a valence space that spans more than one major shell, such as the $sd-fp$ space, and indeed phenomenological interactions for such a space have been successful at describing the island of inversion effects~\cite{Caurier2014}, as well as the charge radii of the calcium isotopes~\cite{Caurier2001}.
Unfortunately, in deriving such an interaction from first principles, one runs into the well-known intruder-state problem which we discuss below.
In fact, attempts to include effects of the continuum---essential for studies near the driplines---suffer from the same problem~\cite{HuPrivComm}.
Understanding and solving this problem, particularly in the context of a nonperturbative approach, is clearly of great interest.

\subsubsection{The intruder-state problem in perturbation theory}
As demonstrated by Schucan and Weidenm{\"u}ller~\cite{Schucan1972a,Schucan1973}, there are serious reasons to doubt the convergence of the perturbative expansion for the effective interaction.
To illustrate this, we split up the Hamiltonian as before into a zero-order piece and a perturbation, with a dimensionless power-counting parameter~$x$
\begin{equation}
    H(x) = H_0 +xV
\end{equation}
such that $H(0)$ is the zero-order Hamiltonian and $H(1)$ is the full Hamiltonian.
The perturbative expansion of $H_{\mathrm{eff}}$ can therefore be seen as a Taylor expansion about $x=0$ evaluated at $x=1$.
The trouble arises if one of the states belonging to the $Q$ space has an energy lower than one of the $P$ space states.
Such a state is called an ``intruder state''.
Assuming the $P$ states are all at lower energy than the $Q$ states at $x=0$, this implies a level crossing for some $x\in[0,1]$.
The value of $x$ at which such a crossing occurs (even if it is an avoided crossing) corresponds to a branch point which places an upper limit on the radius of convergence of the effective Hamiltonian~\cite{Schucan1973}.

Unfortunately such level crossings are the rule, not the exception.
Moreover, if the zeroth-order levels in the valence space are non-degenerate---e.g. if a Hartree-Fock basis is used---then as more particles are added to the valence space, the energy of the highest $P$ space can quickly become higher than the energy of the lowest $Q$ space, even without the residual interaction. 

\subsubsection{The intruder-state problem in IMSRG}
There are a few reasons why one might expect the IMSRG to avoid the intruder-state problem.
First, it is formally a non-perturbative method, so the above argument does not directly apply.
Second, because it is formulated in Fock space, the energies of the $A$-body system do not enter into any energy denominators, and so one would not naively expect divergences due to crossings in the $A$-body system.

\begin{figure}
    \centering
    \includegraphics[width=1.0\textwidth]{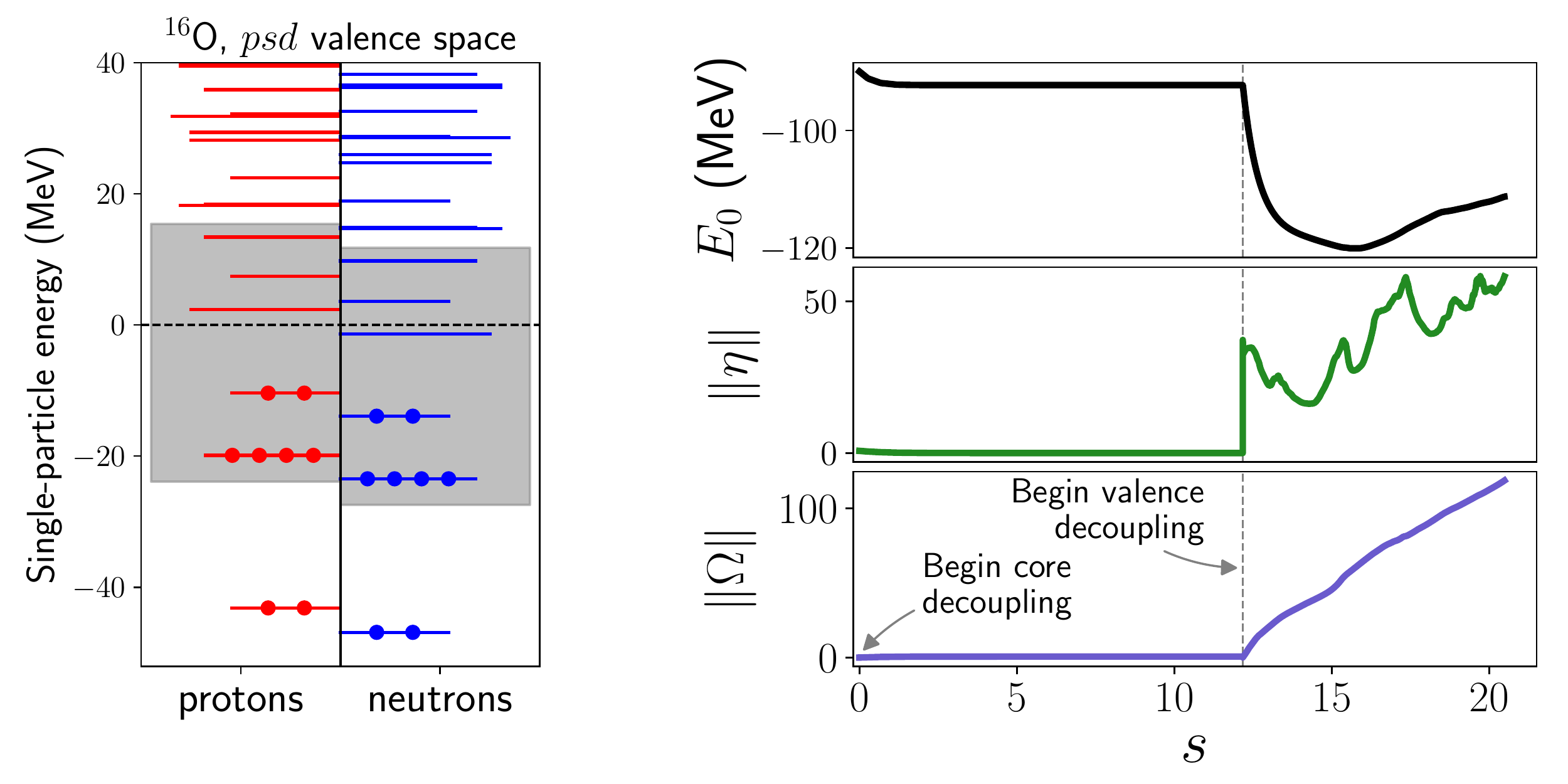}
    \caption{\emph{Left:} Decoupling of the $psd$ valence space using an $^{16}$O reference, shown with the Hartree-Fock single-particle spectrum. \emph{Right:} The zero-body part of the flowing Hamiltonian, the norm of the generator $\eta$ and the norm of the Magnus operator $\Omega$ as a function of the flow parameter $s$.
    At $s \sim 12$, the core is decoupled and the decoupling of the valence space begins.}
    \label{fig:O16psdbadness}
\end{figure}

Unfortunately, the IMSRG suffers from a related, but distinct intruder-state problem.
An illustrative example of the type of behavior encountered is shown in Fig.~\ref{fig:O16psdbadness}.
Here we aim to decouple a valence space consisting of the $p$ and $sd$ major shells from a large space constructed from 7 major harmonic oscillator shells ($e_{max}=6$).
We use an $^{16}$O Hartree-Fock reference state, which is indicated schematically in Fig.~\ref{fig:O16psdbadness}.
Also shown in Fig.~\ref{fig:O16psdbadness} are the zero-body term $E_0(s)$, the norm of the generator $\|\eta(s)\|$ and the norm of the Magnus operator $\|\Omega(s)\|$ as a function of the flow parameter $s$.
As usual, we do this in two steps, first decoupling excitations out of the core ($^{4}$He in this case), followed by a decoupling of the valence space (cf.~Sec.~\ref{sec:imsrg}).
The core decoupling is achieved at $s\approx 12$.
At this point, we observe a jump in $\|\eta(s)\|$ because our new definition of `off-diagonal'' now includes many more matrix elements.
In a well-behaved calculation, these terms would then be suppressed by the IMSRG flow.
Indeed, the size of $\eta$ initially decreases, but it soon begin to grow again, and the caclulation fails to converge.
We also observe that the flow of the zero-body term $E_0$ turns around and diverges, and the Magnus operator $\Omega$ grows indefinitely.
At some point, $\Omega$ grows beyond the radius of convergence of the BCH expansion.
As a result, no effective interaction is obtained.

We mention in passing that there has been some success using the IMSRG to decouple valence spaces other than those defined by a single major harmonic oscillator shell, so long as they are reasonably well separated by a shell gap.
These spaces correspond to the ``extruded-intruded'' spaces described by the Strasbourg group~\cite{Caurier2005}, where due to the spin-orbit potential the orbit with the largest-$j$ orbit drops out (is ``extruded'') and the largest-$j$ orbit from the next shell up comes down (it ``intrudes'').
An example is the space consisting of the orbits $1p_{3/2}$, $1p_{1/2}$, $0f_{5/2}$, $0g_{9/2}$.
This space (for neutrons), was used to treat heavy chromium isotopes~\cite{Mougeot2018}.
However, the results obtained there suggested that this space was not sufficient to describe the ground states of those isotopes.

The connection between intruders and failed convergence of the IMSRG flow can be understood schematically by considering the flow equation formulation $\tfrac{d}{ds}H=[\eta,H]$.
Imagine we have two levels labeled $p$ and $q$, both with degeneracy greater than 2, and with single particle energies $\epsilon_q > \epsilon_p$.
We intend to decouple the $q$ orbit from the Hilbert space, which means suppressing terms like $V_{qqpp}a^{\dagger}_qa^{\dagger}_qa_pa_p$ which excite particles from the $p$ level to the $q$ level.
The flow equation for $V_{qqpp}$ is, schematically,
\begin{equation}\label{eq:SchematiclIntruder}
    \frac{d}{ds} V_{qqpp} \sim V_{qqpp} \frac{2\epsilon_p-2\epsilon_q + V_{pppp} - V_{qqqq}}{2\epsilon_q-2\epsilon_p} + \ldots
\end{equation}
If the one-body terms dominate the right hand side of (\ref{eq:SchematiclIntruder}), then $\tfrac{d}{ds}V_{qqpp}\sim -V_{qqpp}$ and the off-diagonal term is suppressed exponentially.
On the other hand, if the interaction terms $V$ are larger than the one-body terms, and of opposite sign, then $V_{qqpp}$ will be exponentially enhanced.
This can be achieved if $V_{qqqq}$ is negative (attractive) and $V_{pppp}$ is positive (repulsive), and a positive numerator in (\ref{eq:SchematiclIntruder}) corresponds to an inversion of the states $|pp\rangle$ and $|qq\rangle$.

In fact, terms like $V_{pppp}$ and $V_{qqqq}$ can be included in the denominator by repartitioning the Hamiltonian so that the diagonal (i.e. bra=ket) parts of $V$ are included in $H_0$, avoiding the exponential growth.
However, intruders can also be driven by collective effects which cannot be tamed by a straightforward repartitioning.
Consider the case where we have multiple included and excluded levels $p,p'\ldots q,q'\ldots$
In this case, we should also consider contributions like
\begin{equation}
    \frac{d}{ds}V_{qqpp} \sim \frac{V_{qqp'p'}}{2\epsilon_q-2\epsilon_{p'}}V_{p'p'pp}
    - V_{qqq'q'} \frac{V_{q'q'pp}}{2\epsilon_{q'}-2\epsilon_{p}} + \ldots
\end{equation}
If there are many such terms involving $V_{p''p''pp}$, etc. and these terms add coherently, with the $V_{qqq''q''}$ type terms having opposite sign, they can compete with the contributions in (\ref{eq:SchematiclIntruder}) and potentially lead to growth of the off-diagonal terms.
Such a situation will also lead to a crossing of collective levels.
Clearly, this situation and the previous one will be exacerbated by the small energy denominators which occur in multi-shell valence spaces.

There is another way in which intruders can cause trouble, and this is by spoiling the cluster hierarchy. We illustrate this with a toy system in the next section.

\subsubsection{Toy model for the intruder-state problem}
To illustrate how intruders and level crossings can lead to large induced many-body terms, we consider the problem of three kinds of fermion---which could be, say, spin-up neutron, spin-down neutron, spin-up proton--- living in a three-level Hilbert space.
We require three particles because we wish to monitor induced three-body forces.

The initial Hamiltonian is $H(x)=H_0+xV$ where
\begin{equation}\label{eq:ToyFockH}
    H_0\equiv \sum_i \epsilon_i a^{\dagger}_ia_i
    \hspace{1em},\hspace{2em}
    V\equiv \tfrac{1}{4}\sum_{ijkl} V_{ijkl} a^{\dagger}_ia^{\dagger}_ja_la_k.
\end{equation}
Additionally, three-body terms will be induced by the transformation.
All three species have the same single-particle energies: $(\epsilon_1,\epsilon_2,\epsilon_3) = (0,1,20)$.
We define the valence space to consist of the lower two orbits for each flavor.
The antisymmetrized matrix elements of the perturbation $V$ are:
\begin{equation}
\begin{aligned}
    v_{QQ} &=  V_{1323} =  V_{1331}  = V_{1332} = V_{2331} =  V_{2332} =  V_{3132} \\
    v_{PP} &=  V_{1122} \\
    v_{PQ} &= V_{2213} = V_{2231} = V_{1113}  =  V_{1131}.
\end{aligned}
\end{equation}
We take $v_{PP}$=8, $v_{QQ}$=$-8$, $v_{PQ}$=1.
Essentially $v_{PP}$ and $v_{QQ}$ mix configurations within the $P$ and $Q$ spaces, respectively, leading to a collective $Q$ state coming down in energy as the interaction is turned on, while a collective $P$ state is pushed up, and eventually the states cross.
The remaining term $v_{PQ}$, is initially the term we want to suppress. It couples the $P$ and $Q$ states and makes the level crossing an avoided crossing.
To reduce somewhat the size of the problem, we exclude the highest level for the third particle (call it the proton), reducing the three-body Hilbert space to $3\times 3 \times 2=18$ configurations.
We take $v_{QQ}$ and $v_{PQ}$ to act only between neutrons, while $v_{PP}$ acts on all species.

The eigenstates of this problem may easily be found by forming the $18\times 18$ Hamiltonian matrix and diagonalizing.
However, our aim here is to first decouple the $P$ and $Q$ spaces, and then diagonalize within the decoupled spaces.

We perform a non-perturbative decoupling in the 3-body Hilbert space using the iterative method outlined in section~\ref{sec:FockCanonicalPT} (this is essentially the approach proposed by Suzuki~\cite{Suzuki1977} to deal with the intruder-state problem).
The first step is to construct the matrix $\mathbb{H}_{0}$, where the subscript denotes iterations.
Next, we form an anti-hermitian generator $\mathbb{G}_{n}$ which is defined as
\begin{equation}\label{eq:SuzukiIterationG}
    \langle q | \mathbb{G}_{n} | p \rangle  = \frac{\langle q|\mathbb{H}_{n} | p\rangle }{\langle q|\mathbb{H}_{n}|q\rangle-\langle p|\mathbb{H}_{n}|p\rangle}
\end{equation}
(here $p$ and $q$ label $A$-body configurations belonging to the $P$ and $Q$ spaces, respectively) and obtain the next iteration of $\mathbb{H}$ by the Baker-Campbell-Hausdorff expansion
\begin{equation}\label{eq:SuzukiIterationH}
    \mathbb{H}_{n+1} = \mathbb{H}_{n} + \left[ \mathbb{G}_{n},\mathbb{H}_{n}\right]
    + \tfrac{1}{2}\left[\mathbb{G}_{n}, \left[\mathbb{G}_{n},\mathbb{H}_{n}\right] \right]
    +\ldots
\end{equation}
The nested commutators are evaluated until the norm of the last nested commutator falls below $10^{-7}$.
The iteration in $n$ is performed until the norm of $\mathbb{G}_{n}$ falls below $10^{-6}$, at which point the $P$ and $Q$ spaces are decoupled.

\begin{figure}
    \centering
    \includegraphics[width=0.6\textwidth,]{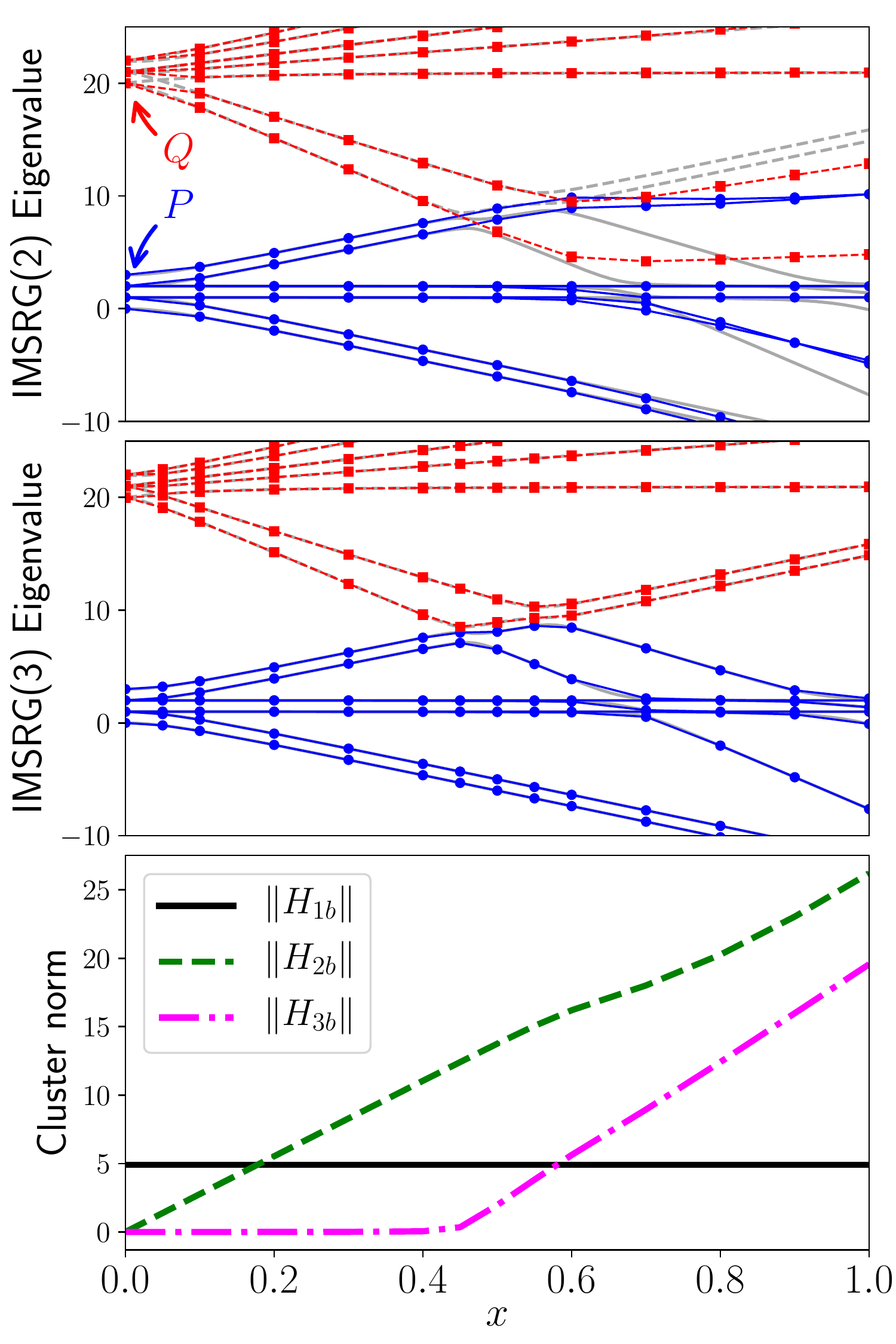}
    \caption{\emph{Top:} The eigenvalues after decoupling the $P$ and $Q$ spaces in the IMSRG(2) approximation (the gray lines indicate the exact values), \emph{Middle:} Eigenvalues after decoupling with the full IMSRG(3), \emph{Bottom:} The cluster decomposition of the $P$-space component of the transformed Hamiltonian, all as a function of the interaction strength parameter $x$.}
    \label{fig:Intruder3level}
\end{figure}

We also perform an IMSRG decoupling\footnote{Strictly speaking, because there is no core, and no normal ordering is performed, there is no ``medium'' and so this is really just an SRG calculation.}, using the flow equation formulation, directly on the Fock space representation of the Hamiltonian (\ref{eq:ToyFockH}).
We use the flow equation formulation because in the Magnus formulation for $x\gtrsim 0.5$, the Magnus operator $\Omega$ grows sufficiently large that the Baker-Campbell-Hausdorff expansion does not converge.
We perform an IMSRG(2) calculation, discarding 3-body terms, and we also perform an IMSRG(3) calculation, including the full three-body commutators, so the calculation is exact for the three-body problem.
The results are presented in Fig.~\ref{fig:Intruder3level} as a function of the perturbation strength parameter $x$.
The gray lines in panels (a) and (b) are the results of the decoupling in the $A$-body space following the iteration procedure in Eqs.~(\ref{eq:SuzukiIterationG}) and (\ref{eq:SuzukiIterationH}).
The purple and red lines and symbols correspond to the IMSRG(2) and IMSRG(3) solution in Fock-space.

We immediately make two observations.
First, as shown in the top panel of Fig.~\ref{fig:Intruder3level}, before the level crossings the IMSRG(2) eigenvalues are in agreement with the exact ones, while after the level crossings they go astay.
Second, as shown in the bottom panel of Fig.~\ref{fig:Intruder3level}, at the first level crossing near $x\approx 0.5$ the size of the three-body contribution to the Hamiltonian begins to grow rapidly.
These two observations are clearly connected.
We may understand the error of the IMSRG(2) calculation straightforwardly as the consequence of discarding important three-body terms.
Alternatively, one can compare the IMSRG(2) results with those presented in e.g. Fig. 1 of Ref.~\cite{Ellis1994}, where a two-level system with an avoided crossing was studied in perturbation theory.
At low orders of perturbation theory, the avoided crossing is not reproduced.
With increasing orders in the perturbation series, the levels slowly bend back to reproduce the avoided crossing.
One must go to very high orders (e.g. 50th in Ref~\cite{Ellis1994}) to fully recover the avoided crossing.

The generation of three-body terms can be understood by considering the flow equation for a $P$-space three-body contribution:
\begin{equation}\label{eq:Leaky3body}
    \frac{d}{ds} V_{111222} \sim \eta_{1123} V_{1322} - V_{1123}\eta_{1322} + \ldots
\end{equation}
The level crossing prevents the off-diagonal two-body matrix elements like $V_{1322}$ from being rapidly suppressed, and keeps the door open for strength to leak into the three-body sector via terms like (\ref{eq:Leaky3body}).

This investigation of a toy problem illustrates how the effects which lead to intruder configurations also cause problems with decoupling within the IMSRG framework, even without small denominators, and even if the flow equation converges.
Presumably, such effects will also arise in the SMCC framework because of the strong similarity between the two approaches.
This remains an open and important problem, and we hope that a better understanding of these effects will lead to a solution in the near future.

\section{\label{sec:OtherDev}Other developments}

\subsection{EFT for the shell model}
As we mentioned in the introduction, the success of shell model phenomenology strongly indicates that the shell model provides the relevant degrees of freedom for nuclear structure.
It is therefore tempting to formulate the shell model as an effective theory with some scheme for systematic improvement.
The main difficulty is in identifying a separation of scales which one can use to form an expansion.

20 years ago, Haxton et al. put forward the idea of formulating the shell model as an effective theory~\cite{Haxton2000,Haxton2001,McElvain:2016zb,McElvain:2019lt}.
The method presented in that work amounts to an effective theory for the NN interaction with the harmonic oscillator basis serving as the regulator, and the Bloch-Horowitz effective interaction (\ref{eq:HBlochHorowitz}) cast as an RG flow equation.
Similar ideas have been pursued in Refs.~\cite{Stetcu2007,Stetcu:2010ve,Yang:2016sf}, and by the Oak Ridge group~\cite{Binder2016,Bansal2018}.
This type of approach is appealing because it is formulated in the harmonic oscialltor basis and so yields an interaction well-suited to a number of popular many-body methods.

Using the harmonic oscillator basis as a regulator is conceptually distinct from formulating the standard shell model directly as an EFT.
In the latter case, one should use the shell model to define the degrees of freedom, and write down all possible terms in the Hamiltonian consistent with the relevant symmetries (parity, rotational invariance, charge, baryon number, etc.).
Then one should assign an importance to those terms based on some power counting.
A recent attempt more along these lines~\cite{Huth2018} employs a Weinberg chiral power counting in a shell-model basis, modified by the Galilean invariance breaking terms due to the presence of the core.
While a rigorous basis for the use of Weinberg's power counting is still lacking-- core excitations introduce a new scale, for instance, that might make it more natural to treat the Fermi momentum as a hard scale-- very encouraging order-by-order convergence was obtained.

Another possibility might be, a-la Landau-Migdal theory~\cite{MigdalBook}, to exploit the similarities between the valence shell model philosophy and Landau's Fermi liquid theory for infinite systems. In modern parlance, the latter can be viewed as an effective field theory for low-lying excitations (``quasi-particles'') in the vicinity of the Fermi surface. As with any EFT, the effective Hamiltonian of Landau's theory incorporates the underlying symmetries of the system, and the low-energy couplings-- the Landau parameters-- can either be fixed by experiment or calculated microscopically based on the underlying theory. Lending credence to the analogy with the valence shell model, Shankar, Polchinski, and others   have shown that Landau's theory can be understood as an IR fixed point of the RG as one integrates out modes away from the Fermi surface~\cite{Shankar:1993pf,Polchinski:1992ed,Chen:1995nj}. Intriguingly, their analysis shows that \emph{(i)} three- and higher-body quasi-particle interactions are irrelevant in the sense of the RG, which might shed light as to why phenomenological shell model interactions with 1- and 2-body terms are so effective, and \emph{(ii)} the natural small parameter is the ratio of excitation energy to the Fermi energy, which could provide guidance for formulating an appropriate powercounting for an EFT tailored to the shell model.

\subsection{Uncertainty quantification}
A major deficiency in essentially all shell model applications to date is the absence of quantified theoretical uncertainties.
This is no small thing; at a fundamental level, a theoretical prediction without some confidence interval cannot be falsified.
In particular, since experimental binding energies and excitation energies can routinely be measured at parts-per-million precision, whether or not theory and experiment are compatible is entirely dependent on the theoretical uncertainty.

In fact, there are various meanings one can assign to error bars.
Perhaps the most straightforward form of shell model uncertainty is the root-mean-squared deviation from experiment (see Fig.~\ref{fig:rmsBare}).
As mentioned above, the phenomenological USDB interaction \cite{Brown2006a} has a very small rms deviation of 130~keV throughout the $sd$ shell, and this may be interpreted in the following way.
There exist a large number of states which can be interpreted as shell model configurations mixed by the same universal residual interaction.
In the context of effective interaction theory these are the states which get mapped to the $P$ space.
We may then interpret levels where the deviation is much larger than 130~keV as not belonging to the $P$ space.
A clear illustration of this can be found in Ref.~\cite{Brown2006a}, where the ground state energies of $^{29,30}$Ne, $^{30,31}$Na, and $^{31,32}$Mg have conspicuously large deviations---a clear signal that these ``island of inversion'' nuclei have a different character than the others.

This notion of uncertainty has been recently made more quantitatively rigorous by Yoshida et al.~\cite{Yoshida2018}, who explored the various possible effective interactions in the $p$ shell and obtained marginal distributions for each parameter based on the deviation from experiment.
These distributions were then propagated to the calculated spectra, yielding an error bar indicating the range of values that could be obtained by a universal $p$-shell interaction.
Comparison to experiment then yields a well-defined quantification of how appropriate the naive shell model picture of a given state is.

A different type of uncertainty is sought by \emph{ab initio} approaches employing chiral interactions.
There, one should quantify and propagate three sources of uncertainty: \emph{(i)} contributions from truncated higher orders in the EFT expansion, \emph{(ii)} the experimental uncertainty in data used to fit the parameters of the EFT, and \emph{(iii)} uncertainty due to approximations made in solving the many-body problem.
In this case, because one has begun with the most general Lagrangian compatible with the symmetries of the Standard Model, the resulting theoretical error bar would then indicate compatibility with the Standard Model.
While we certainly expect such compatibility from the known nuclear energy levels, this becomes very important for testing extensions to the Standard Model through e.g. searches for neutinoless double beta decay~\cite{Engel2017} or dark matter~\cite{Menendez2012}.
While uncertainty quantification is feasible for quantum Monte Carlo or no-core shell model calculations, there is as yet no rigorous means for uncertainty quantification of \emph{ab initio} shell model effective interactions.
This is an important avenue of future work.

\subsection{Coupling to the Continuum}

Throughout the previous sections we have briefly touched upon the need to account for continuum effects in theoretical calculations. Of course, this will be especially relevant as we seek to understand the structure for increasingly neutron-rich nuclei. While a variety of approaches for coupling the shell model and other many-body methods to the continuum exist (see, e.g., \cite{Volya:2006hb,Papadimitriou:2015rw,Baroni:2013qf}, as well as the reviews \cite{Michel:2009oz,Myo:2014zp}), methods based on the Berggren basis \cite{Berggren:1968hb} appear to be most suitable in the context of VS-IMSRG and SMCC.

The Berggren basis adds resonant and scattering states to the single-particle basis from which many-body states are constructed. In valence-space configuration interaction calculations, one obtains what is colloquially known as the Gamow Shell Model (GSM), which entails the large-scale diagonalization of a complex symmetric Hamiltonian \cite{Michel:2009oz}. Applications of this method to weakly bound nuclei have been quite successful \cite{Betan:2012gd,Jaganathen:2014zp,Fossez:2015if,Fossez:2016kb,Fossez:2016wq,Fossez:2017lw,Fossez:2017ty,Jones:2017fr,Fossez:2018ec}, and there is a push to move from the commonly used phenomenological interactions to  fundamental ones \cite{Fossez:2018ec}. The Berggren basis has been used successfully in ground- and excited-state CC calculations (see, e.g., \cite{Hagen2012a,Hagen:2016xe}), hence the inclusion in VS-IMSRG and SMCC is technically straightforward. However, the proliferation of states due to the inclusion of the continuum aggravates the intruder-state problem discussed in Sec.~\ref{sec:Challenges}. If this issue can be solved, it would allow us to properly account for the continuum coupling in the derivation of effective interactions. For the time being, work is underway to at least treat the impact of the continuum on the dynamics of the valence particles via the GSM. 

\section{\label{sec:Conclusions}Concluding remarks}

In this work, we have reviewed the current state of efforts to derive effective interactions for the shell model from modern nuclear forces, with an emphasis on the impact of RG and EFT ideas on our understanding of the shell model itself.
We have summarized the popular approaches, and discussed their relations at a formal level. We emphasized the importance of three-body forces in eliminating the need for phenomenological adjustments and presented the ensemble normal ordering (ENO) approach to efficiently including three-body effects. We presented highlights from recent applications of \emph{ab initio} shell model calculations, and discussed the current challenges of low-lying collective excitations and intruder states.

Before concluding, we would like to make some remarks and clarify some common misconceptions about \emph{ab initio} valence space methods. 
\begin{summary}[SUMMARY POINTS]
\begin{itemize}
    \item \textbf{The shell model picture is inherently a low-momentum description of nuclear structure.}
    The basic assumption of the shell model is that nucleons are (almost) independent particles moving in a mean field potential, and that nuclear spectra can be explained by the mixing of a few valence configurations above an inert core via a residual interaction. The bound mean-field solution and weak (possibly even perturbative) residual interaction that are the foundation of this intuitive picture can only be obtained if low and high momenta are decoupled in the Hamiltonian \cite{Bogner2010,Tichai:2016vl,Hoppe:2017fm}. 
    
    Of course, nuclear observables --- energies, radii, transition rates --- must be independent of the resolution scale at which a theory operates. In principle, there is nothing that prevents one from microscopically constructing a valence shell model Hamiltonian starting from a high-resolution description, e.g., using an input interaction with a highly repulsive core. However, not only does such a choice make computations more difficult, but it complicates interpretations as the resulting shell model wave functions bear little resemblance to the exact ones, which contain sizable contributions from a vast number of configurations and defy a simple interpretation. In contrast, the exact wave functions of a low-resolution Hamiltonian at least qualitatively resemble those that come out of the shell model diagonalization, providing a simple and intuitive picture. 

    \item \textbf{Approaches such as the VS-IMSRG, SMCC, or the $\hat{Q}$-box resummation, are methods for solving the nuclear many-body problem, not new shell model interactions.}
    The approaches described in this article combine the derivation of effective interactions with a shell model diagonalization. This should be understood as an efficient alternative to a large-scale, full no-core configuration interaction (e.g., NCSM) calculation that would yield exact results for nuclear spectra, but is infeasible in most cases. 
    
    \item \textbf{A careful comparison with experimental data or theoretical results \emph{requires} that the method and underlying nuclear interaction be specified.}
    When comparing two phenomenological shell model calculations to experimental data, the interpretation is generally straightforward: the interaction that better reproduces the data is the better interaction.
    With \emph{ab initio} approaches, such an interpretation is no longer appropriate. 
    
    Disagreements between theory and experiment must be caused either by deficiencies in the underlying nuclear interactions, or the approximations employed in deriving the effective interaction (provided the shell model calculation is done without further approximations of its own). It is therefore \emph{crucially important} to specify both the method and the input interaction when comparisons with experimental data or other theoretical results are presented.  
    
    It should also be kept in mind that the major advantage of \emph{ab initio} approaches is the ability to systematically improve the precision of the theoretical result by lifting approximations, or improving the input nuclear interactions.

    \item \textbf{Three-body forces are inevitable and non-negligible in nuclear structure.} So long as we choose to use protons and neutrons as our active degrees of freedom---excluding explicit Deltas, anti-nucleons, etc.---there will be ``genuine'' (in the traditional language) many-body forces accounting for these integrated out degrees of freedom. So long as we wish to work in a valence space, there will be ``effective'' many-body forces accounting for excitations outside the valence space. The relative importance of these many-body forces will depend on the details of the implementation (scheme and scale). Indeed, an excellent description of a local region of the chart can be obtained with a purely two-body interaction, like USDB for the $sd$-shell \cite{Brown2006a}. But such an interaction will only work locally, and will need modification (e.g., scaling of matrix elements with mass --- again think of USDB) in order to be used over a wider range of nuclei. The theoretical evidence supports the expectation that three-body forces are the underlying source of such ad hoc modifications (see Sec.~\ref{sec:3body}).
    \item \textbf{The mass-dependence of modern effective interactions does not imply a loss of predictive power.} Effective interactions for different target nuclei are derived from the \emph{same} two- plus three-nucleon force, and there are \emph{no parameter refits or phenomenological modifications}. Again, the derivation of the effective interaction and subsequent shell model diagonalization are merely an efficient alternative to a full no-core configuration interaction calculation.    
    \item \textbf{All observables, not just the Hamiltonian, must be treated consistently to produce a true \emph{ab initio} result.} As discussed in section~\ref{sec:Comparison}, the effective interaction corresponds to a similarity transformation of the original Hamiltonian, and in order to perform a consistent calculation, \emph{all operators} must also be transformed. 
    
    As a simple example, consider computing the deuteron ground state by generating an effective interaction for the $0s$ shell. By construction, the energy obtained by a (trivial) diagonalization in the $0s$ shell would be identical to the result from a diagonalization in the full space with the bare Hamiltonian. Now, if one were to calculate the deuteron quadrupole moment using the bare $E2$ operator with the valence space wave functions (which would be pure $s$-wave), the result would be zero. Using a consistently transformed $E2$ operator will give the full-space result (for an illustration, see Ref.~\cite{Parzuchowski2017a}).
    
    One could argue that the bare operator gives the leading-order contribution to the effective operator (cf.~Eq.~\eqref{eq:MagnusOeff}). Trouble arises when the contribution of the leading term is suppressed by a symmetry, or missing degrees of freedom. Examples are the case for the $E2$ operator in the deuteron or in $sd$-shell oxygen isotopes, which only have neutrons in the valence space. Along these same lines, the use of phenomenological effective charges in conjunction with an \emph{ab initio} shell model interaction should be considered inappropriate. In general, it is difficult to make meaningful conclusions based on inconsistent calculations.
    \item \textbf{ The use of an inert core does not constitute an {\it ad hoc} approximation.}
    Calculations based on effective interction theory do not formally rely on an assumption that excitations out of the core are ``negligible''.
    Such excitations are accounted for by the effective interaction.
    Certainly, there will be states in the experimental spectrum that are not generated in the valence space calculation even with a perfect effective interaction---these belong to the excluded $Q$-space.
    However, those states that are generated will not be improved by, e.g., allowing core excitations described by a schematic interaction, as this would amount to double-counting.
    Instead, to include core excitations explicitly, one should re-define the $P$ and $Q$ spaces and derive a new effective interaction.
\end{itemize}
\end{summary}

The shell model has been the primary intellectual and computational framework for low energy nuclear structure for the past 70 years.
While the computational work has been largely phenomenological over that time, an enormous amount of knowledge and intuition has been developed.
At the same time, perhaps no problem in nuclear structure has so stubbornly resisted a satisfactory solution as the microscopic derivation of shell model interactions.
The general path has been more or less known for over half a century, but it is only recently that the available computational power, combined with a more systematic way of thinking about nuclear forces and the many-body problem, has allowed a direct connection between the shell model, the forces applicable to few-body scattering, and the underlying physics of the Standard Model.

We are not quite yet in the promised land.
While there are certainly many details remaining to be worked out (including those mentioned here), and several clear extensions to be made (continuum effects, reactions), there are still two major hills to climb: a fully consistent and satisfactory power-counting for the interaction, and a rigorous uncertainty quantification for our many-body methods.
We hope that progress can be made on these fronts in the near future, enabling a broadly applicable, quantitatively predictive theory of nuclear structure.



 \section*{ACKNOWLEDGMENTS}
We would like to thank B.~Alex Brown, Takayuki Miyagi, Titus Morris, Petr Navr\'atil and Zhonghao Sun for helpful discussions, and Gaute Hagen for providing coupled cluster results.
S.R.S is supported by the U.S. DOE under contract DE-FG02-97ER41014. H.H. acknowledges support by the National Science Foundation under Grant No.
PHY-1614130, as well as the U.S. Department of Energy, Office of Science, Office of Nuclear Physics under Grants No. DE-SC0017887 and DE-SC0018083 (NUCLEI SciDAC Collaboration). S.K.B. acknowledges support by the National Science Foundation under Grant No. PHY-1713901, as well as as the U.S. Department of Energy, Office of Science, Office of Nuclear Physics under Grant No. DE-SC0018083 (NUCLEI SciDAC Collaboration). 
\section*{APPENDIX}
\appendix
\section{\label{app:IMSRGFlow}IMSRG flow equations}
For reference, we now present the IMSRG(2)/VS-IMSRG(2) flow equations \cite{Tsukiyama:2011uq,Hergert2016,Hergert2017}. Ground-state and valence-space decoupling only differ by the choice of the generator $\eta$ (see Refs.~ \cite{Tsukiyama:2012fk,Hergert2017}).
   
The system of flow equations for the zero-, one-, and two-body parts of $H(s)$ result from evaluating
\begin{equation}
    \frac{dH}{ds} = [\eta(s),H(s)]
\end{equation}
with normal-ordered Fock-space operators that are truncated at the two-body level\footnote{These expressions can be easily adapted to evaluate the nested commutators appearing in the Magnus formulation of the IMSRG.}:
\begin{equation}
    H = E_0 + \sum_{ij} f_{ij}\{ a^{\dagger}_i a_j\} + \frac{1}{4}\sum_{ijkl}\Gamma_{ijkl}\{a^{\dagger}_ia^{\dagger}_ja_la_k \}
\end{equation}
\begin{equation}
    \eta = \sum_{ij}\eta_{ij}\{a^{\dagger}_ia_j \}+ \frac{1}{4}\sum_{ijkl}\eta_{ijkl}\{a^{\dagger}_ia^{\dagger}_ja_la_k \}.
\end{equation}
The flow equations are then
\begin{align}
     \frac{d E_0}{ds} &= \sum_{ab}n_a\bar{n}_b (\eta_{ab}f_{ba}-f_{ab}\eta_{ba}) 
    +\frac{1}{4} \sum_{abcd}n_an_b\bar{n}_c\bar{n}_d(\eta_{abcd}\Gamma_{cdab}-\Gamma_{abcd}\eta_{cdab})   \\
    \frac{df_{ij}}{ds} &= \sum_{a}(\eta_{ia}f_{aj}-f_{ia}\eta_{aj})
    +\sum_{ab}(n_a-n_b)(\eta_{ab}\Gamma_{biaj} -f_{ab}\eta_{biaj}) \nonumber \\
    &~~+\frac{1}{2}\sum_{abc}(n_an_b\bar{n}_c+\bar{n}_a\bar{n}_bn_c)\left(\eta_{ciab}\Gamma_{abcj}-\Gamma_{ciab}\eta_{abcj}\right)  \\
        \frac{d\Gamma_{ijkl}}{ds} &= \sum_{a} (1-P_{ij})(\eta_{ia}\Gamma_{ajkl}-f_{ia}\eta_{ajkl})-(1-P_{kl})(\eta_{ak}\Gamma_{ijal}-f_{ak}\eta_{ijal}) \nonumber \\
    &~~+ \frac{1}{2}\sum_{ab}(n_an_b-\bar{n}_a\bar{n}_b)(\eta_{ijab}\Gamma_{abkl}-\Gamma_{ijab}\eta_{abkl}) \nonumber \\
    &~~- \sum_{ab}(n_a-n_b) (1-P_{ij})(1-P_{kl})\eta_{bjal}\Gamma_{aibk}
\end{align}
where $P_{ij}$ exchanges indices $i$ and $j$, $n_a$ is the occupation of orbit $a$ and ${\bar{n}_a\equiv1-n_a}$.

\section{\label{app:Perturbative}Canonical perturbation theory to second order}
We partition the Hamiltonian into a zeroth-order piece and a perturbation,
\begin{equation}
    H = H_0 + xV\,,
\end{equation}
and we consider a perturbative expansion in powers of the dimensionless order parameter $x$, where in the end we will take $x=1$.
We further distinguish between ``diagonal'' and ``off-diagonal'' components, $V=V^{d}+V^{od}$, where ``off-diagonal'' generically means the terms we wish to suppress by the transformation
\begin{equation}
    H_{\mathrm{eff}} = e^{\gen}He^{-\gen}\,.
\end{equation}
As in the discussion leading to Eq.~(\ref{eq:GperturbativeSeries}), we use the superoperator notation to express a commutator with $H_0$ in terms of an energy denominator $\Delta$.
Through second order in $x$, we obtain for $\gen$
\begin{equation}
    \begin{aligned}
    \gen^{[]1]} &= \frac{V^{od}}{\Delta} \\
    \gen^{[2]} &= \left[\gen^{[1]},V^{d} \right]^{od}/\Delta + \tfrac{1}{2} \left[\gen^{[1]},V^{od} \right]^{od}/\Delta
    \end{aligned}
\end{equation}
The second term in $\gen^{[2]}$ will vanish for $A$-body Hilbert space formulations, but not in general for a Fock space formulation.
This is related to the different meanings of ``off-diagonal'' in the two formulations.
The transformed Hamiltonian through second order is
\begin{equation}
\begin{aligned}
    H_{\mathrm{eff}}^{[0]} &= H_0 \\
    H_{\mathrm{eff}}^{[1]} &= V^{d} \\
    H_{\mathrm{eff}}^{[2]} &= \left[ \gen^{[1]},V^{d} \right]^{d} + \tfrac{1}{2}\left[\gen^{[1]},V^{od} \right]^{d}
    \end{aligned}
\end{equation}

\section{\label{app:PerturbativeMagnus}Integration of Magnus flow equation to second order}
Here, we integrate the flow equation (\ref{eq:MagnusFlow}) explicitly to second order.
As in appendix~\ref{app:Perturbative}, we partition the Hamiltonian into a zero order piece $H_0$, and a perturbation $V$, and we split up the perturbation into ``diagonal'' and ``off-diagonal'' pieces.
We use the White generator, which we write as
\begin{equation}
    \eta(s) \equiv \frac{H^{od}(s)}{\Delta}
\end{equation}
using the super-operator notation introduced in Sec.~\ref{sec:imsrg}.
To first order in $x$, the flow equation for $\Omega$ is
\begin{equation}
    \frac{d\Omega^{[1]}}{ds} = \eta^{[1]}(s) = \frac{H^{[1]od}(s)}{\Delta}
    = \frac{V^{od}}{\Delta} + [\Omega^{[1]}(s),H_0]^{od}/\Delta = \frac{V^{od}}{\Delta}-\Omega^{[1]}(s).
\end{equation}
A differential equation for $\Omega^{[2]}(s)$ may be obtained in a similar manner. The solutions given the initial condition $\Omega(0)=0$ are
\begin{equation}
    \begin{aligned}
    \Omega^{[1]}(s) &= (1-e^{-s})\frac{V^{od}}{\Delta} \\
    \Omega^{[2]}(s) &= (1-e^{-s}-se^{-s})\left[\frac{V^{od}}{\Delta},V^{d} \right]^{od}/\Delta + \tfrac{1}{2}(1-e^{-2s})\left[\frac{V^{od}}{\Delta},V^{od} \right]^{od}/\Delta.
    \end{aligned}
\end{equation}
The flowing Hamiltonian through second order is
\begin{equation}
    \begin{aligned}
    H^{[0]} &= H_0 \\
    H^{[1]} &= V^{d}+e^{-s}V^{od} \\
    H^{[2]} &= (1-e^{-s})\left[\frac{V^{od}}{\Delta},V^{d} \right]^{d}
    +\tfrac{1}{2}(1-e^{-2s})\left[\frac{V^{od}}{\Delta}, V^{od} \right]^{d} \\
    &~~+ se^{-s}\left[\frac{V^{od}}{\Delta},V^{d} \right]^{od}
    + e^{-s}(1-e^{-s})\left[\frac{V^{od}}{\Delta},V^{od} \right]^{od}.
    \end{aligned}
\end{equation}
We see that at first order, the off-diagonal part of the perturbation is exponentially suppressed.
At $s=0$, the second order piece is by definition zero.
As $s$ increases, we initially induce both diagonal and off-diagonal second-order terms, and eventually, the induced second-order terms are suppressed exponentially, leaving a purely diagonal the second-order correction.

Taking the limit $s\rightarrow\infty$, we obtain the same generator and effective Hamiltonian as the canonical perturbation theory in appendix~\ref{app:Perturbative}.
Note that this equivalence requires the same definition of ``off-diagonal'', and that the results will differ at higher orders if we include the commutator terms on the right hand side of the Magnus flow equation in (\ref{eq:MagnusFlow}).

\bibliographystyle{ar-style5}
\bibliography{references}

\end{document}